\documentclass[a4paper, 11pt]{article}
\usepackage{fullpage}
\usepackage{indentfirst}
\usepackage[titletoc,toc,title]{appendix}
\usepackage{amssymb, amsmath, amsthm, mathdots}
\numberwithin{equation}{section}
\usepackage{bm}
\usepackage{float}
\usepackage{graphicx}
\usepackage{adjustbox}
\usepackage[font=small,labelfont=bf]{caption}
\usepackage{hyperref}
\hypersetup{
    colorlinks=true,
    linkcolor=blue,
    urlcolor=blue,
}

\usepackage{ulem}
\usepackage{subcaption}
\usepackage{tabu}
\usepackage{slashed}
\usepackage{mathrsfs}
\usepackage{authblk}

\usepackage[compat=1.1.0]{tikz-feynman}

\usepackage{multirow}

\begin{document}
\title{\large \textbf{Probing B-Anomalies via Dimuon Tails at a Future Collider}}
\author[1]{Bradley Garland}
\author[1]{Sebastian J\"ager}
\author[2,3]{Charanjit K. Khosa}
\author[4]{Sandra Kvedarait\.e\thanks{Email: b.garland@sussex.ac.uk, s.jaeger@sussex.ac.uk, charanjit.kaur@bristol.ac.uk, kvedarsa@ucmail.uc.edu}}
\affil[1]{\small Department of Physics and Astronomy, University of Sussex, Brighton BN1 9QH, UK.}

\affil[2]{\small H.H. Wills Physics Laboratory, University of Bristol, Tyndall Avenue, Bristol BS8 1TL, UK.}

\affil[3]{\small Dipartimento di Fisica, Universit\`a di Genova and INFN, Sezione di Genova, Via Dodecaneso 33, 16146, Italy.}

\affil[4]{\small Department of Physics, University of Cincinnati, 
Cincinnati, OH 45221, USA.}

\date{}

\maketitle

\begin{abstract}

{We investigate the sensitivity of future proton-proton colliders to a contact interaction of the form $1/\Lambda^2 (\bar b_L \gamma_\mu s_L)(\bar \mu_L \gamma^\mu \mu_L)$ as indicated by the long-standing rare $B$-decay anomalies. We include NLO QCD and electroweak effects and employ an optimized binning scheme, and carefully validate our background calculation against ATLAS and CMS data.
We find that the FCC-hh with $40$ ab$^{-1}$ of luminosity is able to exclude
scales $\Lambda$ up to 26 TeV at $95 \%$ CL, and discover
$\Lambda$ up to 20 TeV. While this is not quite enough to exclude
or discover the current best-fit value of $39$ TeV, this can
in principle be achieved with more luminosity and/or higher energy, as we study quantitatively.
Our analysis is conservative in that it assumes
\textit{only} a $\bar b s \mu \mu $ contact interaction.}

\end{abstract}

\section{Introduction} \label{Intro}
In recent years, a number of measurements of rare semileptonic $b \to s \ell^+ \ell^-$
transitions have shown significant discrepancies with Standard Model (SM) predictions (see \cite{Cerri:2018ypt,London:2021lfn}
for reviews).
A particularly clear picture is provided by measurements of the lepton-flavour-universality (LFU) ratios \cite{Hiller:2003js}
\begin{equation}
    R_{K^{(*)}}=\frac{\text{BR}(B\to K^{(*)}\bar{\mu}\mu)}{\text{BR}(B\to K^{(*)}\bar{e}e)},
\end{equation}
together with the purely leptonic branching fraction $BR(B_s \to \mu^+ \mu^-)$, for which theoretical
uncertainties are currently negligible compared to experimental statistical ones (Table \ref{tab:bphysdata}).
This ``clean" dataset on its own is currently at variance with the SM at $4.2 \sigma$ significance (\cite{Geng:2021nhg},
October 2021 update). It is the goal of this paper to study the implications of the rare $B$ decay anomalies for future $pp$ colliders,
in a manner as model-independent and conservative as possible. We will be focusing on the inclusive dimuon signal
$pp \to \mu^+ \mu^-$.
\begin{table}[t]
	\centering
	{\tabulinesep=1.0mm
		\begin{tabu}{||c|c|c|c||}
			\hline
			Observable & SM & Measurement & Experiment \\
			\hline
			$BR(B_s \to \mu^+ \mu^-)$ & $(3.63 \pm 0.13) \times 10^{-9}$  & $(2.8 \pm 0.3) \cdot 10^{-9}$ & \parbox{4cm}{average \cite{Geng:2021nhg} of ATLAS \cite{ATLAS:2018cur}, CMS \cite{CMS:2019bbr} and LHCb \cite{LHCb:2021awg}} \\
			\hline
			$R_K[1.1,6]$        & \multirow{2}{*}{ $1.0004^{+0.0008}_{-0.0007}$ } & $0.85 \pm 0.04$ & LHCb \cite{Aaij:2021vac} \\
			$R_K[1,6]$          & & $1.03 \pm 0.28$ & Belle \cite{BELLE:2019xld} \\
			\hline
			\multirow{2}{*}{ $R_{K^*}[0.045,1.1]$ } &  \multirow{2}{*}{ $0.920^{+0.007}_{-0.006}$} & $0.66 \pm 0.11$ & LHCb \cite{Aaij:2017vbb} \\
			                    & & $0.52 \pm 0.37$ & Belle \cite{Belle:2019oag} \\
			\hline
			\multirow{2}{*}{$R_{K^*}[1.1,6]$ }  & \multirow{2}{*}{ $0.996 \pm 0.002$ } & $0.69 \pm 0.12$ & LHCb \cite{Aaij:2017vbb} \\
                                & & $0.96 \pm 0.45$ & Belle \cite{Belle:2019oag} \\
            
            \hline
 	\end{tabu}}
 	
	\caption{Selection of $b \to s \ell^+ \ell^-$ data. The notation $R_{K^{(*)}}[a,b]$ refers to a dilepton mass bin
	  $a~ \mbox{GeV}^2 < q^2 < b~ \mbox{GeV}^2$. Theory predictions for $R_K^{(*)}$ are taken from \cite{Geng:2017svp} and
	for $BR(B_s \to \mu^+ \mu^-)$ from \cite{Beneke:2019slt}. Predictions for $R_{K^{(*)}}$ do not include effects of electromagnetic
	radiation. These are on the order of a few percent, and corrected for by experiments, with the residual error currently
	negligible compared to experimental statistical uncertainties \cite{Bordone:2016gaq,Isidori:2020acz}.
	\label{tab:bphysdata} }
\end{table}
The rare $B$-decay data can be model-independently described by new beyond-SM (BSM) four-fermion contact interactions
in the low-energy effective weak Hamiltonian involving muons only,
\begin{eqnarray}
\mathcal{L}^{\rm BSM}_{\text{eff}} = - \mathcal{H}_{\text{eff}}^{\rm BSM} &=& \frac{4 G_F}{\sqrt{2}} V_{ts} V^*_{tb} \frac{e^2}{16\pi^2}
  \left\{ C_9^{\rm NP} \left( \bar{b}_L\gamma_\mu s_L\right)\left(\bar{\mu}\gamma^\mu \mu \right)
       + C_{10}^{\rm NP} \left( \bar{b}_L\gamma_\mu s_L\right)\left(\bar{\mu} \gamma^\mu \gamma^5 \mu \right) + \dots
  \right\} \nonumber \\
  &\equiv& \frac{1}{\Lambda_{\rm LL}^2 } \left( \bar{b}_L\gamma_\mu s_L\right)\left(\bar{\mu}_L\gamma^\mu \mu_L \right)
  + \frac{1}{\Lambda_{\rm LR}^2 } \left( \bar{b}_L\gamma_\mu s_L\right)\left(\bar{\mu}_R\gamma^\mu \mu_R \right) + \dots .
\end{eqnarray}
In Eq.~\ref{eq:Lminimal} the ellipses refer to additional operators with different Dirac structures which are not favoured by the data.
In fact, an excellent description of all $b \to s \ell^+ \ell^-$ data is obtained by a purely left-handed interaction,
$C_9^{\rm NP} = - C_{10}^{\rm NP} \equiv C_L$ \cite{Geng:2021nhg,Altmannshofer:2021qrr,Alguero:2021anc}. In \cite{Geng:2021nhg}, a fit to the ``clean'' data alone resulted in the $1\sigma$ range $C_L = -0.40^{+0.09}_{-0.08}$; a fit also including angular data
in $B \to K^* \ell^+ \ell^-$ gave a very similar value
$C_L = -0.39^{+0.08}_{-0.09}$.
Fitting the global dataset jointly
to $C_9$ and $C_{10}$ and profiling over the combination $C_R = (C_9 + C_{10})/2$ gave instead
$C_L = -0.43^{+0.10}_{-0.10}$. Refs.~\cite{Altmannshofer:2021qrr,Alguero:2021anc} find very similar $C_L$ ranges.
In other words, the value of $C_L$ is well determined
and rather robust against the choice of dataset and whether a coupling to right-handed muons
is allowed or not. In summary, the rare $B$ decay dataset points to the presence of a left-handed contact interaction
\begin{equation}  \label{eq:Lminimal}
  \mathcal{L}_{\rm eff} \supset \frac{1}{\Lambda^2} \mathcal{O}_{LL} ,
\end{equation}
where  $\mathcal{O}_{LL} = \left(\bar{b}_L\gamma_\mu s_L\right)\left(\bar{\mu}_L\gamma^\mu \mu_L\right)$
and
\begin{equation}
  \Lambda = (39 \pm 4) \mbox{TeV},
  \label{Lambda 39}
\end{equation}
and allows for the possible presence of an additional coupling to right-handed muons.

The interaction Eq.~\ref{eq:Lminimal} provides a minimal description of all rare $B$-decay anomalies and in the following we will
assume it is the only BSM interaction present. While it is possible that further interactions are present in the effective Hamiltonian, they do
not improve the description of the data further \cite{Alguero:2021anc,Hurth:2021nsi}. Importantly, they will tend to increase
the signal in $pp \to \mu^+ \mu^-$ further, either because the $C_L$ value in the fit remains the same or because  a comparable or larger Wilson coefficient appears instead. For example, a
model employing operators with electrons instead of muons still fits the data significantly better than the SM (though not as
well as Eq.~\ref{eq:Lminimal}, as it does not contribute to the $B \to K^* \mu^+ \mu^-$ angular observables, $B_s \to \mu^+ \mu^-$, etc.). In addition, the fitted interaction strength is close in magnitude to the muonic case considered here \cite{Altmannshofer:2021qrr,Hurth:2021nsi}
and would generate a comparable $pp \to e^+ e^-$ signal at a collider. Altogether, our assumption of a minimal interaction
Eq.~\ref{eq:Lminimal} is a conservative one, in that it will lead to conservative results for the sensitivity at a future collider.

\begin{figure}[t]
\centering
\includegraphics[scale=0.45]{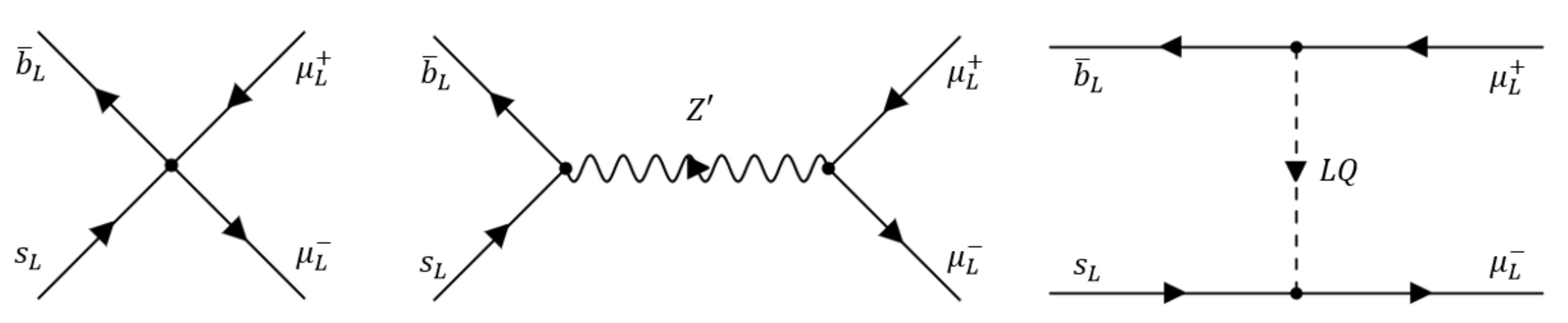}
\caption{Feynman diagrams for a purely left-handed effective $bs\mu\mu$ operator (left) and the two possible tree-level mediators of contact interactions of this type; $Z'$ (middle) and leptoquark (right).}
\label{CI_Zp_LQ}
\end{figure}

The minimal effective interaction Eq.~\ref{eq:Lminimal} can be interpreted as a low-energy effective description
of an extension of the SM by new degrees of freedom. A plethora of such simplified models and possible UV completions
has been constructed. The simplest and most studied involve tree-level exchange of a neutral vector ($Z'$ ) or
a (spin-1 or spin-0) leptoquark (see Fig.~\ref{CI_Zp_LQ}).
Such mediators can be directly searched for at the LHC. For example, a $Z'$ mediator would
cause a bump in the inclusive dimuon signal \cite{ATLAS:2019erb, CMS:2019kaf}. The sensitivity of the LHC and future colliders to tree-level mediators that can underpin the rare $B$-decay anomalies has been studied for the $Z'$ case in \cite{Allanach:2017bta,Allanach:2018odd,Allanach:2019mfl,Allanach:2021gmj} and for leptoquarks in \cite{Bar-Shalom:2018ure,Allanach:2019zfr, Hiller:2021pul} (see also \cite{Altmannshofer:2020axr,BhupalDev:2021ipu}); for the collider signatures of other models of the $B$-physics anomalies
see \cite{Cerri:2018ypt,London:2021lfn} and references therein. Such searches are necessarily model-dependent, and,
even at a 100 TeV machine, do not cover the entire parameter space. For example, the mediator could simply be too heavy: perturbative
unitarity alone would allow, for $\Lambda = 39$ TeV, mediators as heavy as 105 TeV \cite{DiLuzio:2017chi}.

Another, more model-independent approach is to look for the effects of heavy new physics in the ``low-energy'' tails of
the dilepton invariant mass and other distributions within the context of effective field theory \cite{Greljo:2017vvb, Alioli:2017nzr}.  Ref. \cite{Greljo:2017vvb} employed the ATLAS inclusive dimuon search
\cite{Aaboud:2017buh} to obtain a 95\% confidence level (CL) bound of $\Lambda > 2.5$ TeV, and a projected bound
of $\Lambda > 4.1$ TeV at the HL-LHC for $L = 3000$ fb$^{-1}$ at $\sqrt{s} = 13$ TeV.
Ref.~\cite{Afik:2018nlr} (see also \cite{Afik:2019htr}) investigated final states including a dimuon and a $b$-jet, obtaining a projected bound $\Lambda > 7.7$ TeV for $L = 3000$ fb$^{-1}$ at $\sqrt{s} = 13$ TeV. 
More recent LHC dimuon searches and the ATLAS search have been published in \cite{Aaboud:2017buh, CMS:2021ctt, ATLAS:2021mla}. Ref.~\cite{Huang:2021biu}
considered the prospects at a muon collider, with encouraging results.

With the LHC and HL-LHC falling short of being able to detect Eq.~\ref{eq:Lminimal} for $\Lambda \sim 40$ TeV, the question is, to what extent does increased collider centre of mass (c.o.m) energy $\sqrt{s}$ improve the sensitivity to $bs\mu\mu$ contact interactions?
In the present paper, we investigate this using the tails of the inclusive dimuon invariant mass $m_{\bar{\mu}\mu}$ distributions. We take the proposed $\sqrt{s}=100$~TeV FCC-hh as a baseline whilst also providing updated limits at the $\sqrt{s}=14$~TeV HL-LHC. We derive both 95\% CL exclusion limits and expected $5\sigma$ discovery sensitivities. In doing so, we include the NLO QCD and EW corrections to our EFT signal. We also consider the validity of our limits from the perspective of partial-wave unitarity and the EFT expansion.

The remainder of this paper is organised as follows. In Sec. \ref{SMEFT} we describe our basis and notations for the $bs\mu\mu$ contact interaction within the SMEFT framework. Sec. \ref{set-up} is devoted to our analysis set-up where we discuss event simulation, statistical methods, event selection as well as commenting on unitarity constraints and the validity of EFT approach. Our findings are detailed in Sec. \ref{results} where we give limits at $\sqrt{s}=14$ TeV, $\sqrt{s}=100$ TeV and beyond. We conclude in Sec. \ref{conclusions}.

\section{Standard Model Effective Field Theory}\label{SMEFT}

At energies above the electroweak scale, but below the mass scale of the underlying UV physics, 
the interaction Eq.~\ref{eq:Lminimal} is appropriately described within the framework of the SMEFT.
The SMEFT Lagrangian is an expansion in local operators of increasing mass dimension,
\begin{equation}
     \mathcal{L}^\text{SMEFT}=\mathcal{L}^{\text{SM}}+\sum_ic^{(6)}_i\mathcal{O}^{(6)}_i+\sum_jc^{(8)}_j\mathcal{O}^{(8)}_j+\cdots
     \label{SMEFT exp}
\end{equation}
constructed out of the SM fields, where $\mathcal{O}^{(D)}_i$ is a gauge invariant operator of dimension $D$ and $c_i^{(D)}$ is its corresponding Wilson coefficient. 
For later convenience, we note that the Wilson coefficients $c_i^{(D)}$ can be expressed as the ratio of a dimensionless
coupling $g^{(D)}_i$ and an arbitrary mass scale $\hat{\Lambda}$ such that 
\begin{equation}
    c_i^{(D)}=\frac{g_i^{(D)}}{\hat{\Lambda}^{D-4}}.
\end{equation}

The predictivity of the SMEFT rests on the higher-dimensional operators in Eq. \ref{SMEFT exp} being suppressed such that the expansion
can be truncated at some dimension such that the remainder can be neglected;
in our work we will truncate at the leading BSM dimension $6$. A separate
requirement is that the truncated amplitude satisfies $S$-matrix unitarity constraints. These questions will be discussed in more detail in Sec. \ref{EFT Validity}. 

The relevant dimension-$6$ gauge invariant operators that contribute to purely left-handed four-fermion interactions can be found in \cite{Buchmuller:1985jz,Grzadkowski:2010es}. Ignoring flavour indices, there are two dimension-$6$ semileptonic operators
with a $(\bar{\bm{L}}\bm{L})(\bar{\bm{L}}\bm{L})$ chirality structure. In the Warsaw basis \cite{Grzadkowski:2010es}, and including flavour indices,
we have
\begin{eqnarray}
    \mathcal{L}^\text{SMEFT} &\supset& \frac{c^{(3)}_{Q_{ij}L_{\mu\mu}}}{\hat{\Lambda}^2}(\bar{Q}_j\gamma_\rho\sigma^aQ_i)(\bar{L}_2\gamma^\rho\sigma_aL_2)+
    \frac{c^{(1)}_{Q_{ij}L_{\mu\mu}}}{\hat{\Lambda}^2}(\bar{Q}_j\gamma_\rho Q_i)(\bar{L}_2\gamma^\rho L_2)
\nonumber \\
    &\equiv&  \frac{c^{(3)}_{Q_{ij}L_{\mu\mu}}}{\hat{\Lambda}^2} {\cal O}^{(3)}_{ij}
              + \frac{c^{(1)}_{Q_{ij}L_{\mu\mu}}}{\hat{\Lambda}^2} {\cal O}^{(1)}_{ij} ,
\label{L smeft warsaw basis}
\end{eqnarray}
where $i,j$ run over the generations of quarks and leptons, $Q_i=(V^*_{ji}u^j_L,d^i_L)^{\text{T}}$, $L_i=(\nu^i_L,l^i_L)^{\text{T}}$ and $\sigma^a$ are the Pauli matrices. 

It is convenient to express Eq. \ref{L smeft warsaw basis} in a different operator basis,
\begin{equation}
  {\cal O}^\pm_{ij} = \frac{1}{2} \left( {\cal O}^{(1)}_{ij} \pm {\cal O}^{(3)}_{ij}  \right), 
  \qquad 
  C^{\pm}_{ij} = \frac{c^{(1)}_{Q_{ij}L_{\mu\mu}}\pm c^{(3)}_{Q_{ij}L_{\mu\mu}} }{\hat{\Lambda}^2} ,
\end{equation}
such that 
\begin{equation}
    \begin{split}
        \mathcal{L}^\text{SMEFT}\supset\:&C^+_{ij}(\bar{d}^{\;j}_L\gamma_\rho \;d^i_L)\left(\bar{\mu}_L\gamma^\rho \mu_L\right)\:+
        C^-_{ij}(\bar{d}^{\;j}_L\gamma_\rho \;d^i_L)\left(\bar{\nu}_\mu\gamma^\rho \nu_\mu\right)\:+\\
        &\sum_{k,l}V^*_{ki}C^+_{ij}V_{lj}(\bar{u}^{\;l}_L\gamma_\rho u^k_L)(\bar{\nu}_\mu \gamma^\rho \nu_\mu)\:+
        \sum_{k,l}V^*_{ki}C^-_{ij}V_{lj}(\bar{u}^{\;l}_L\gamma_\rho u^k_L)(\bar{\mu}_L\gamma^\rho \mu_L) .
        \label{L smeft pm basis}
    \end{split}
\end{equation}
The benefit of working in the $(C^+_{ij},C^-_{ij})$ basis is that the ${\cal O}^+$ operators couple muons to down-type quarks only, and
the ${\cal O}^-$ operators to up-type quarks only. In particular, the $b \to s \mu^+ \mu^-$ transitions are governed by
a single Wilson coefficient $C^+_{sb}$.

In the present paper, we consider the minimal scenario in which our EFT signal is given by Eq.~\ref{eq:Lminimal}, which
in the SMEFT corresponds to
\begin{equation}
C^+=\left(\begin{matrix}
0 & 0 & 0 \\
0 & 0 & C^+_{sb} \\
0 & {C^+_{sb}}^* & 0 \\  
\end{matrix}\right)
\quad\quad\quad
C^-=\left(\begin{matrix}
0 & 0 & 0 \\
0 & 0 & 0 \\
0 & 0 & 0 \\  
\end{matrix}\right).
\end{equation}
We then have 
\begin{equation}
        \mathcal{L}^\text{SMEFT}\supset\:C^+_{sb}(\bar{b}_L\gamma_\rho \;s_L)\left(\bar{\mu}_L\gamma^\rho \mu_L\right)\:+
        \sum_{k,l}V^*_{ks}C^+_{sb}V_{lb}(\bar{u}^{\;l}_L\gamma_\rho u^k_L)(\bar{\nu}_\mu \gamma^\rho \nu_\mu)\:+\text{h.c.} ,
        \label{L smeft Csb}
\end{equation}
and by comparing to Eq.~\ref{eq:Lminimal} we have
\begin{equation}
    C^+_{sb}=\frac{1}{\Lambda^2}.
    \label{C_bs}
\end{equation}
We stress that this is a conservative assumption: in a UV model, there will typically be additional nonzero entries in the matrices
$C^+$ and $C^-$, such as for example a $\bar b b \bar \mu \mu$ coupling. In the $p p \to \mu^+ \mu^-$ process we will consider, such additional
couplings will generate additional contributions to the signal, enhancing the sensitivity to the interaction. The signal we will consider
is the minimal one required by the rare $B$-decay anomalies, and therefore the exclusion and discovery reach we will find
are conservative and universally applicable, subject only to EFT validity/applicability requirements.

\section{Analysis Set-up\label{set-up}}
In this section we describe the set-up of our analysis. We start by discussing the methods used to obtain the contributions of the SM background and EFT signal to the dimuon invariant mass $m_{\bar{\mu}\mu}$ spectrum. We then describe the statistical methods used to assess the sensitivity of a future collider with c.m. energy $\sqrt{s}$ to the EFT signal. In our analysis, we are principally concerned with the $\sqrt{s}=14$ TeV HL-LHC and the proposed $\sqrt{s}=100$ TeV FCC-hh. Finally, we detail the event selection and binning schemes used at different c.o.m energies before discussing the constraints on our sensitivity calculations arising from tree-level unitarity and the validity of the EFT expansion.

\subsection{Event Simulation} \label{Event Simulation}

In this subsection we detail how we obtain the dimuon invariant mass $m_{\bar{\mu}\mu}$ distributions used in our sensitivity calculations. We first describe how we model the SM background, validating our modeling of the SM background by comparing it to the latest ATLAS \cite{Aad:2020otl} and CMS \cite{CMS:2021ctt} searches for non-resonant phenomena in high-mass dilepton final states at $\sqrt{s}=13$ TeV. This validation allows us to proceed with obtaining the background at the future collider. We then discuss the modeling of the EFT signal where we include the next to leading order effects in both the QCD and EW couplings.  

The relevant SM processes that contribute to $pp\to \mu^+\mu^-$ are Drell–Yan (DY) via $Z/\gamma^*$ exchange, diboson ($ZZ$, $WZ$ and $WW$) production and top-quark production ($\bar{t}t$ \& $tW$). The dominant contribution is DY which, at $\sqrt{s}=13$ TeV, makes up $\sim80\%$ of all events in the region where $m_{\bar{\mu}\mu}>600$ GeV and $\sim90\%$ in the region where $m_{\bar{\mu}\mu}>1300$ GeV. To model the DY process we use \textsc{MadGraph5\textunderscore aMC$@$NLO} \cite{Alwall:2014hca, Frederix:2018nkq}. Specifically, we perform combined NLO QCD and EW fixed order calculations to obtain the cross section of the DY process in chosen bins in the dimuon invariant mass. To model the top quark and diboson background we again use \textsc{MadGraph5\textunderscore aMC$@$NLO}. However, since these effects are subleading, we only consider the LO contributions. 

Our calculations are performed using the
\textsc{NNPDF31\_nlo\_as\_0118\_luxqed} PDF set \cite{NNPDF:2017mvq} via \textsc{LHAPDF6} \cite{Buckley:2014ana} in the 5 flavour scheme\footnote{It is noted that we use the 4 flavour scheme when modeling the background arising from top quark production.}. In regards to the EW calculations to the EFT signal and Drell-Yan process, we use the $G_{\mu}$-scheme as an input scheme. We also consider dressed muons in the final state. Here collinear muon-photon pairs arising from real photon emission are recombined if they lie within a cone of radius $R_{\text{rec}}=0.1$ around the muon. This ensures IR insensitivity and reliability of fixed-order results.

The number of events in a given bin $N_{\text{bin}}$ of the invariant mass distribution is calculated using 
\begin{equation}
    N_{\text{bin}}=\varepsilon^2 L\;\sigma_{\text{bin}}.
    \label{Nbin}
\end{equation}
Here $\sigma_{\text{bin}}$ is the cross section of the given process in a given bin (this applies to the EFT signal also), $L$ is the total integrated luminosity and $\varepsilon$ is the combined muon identification and reconstruction efficiency of the given $pp$-collider considered. A detector's ability to identify and  reconstruct muons produced in collisions can significantly affect our sensitivity calculations and such detector effects are especially important when comparing to experimental analyses. We address the problem of muon identification and reconstruction by introducing $\varepsilon$ in Eq. \ref{Nbin}. When comparing to experimental analysis at $\sqrt{s}=13$ TeV or preforming sensitivity calculations at the $\sqrt{s}=14$ TeV HL-LHC, we take $\varepsilon=0.75$ ($\varepsilon=0.965$) for the ATLAS \cite{ATLAS:2019erb, ATLAS:2020auj} (CMS \cite{CMS:2021ctt}) detector. For a future collider with $\sqrt{s}=100$ TeV we use a muon identification efficiency $\varepsilon=0.95$ in line with Ref. \cite{Jamin:2019mqx}.

\begin{figure}[t]
\centering
\includegraphics[scale=0.51]{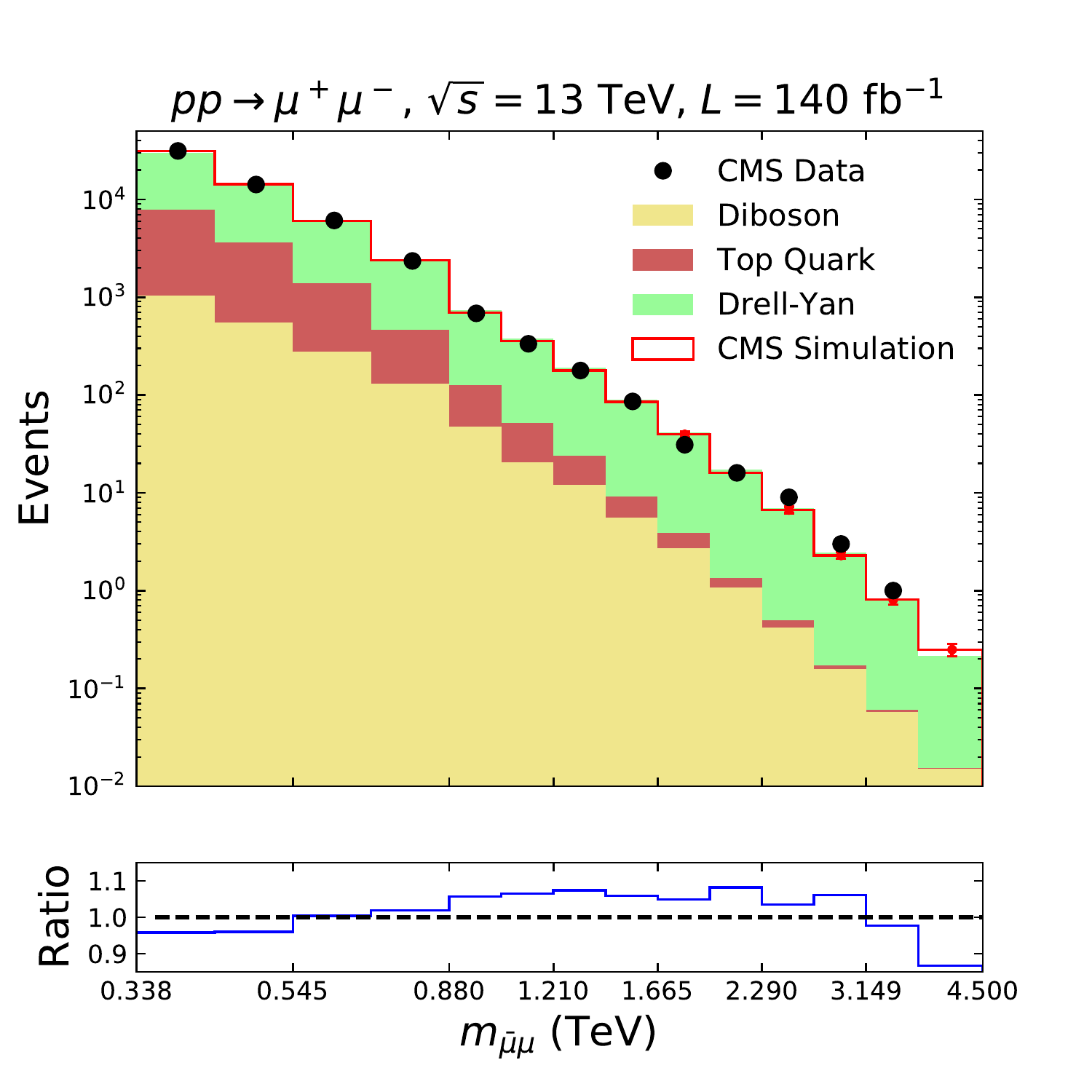}
\includegraphics[scale=0.51]{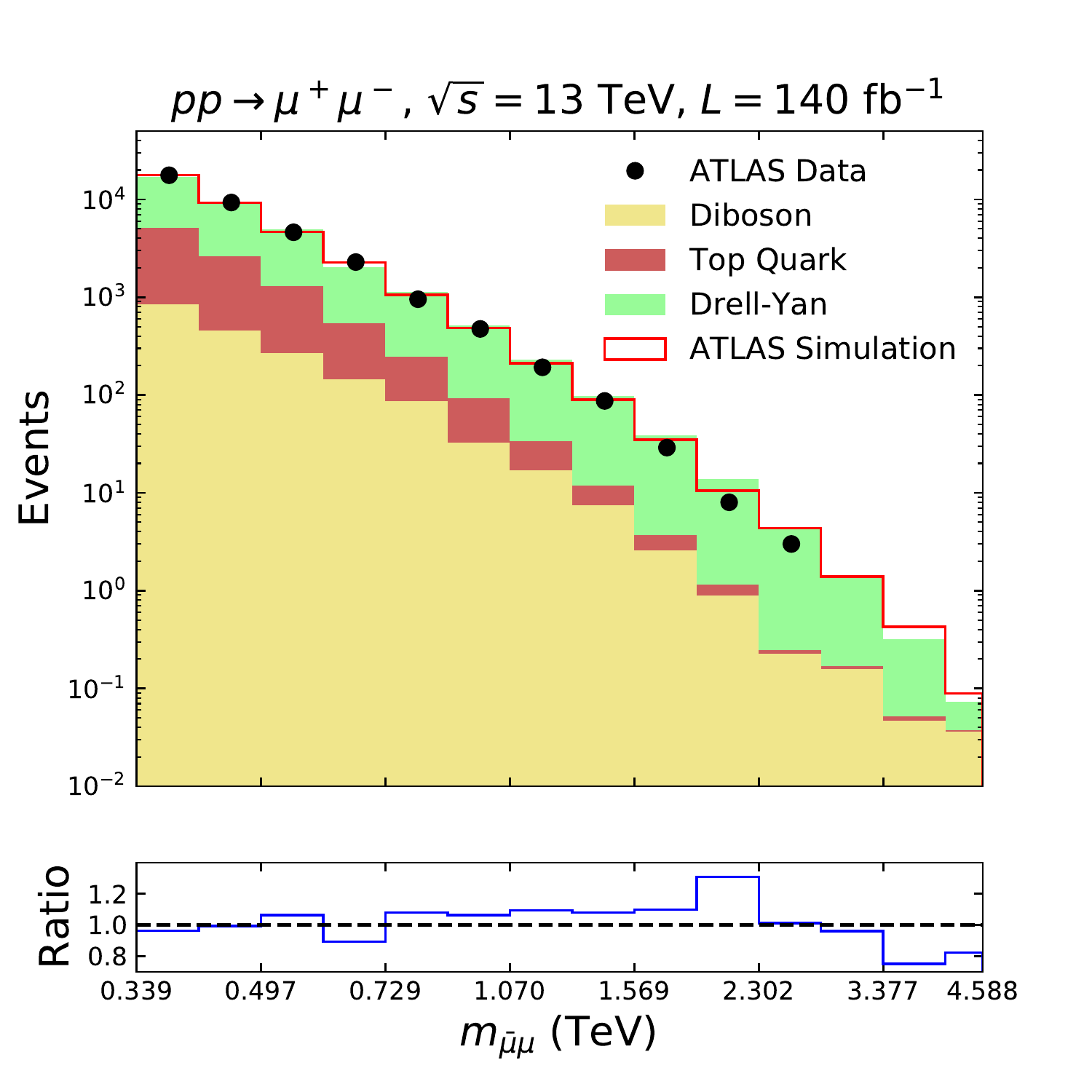}
\caption{Comparison of our SM background, generated using the methodology outlined in Sec. \ref{Event Simulation}, and the reported backgrounds and data from the recent CMS \cite{CMS:2021ctt} (left) and ATLAS \cite{Aad:2020otl} (right) searches. The observed data and predicted background from Monte Carlo simulations are directly taken from Fig.~2 in \cite{CMS:2021ctt} and Fig.~5 of the auxiliary material of \cite{Aad:2020otl}. In the left plot we have included the uncertainty in CMS background calculation. The lower panel of both plots gives the ratio of the number of events in each bin from our background calculation to the number calculated in the CMS or ATLAS simulations.}
\label{Background_Validation}
\end{figure}

Due to the inclusiveness of the dimuon final state, the need to conduct a full Monte-Carlo collider simulation with parton shower is largely unnecessary. In Fig. \ref{Background_Validation}, it can be seen that our SM background is in excellent agreement with the latest ATLAS and CMS searches. To generate the distributions shown in Fig. \ref{Background_Validation} we apply the same cuts on the transverse momentum $p_T$ and pseudo-rapidity $\eta$ of the muons as done in the ATLAS \cite{Aad:2020otl} and CMS analyses \cite{CMS:2021ctt}. When comparing to the ATLAS search, we use $p_T>30$ GeV and $|\eta|<2.5$ and, for the CMS search, we use $p_T>53$ GeV and $|\eta|<2.4$. 

\begin{figure}[t]
\centering
\includegraphics[scale=0.51]{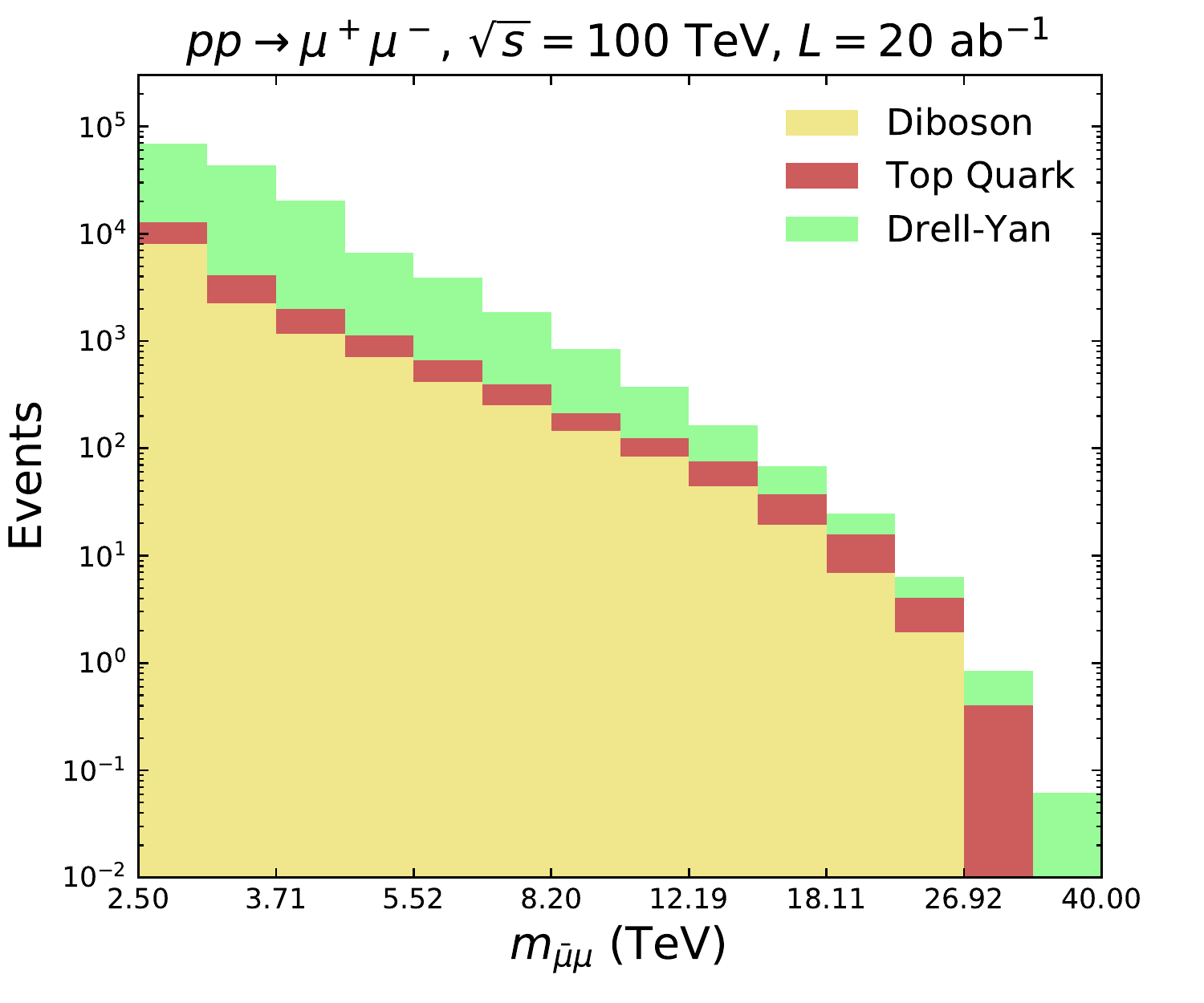}
\caption{The Standard Model background at $\sqrt{s}=100$ TeV calculated with a total integrated luminosity of $L=20$ ab$^{-1}$ and muon identification efficiency $\epsilon=0.95$.}
\label{Background_100TeV}
\end{figure}

In Fig. \ref{Background_Validation} it can be seen that our SM background lies within $\sim10\%$ of the CMS simulation. Our accuracy to the ATLAS simulation is slightly worse; however, this can be put down to the more complex selection criteria used in the ATLAS search and the method used to extract the data from Fig. 5 of the auxiliary material of \cite{Aad:2020otl}. Having validated our modeling of SM backgrounds at $\sqrt{s}=13$ TeV we can use this to obtain a good estimate of the SM background at a $\sqrt{s}=100$ TeV collider. The invariant mass of dimuon pairs originating from the SM background at $\sqrt{s}=100$ TeV is shown in Fig. \ref{Background_100TeV} over the range $m_{\bar{\mu}\mu}=[2.5,40]$ TeV.

Similarly to the SM background,  \textsc{MadGraph5\textunderscore aMC$@$NLO} is used to compute the EFT signal cross-section. We again perform combined NLO fixed order calculations including both the QCD and EW couplings. To this end, a \textsc{UFO} model \cite{Degrande:2011ua} was created in which the dim-$6$ terms in Eq. \ref{L smeft Csb} were added to the default SM \textsc{UFO} model used by \textsc{MadGraph5\textunderscore aMC$@$NLO}. For NLO calculations of the EFT signal to yield physical gauge-invariant results, rational terms, specifically $R_2$ terms, must be added to the model file by hand. The necessary $R_2$ terms for the EFT signal have been calculated for the first time here and were thus included in the \textsc{UFO} model file used in this study. For more details on the calculation of the rational $R_2$ terms see Appendix~\ref{NLO-signal}.

The motivation for including the NLO corrections to the EFT signal comes from the fact that the DY process receives large EW corrections in the high dilepton invariant mass region. Specifically, the NLO-EW corrections contain large negative Sudakov double logarithms that have been seen to reduce the differential cross section $d\sigma/dm_{\bar{\mu}\mu}$ of the inclusive DY process at large values of $m_{\bar{\mu}\mu}$. At $\sqrt{s}=13$ TeV it is seen that the differential cross section reduces by more that $30\%$ for values of $m_{\bar{\mu}\mu}>1$ TeV  \cite{Baur:2001ze}. The Sudakov double logarithms that cause such large negative corrections to the DY cross section originate from Feynman diagrams in which two external legs exchange a virtual particle. Given this, analogous NLO-EW diagrams exist for the EFT signal where the s-channel mediator is replaced by an effective vertex (see diagrams of Type A, B, and C in Appendix \ref{NLO-signal}).

In Fig. \ref{NLO vs LO} we plot the ratio of the NLO cross section to the LO cross section as a function of a lower cut on $m_{\bar{\mu}\mu}$ for both the EFT signal and the DY process. Here, we consider three NLO to LO ratios with the NLO cross section corresponding to the pure NLO-QCD, pure NLO-EW or the combined NLO-QCD\&EW cross section. In Fig \ref{NLO vs LO}, it is seen that the inclusion of the pure NLO-QCD corrections yields a consistent $\sim10\%$ increase in both the EFT signal and DY cross section at $\sqrt{s}=13$ TeV and 
$\sqrt{s}=100$ TeV. However, it is a different story for the NLO-EW corrections. For the DY process the NLO-EW corrections are seen to decrease the cross section by as much as $\sim20\%$ at $\sqrt{s}=13$ TeV and $\sim45\%$ at $\sqrt{s}=100$ TeV. Again, this is due to the presence of large negative Sudakov double logarithms. Whilst logarithms of this type reduce the NLO-EW EFT signal cross section, the effect is less dramatic than that seen for DY. In fact, the combined NLO-QCD\&EW signal cross section is always within $10\%$ of the LO cross section. Ultimately, this effect is not as large as may have been expected and a $10\%$ variation in the signal cross section does not have a significant effect on our sensitivity calculations. 

\begin{figure}[t]
\centering
\includegraphics[scale=0.51]{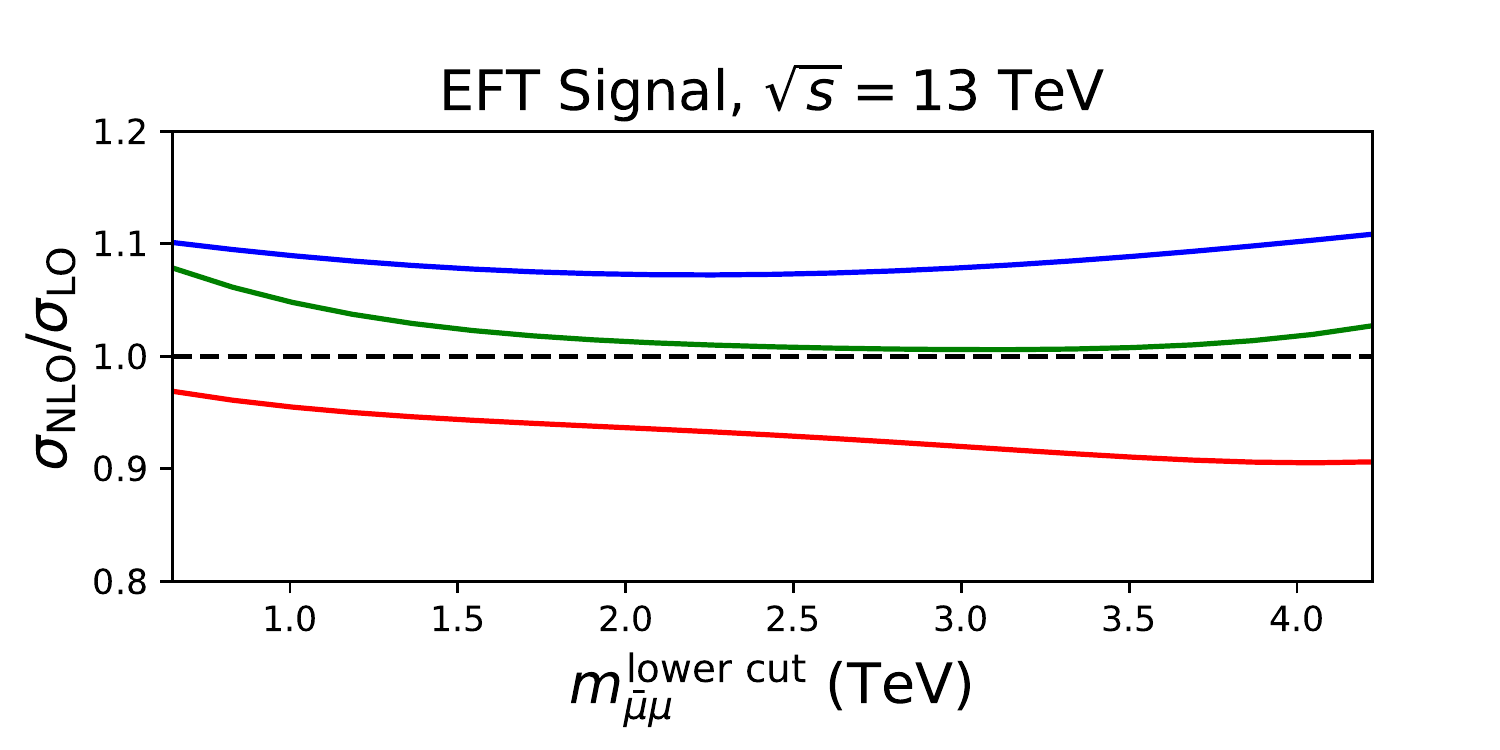}
\includegraphics[scale=0.51]{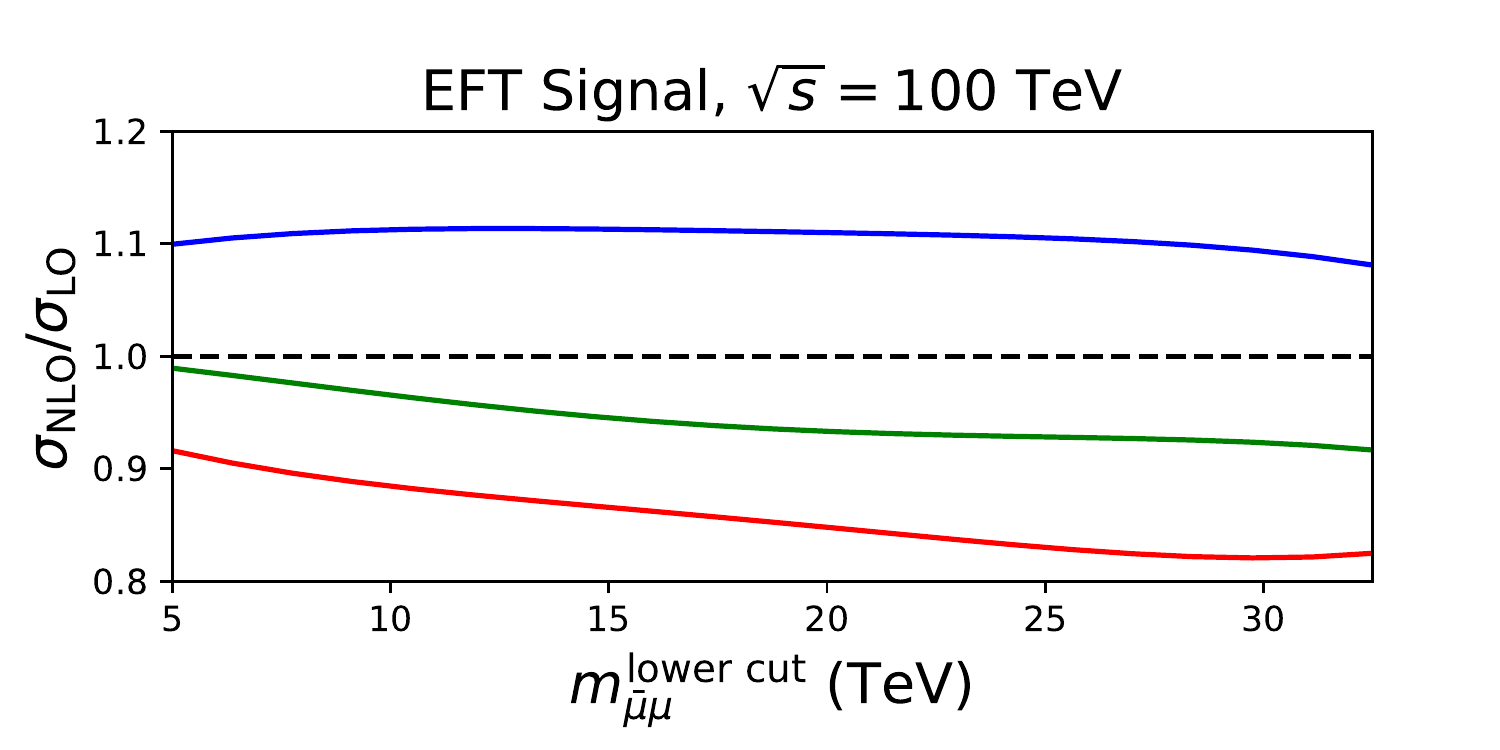}
\includegraphics[scale=0.51]{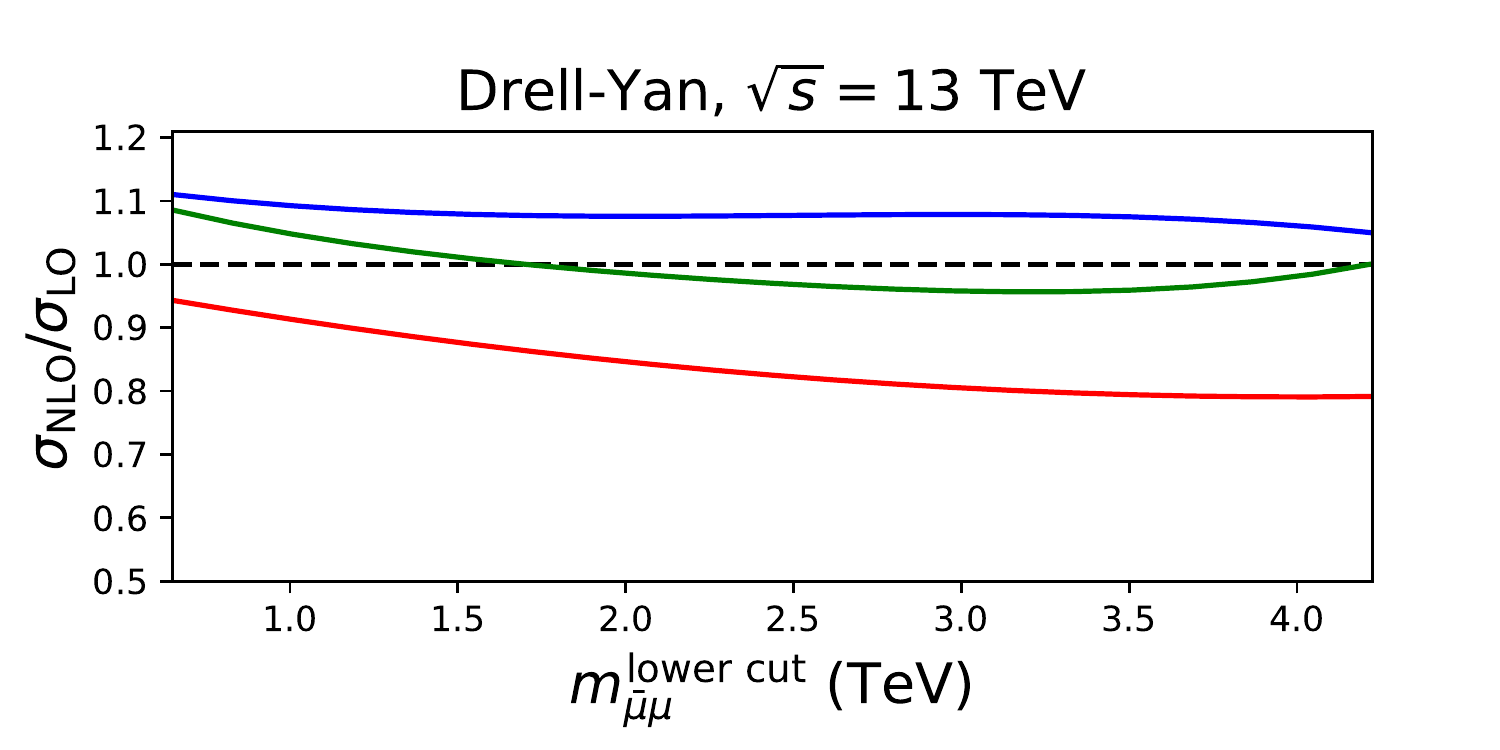}
\includegraphics[scale=0.51]{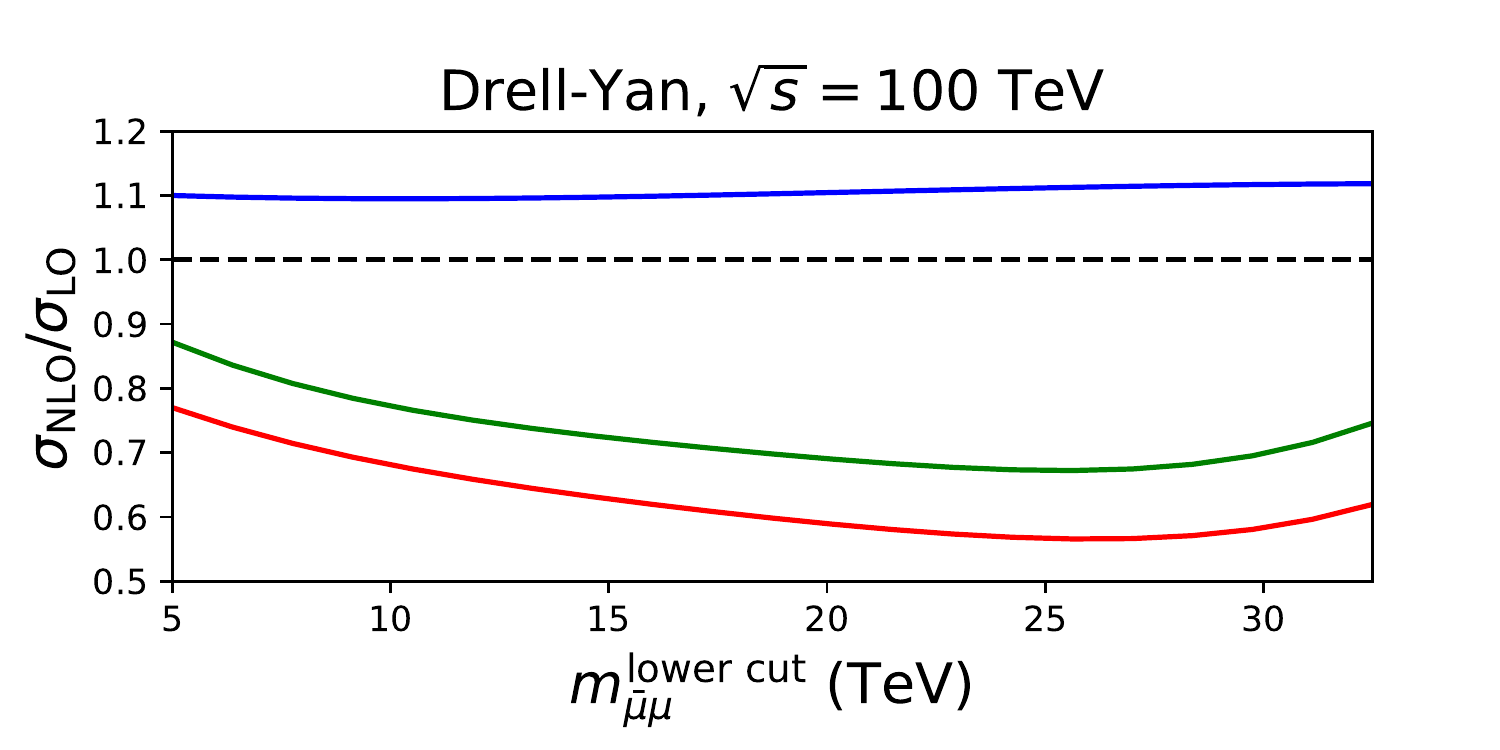}
\caption{The ratios of the NLO-QCD, NLO-EW and NLO-QCD\&EW cross sections to the LO cross section for both the EFT signal and the DY process as a function of a lower cut on the dimuon invariant mass $m^{\text{lower cut}}_{\bar{\mu}\mu}$. The blue lines correspond to NLO-QCD, the red to NLO-EW and the green to NLO-QCD\&EW. The following cuts on muons are used: at $\sqrt{s}=13$~TeV, $p_T=52$~GeV and $|\eta|<2.5$ and at $\sqrt{s}=100$~TeV,  $p_T=400$ GeV and $|\eta|<4.0$.} 
\label{NLO vs LO}
\end{figure}

\subsection{Statistics} \label{Statistics}

Here we characterise the sensitivity of the HL-LHC and a future collider to the EFT signal detailed in Sec. \ref{SMEFT}. To do this we perform two types of statistical tests based on binned dimuon invariant mass distributions. Given a certain signal strength, characterised by the mass scale $\Lambda$ in Eq. \ref{C_bs}, we calculate the expected significance to reject a given null hypothesis $H_0$ in favour of an alternative hypothesis $H_1$. Our statistical methods closely follow those developed and outlined in \cite{Cowan:2010js}.

The first statistical test we perform involves setting exclusion limits on $\Lambda$. Here we define $H_0$ to be the signal+background hypothesis with $H_1$ being the background only hypothesis. We then derive the value of the $\Lambda$ needed to reject the signal+background hypothesis at the $95\%$ CL. Our second statistical test involves the discovery of the EFT signal. Here the roles of $H_0$ and $H_1$ are reversed and we calculate the value of $\Lambda$ needed to reject the background only hypothesis at the $n\sigma$ level.

To calculate the expected significance, we construct a profile likelihood ratio from our binned invariant mass distributions. We first define a test statistic $t_{\bm{\mu}}$ to measure the level of agreement between $H_1$ and $H_0$. Our test statistic is given by
\begin{equation}
t_{\bm{\mu}}=-2\ln\lambda(\bm{\mu}),
\end{equation}
where $\lambda(\bm{\mu})$ is a profile likelihood ratio given by
\begin{equation}
\lambda(\bm{\mu})=\frac{L(\bm{\mu})}{L(\bm{\hat{\mu}})}.
\end{equation}
Here $\bm{\mu}=(\mu_1,\mu_2,\cdots,\mu_N)$, where $\mu_j$ parameterizes the strength of the signal process in the $j^{\text{th}}$ bin. For a histogram with $N$ bins, the likelihood function $\mathscr{L}(\bm{\mu})$ is constructed treating every bin as an independent Poisson variable such that
\begin{equation}
\mathscr{L}(\bm{\mu})=\prod_{j=1}^N\frac{(\mu_j s_j+b_j)^{n_j}}{n_j!}e^{-(\mu_j s_j+b_j)}.
\end{equation}
Here, $s_j$ and $b_j$ are the expected number of signal and background events in the $j^{\text{th}}$ bin respectively and $n_j$ is the total number of events in the $j^{\text{th}}$ bin according to the alternate hypothesis $H_1$. Finally, $\bm{\hat{\mu}}=(\hat{\mu}_1,\hat{\mu}_2,\cdots,\hat{\mu}_N)$ is the maximum-likelihood estimator of $\bm{\mu}$ which, in this instance, is given by $\hat{\mu}_j=(n_j-b_j)/s_j$. 

The significance to reject the null hypothesis is given by
\begin{equation}
Z_{\bm{\mu}}=\sqrt{t_{\bm{\mu}}}.
\end{equation}
To calculate the expected significance $E[Z_{\bm{\mu}}]$ we use the so-called Asimov data set \cite{Cowan:2010js}. We take $n_j\to b_j$ in the case of exclusion and $n_j\to s_j+b_j$ in the case of discovery. Given this, the expected exclusion significance $E[Z_e]$ is given by \cite{Cowan:2010js}
\begin{equation}
E[Z_e]=\sqrt{2\left[\sum_{j=1}^N\left(s_j+b_j\ln\left(\frac{b_j}{b_j+s_j}\right)\right)\right]}.
\label{EZe}
\end{equation}
An expected exclusion significance at the $95\%$ CL corresponds to $E[Z_e]=1.64$. The expected discovery significance $E[Z_0]$ is given by \cite{Cowan:2010js}
\begin{equation}
E[Z_0]=\sqrt{-2\left[\sum_{j=1}^N\left(s_j+(b_j+s_j)\ln\left(\frac{b_j}{b_j+s_j}\right)\right)\right]}.
\label{EZ0}
\end{equation}
An expected discovery significance at the $n\sigma$ level corresponds to $E[Z_e]=n$.

\subsection{Event Selection \& Binning Scheme} \label{Event Selection}

\begin{figure}[t]
\centering
\includegraphics[scale=0.51]{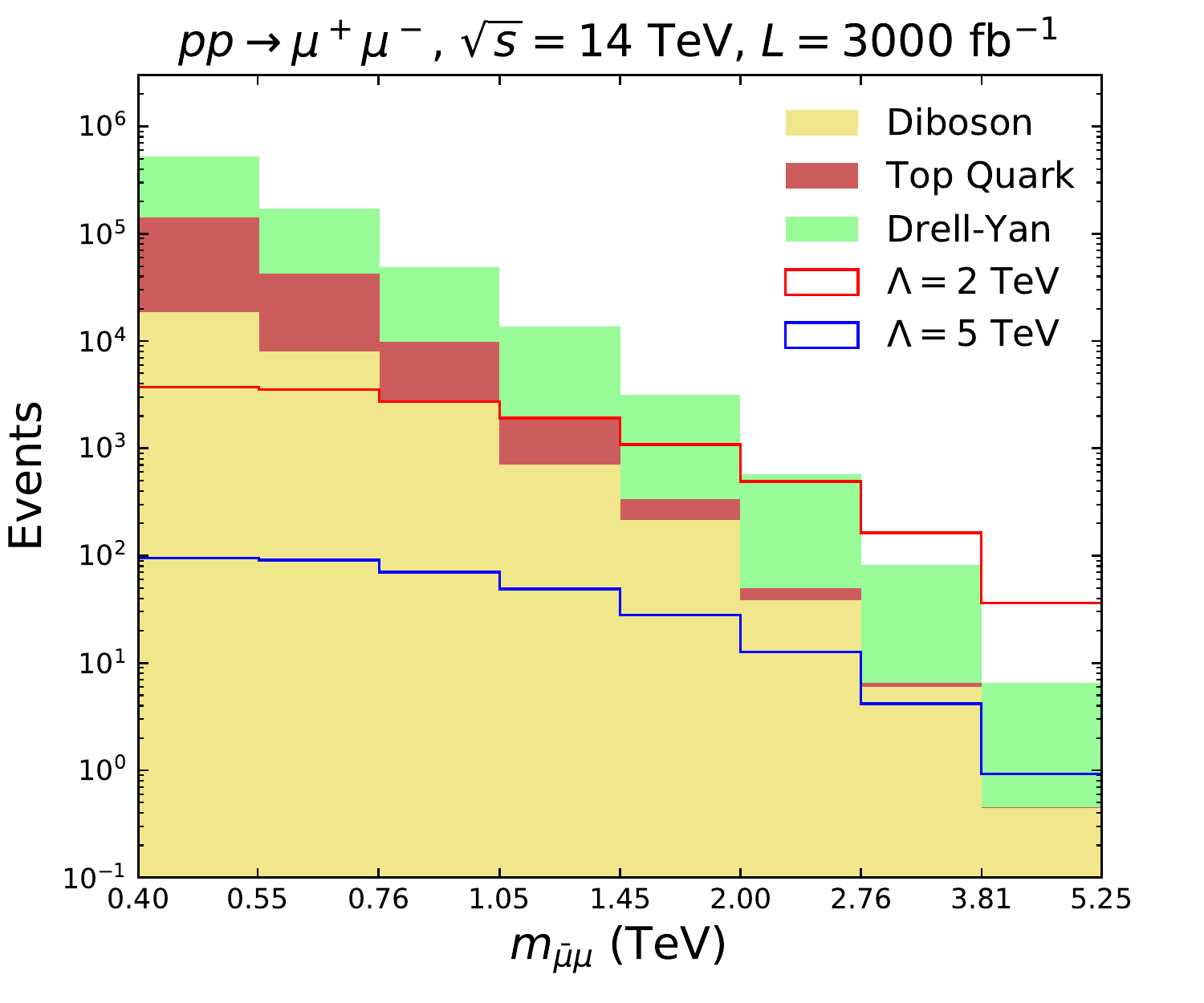}
\includegraphics[scale=0.51]{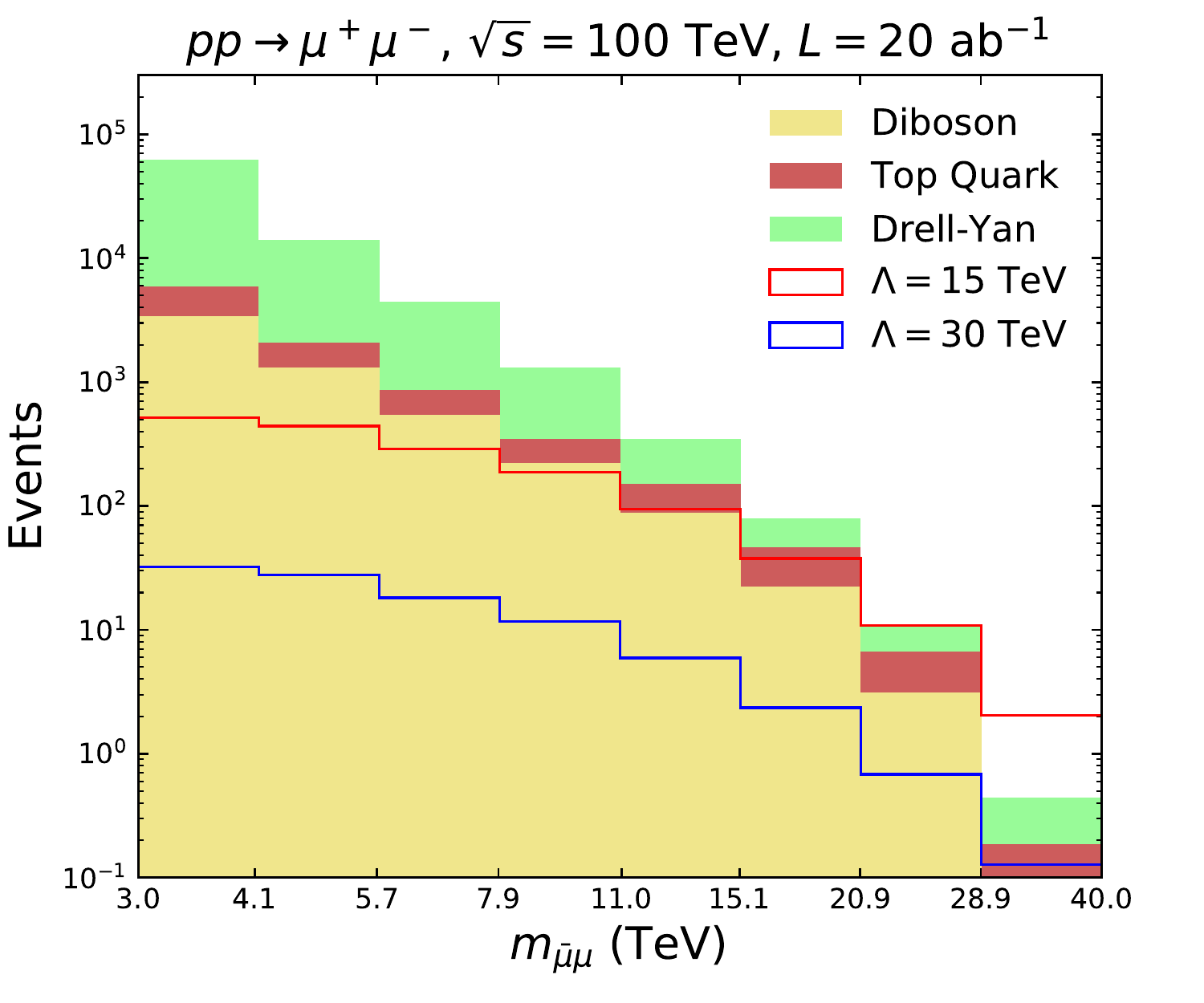}
\caption{Dimuon invariant mass distributions including both the signal and SM background at $\sqrt{s}=14$~TeV (left) and $\sqrt{s}=100$~TeV (right) using the binning scheme detailed in Sec. \ref{Event Selection}.}
\label{Invariant_mass}
\end{figure}

The event selection and binning scheme used in our sensitivity calculations for a future collider is guided by the latest ATLAS \cite{Aaboud:2017buh, Aad:2020otl} and CMS \cite{CMS:2021ctt} searches. Firstly, we scale our cut on the transverse momentum $p_T$ of the muons  linearly with the c.o.m energy of the given collider considered. Thus, we take $p_T=\sqrt{s}/250$ in line with the CMS \cite{CMS:2021ctt} selection at $\sqrt{s}=13$~TeV. Secondly, we alter our cut on  the pseudo rapidity $|\eta|$ of the muons. When considering the $\sqrt{s}=14$~TeV HL-LHC we take $|\eta|<2.5$ in line with ATLAS and CMS detector configurations; however, for a future collider with $\sqrt{s}>14$ TeV we take $|\eta|<4.0$ as is done in \cite{Allanach:2019zfr}.

Both the CMS \cite{CMS:2021ctt} and ATLAS \cite{Aaboud:2017buh} searches for contact interactions at $\sqrt{s}=13$~TeV impose a minimum cut on the dimuon invariant mass of $m^{\text{min}}_{\bar{\mu}\mu}=0.4$~TeV.  Given this, to define our binning scheme we take $m^{\text{min}}_{\bar{\mu}\mu}=0.43$~TeV at $\sqrt{s}=14$~TeV and $m^{\text{min}}_{\bar{\mu}\mu}=3$~TeV at $\sqrt{s}=100$~TeV; here, we have loosely scaled up the minimum cut on $m_{\bar{\mu}\mu}$ with $\sqrt{s}$ with respect to $m^{\text{min}}_{\bar{\mu}\mu}=0.4$~TeV at $\sqrt{s}=13$ TeV. It is important to include as many signal events into our sensitivity calculations as possible. Whilst the EFT signal tends to dominate in the high invariant mass region, if $m^{\text{min}}_{\bar{\mu}\mu}$ is chosen to be too large then large numbers of signal events can be discarded and sensitivity is reduced. Hence, we find the mentioned  minimum cuts on $m_{\bar{\mu}\mu}$ at $\sqrt{s}=14$ and $100$ TeV give the optimal significance and including events with lower invariant masses has no significant effect, as the SM background dominates in this region.

The latest CMS (ATLAS) searches define 8 (6) bins of increasing width above $400$ GeV \footnote{The most recent inclusive ATLAS search \cite{Aad:2020otl} using $139$ ab$^{-1}$ of data uses a single bin above $m_{\bar{\mu}\mu}>2$~TeV. We find that using multiple bins over a larger range in $m_{\bar{\mu}\mu}$ gives better sensitivity.}. Given this, we consider 8 logarithmic spaced bins in the interval $\left[m^{\text{min}}_{\bar{\mu}\mu},m^{\text{max}}_{\bar{\mu}\mu}\right]$ \footnote{Bins with constant widths have also been considered and only a minor reduction in the expected significance is seen.}. Whilst both \cite{Aaboud:2017buh} and \cite{CMS:2021ctt} take $m^{\text{max}}_{\bar{\mu}\mu}=6$~TeV it is noted that, the value of $m^{\text{max}}_{\bar{\mu}\mu}$ cannot be chosen to be arbitrarily large. This point is discussed in more detail in Sec. \ref{EFT Validity}. Fig. \ref{Invariant_mass} shows invariant mass distributions at $\sqrt{s}=14$ TeV and $\sqrt{s}=100$ TeV using the event simulation described in Sec.~\ref{Event Simulation} and the binning scheme and event selection described above. Here we have taken $m^{\text{max}}_{\bar{\mu}\mu}$ to be $5.25$ TeV and $40$ TeV respectively. This is done without regard for the validity of the EFT in order to present a large region of the invariant mass spectrum.

\subsection{Unitarity Constraints and EFT Validity}\label{EFT Validity}

When defining our event selection, it is important to note that the value of $m^{\text{max}}_{\bar{\mu}\mu}$ cannot be taken to be arbitrarily large. Requiring the
EFT amplitude to respect tree-level unitarity implies \cite{DiLuzio:2017chi}
\begin{equation}
m_{\bar{\mu}\mu}<\sqrt{\frac{4\pi}{\sqrt{3}}}\Lambda \equiv \Lambda_* .
\label{UTB EFT}
\end{equation}
Whenever $m_{\bar{\mu}\mu}$ becomes larger than $\Lambda_*$ our description of new physics as a
dimension-$6$ effective operator necessarily breaks down. Here, contributions from higher orders in perturbation theory, higher-dimensional operators or insertions of multiple operators, and/or the appearance of new on-shell degrees of freedom become as important as our tree-level signal.

In concrete UV completions of the EFT, the tree-level unitary bound may be reached at lower
values of $m_{\bar{\mu}\mu}$. For example, in simplified $Z'$ models \cite{DiLuzio:2017chi}
\begin{equation}
	m_{\bar{\mu}\mu}<m_{Z'}<\sqrt{\frac{2\pi}{\sqrt{3}}}\Lambda ,
	\label{UTB Z'}
\end{equation}
where $m_{Z'}$ is the mass of the $Z'$ and the final bound is the tree-level unitarity constraint
calculated in the $Z'$ model. The fact that this bound is below $\Lambda_*$ results from multiple channels
(such as $b \bar s \to s \bar b$) being mediated by the same $Z'$ couplings that cause the interaction
in Eq.~\ref{eq:Lminimal}; at low energies this situation corresponds to additional SMEFT operators not relevant to $b \to s \ell^+ \ell^-$ transitions.

A related question concerns the validity of the EFT expansion in Eq. \ref{SMEFT exp} \cite{Contino:2016jqw,Farina:2016rws,Torre:2020aiz}.
In our case dimension-6 is leading (due to the negligible SM contribution to $\bar b s \to \mu^+ \mu^-$),
and the relevant comparison is with dimension-8 operators.
This is necessarily model-dependent. For tree-level mediators, the EFT expansion of our signal amplitude simply
reproduces the expansion of the mediator propagator in $p^2/M^2$, where $M$ is the mediator mass. For an $s$-channel
mediator ($Z'$), the resulting bound amounts to requiring that the dimuon invariant mass is less than the
mediator mass. For a $Z'$, this simply means that the signal window must exclude the
$Z'$ peak in order to remain in the tail (first inequality in Eq. \ref{UTB Z'}). The corresponding cut for
a leptoquark mediator is on the value of $|t|$, which is a function of both the dimuon mass and the rapidity
and therefore is less intuitive; however, as $|t|\leq s$ for $2 \to 2$ scattering, the cut $m_{\bar{\mu}\mu} < m_{\rm LQ}$,
while conservative, still ensures validity of the EFT expansion.
The $Z'$ case is worked out in Appendix \ref{appendix b} in detail in the EFT language, and
the leptoquark case described qualitatively.

In the following, we give all exclusion and discovery limits as a function of $m^{\text{max}}_{\bar{\mu}\mu}$ and we will highlight the regions in which tree-level unitary is violated in the EFT (new physics described by a dim-$6$ effective operator) and under the assumption
the $bs\mu\mu$ contact interaction is mediated by a $Z'$. 

\section{Results\label{results}}

In this section we investigate the exclusion and discovery potential of a future collider to the $bs \mu \mu$ contact interactions that define our signal Eq.~\ref{eq:Lminimal}. We first give updated bounds at the LHC before moving to the proposed FCC-hh and beyond. We give all $95\%$ exclusion limits and $5\sigma$ discovery sensitivities in terms of $\Lambda$. In accordance with Sec. \ref{EFT Validity} we give the limits and sensitivities to $\Lambda$ as either a function of $m^{\text{max}}_{\bar{\mu}\mu}$ or for a the  value of $m^{\text{max}}_{\bar{\mu}\mu}$ that saturates the unitary constraint Eq.~\ref{UTB EFT}.

\subsection{Sensitivity at the LHC}

The sensitivity of the $\sqrt{s}=13$ TeV LHC to the $bs\mu\mu$ contact interactions Eq.~\ref{eq:Lminimal} using an inclusive dimuon final state has been investigated in~\cite{Greljo:2017vvb}. Here $95\%$ exclusion limits have been obtained from a collider recast based on \cite{Aaboud:2017buh} using $36$ fb$^{-1}$ of data along with a projection for the HL-LHC at $\sqrt{s}=13$ TeV. It is seen that the LHC (HL-LHC) can exclude $\Lambda=2.5 \;(4.1)$~TeV at the $95\%$ CL \cite{Greljo:2017vvb}. To this end, we first rederive the results presented in \cite{Greljo:2017vvb}, here including NLO effects to the EFT signal. We find a good agreement with \cite{Greljo:2017vvb} where the slight difference can be attributed to difference in our binning scheme and inclusion of detector effects. We then update the exclusion limits to include the $\sqrt{s}=13$ TeV LHC with $L=140$ fb$^{-1}$, the most recent LHC data set, as-well as giving projections at the $14$ TeV HL-LHC. In addition to $95\%$ exclusion limits, we also derive the value of $\Lambda$ needed for $5\sigma$ discovery. Our results are detailed in Table \ref{13 and 14 TeV bounds}, where we consider two different muon identification and reconstruction efficiencies corresponding to the CMS or ATLAS detector. 

\begin{table}[t]
	\centering
	{\tabulinesep=1.0mm
		\begin{tabu}{||c||c|c|c|c||c|c|c|c||}

			\hline
			~ & \multicolumn{4}{c||}{$95\%$ Exclusion} & \multicolumn{4}{c||}{$5\sigma$ Discovery} \\
			\hline
			$\sqrt{s}$ (TeV) & \multicolumn{3}{c|}{$13$} & $14$ & \multicolumn{3}{c|}{$13$} & $14$ \\
			\hline
			$L$ (fb$^{-1}$) & $36$ & $139$ & $3000$  & $3000$ & $36$ & $139$ & $3000$  & $3000$\\
			\hline
			$\Lambda$ (TeV) ($\epsilon=0.75$) & $2.3$ & $2.7$ & $4.1$ &  $4.2$ & $1.7$ & $2.1$ & $3.1$ &  $3.2$ \\
			$\Lambda$ (TeV) ($\epsilon=0.965$) & $2.4$ & $2.9$ & $4.3$ &  $4.5$ & $1.8$ & $2.2$ & $3.2$ &  $3.4$ \\
			\hline
	\end{tabu}}
	\caption{$95\%$ exclusion limits and $5\sigma$ discovery sensitivities for $\Lambda$ at the LHC. At $\sqrt{s}=14$ TeV the event selection detailed in Sec. \ref{Event Selection} is used with $m^{\text{max}}_{\bar{\mu}\mu}=5.25$ TeV. For $\sqrt{s}=13$ TeV, the cuts on the muon are taken directly from \cite{Aad:2020otl,CMS:2021ctt} in conjunction with the binning scheme detailed in Sec. \ref{Event Selection} with $m^{\text{max}}_{\bar{\mu}\mu}=4.5$ TeV. Both values of $m^{\text{max}}_{\bar{\mu}\mu}$ have been chosen to maximise  sensitivity whilst still lying comfortably below the constraint imposed by tree-level unitarity of the EFT amplitude Eq.~\ref{UTB EFT} (as well as that of a simplified $Z'$ model Eq.~\ref{UTB Z'}).}
	\label{13 and 14 TeV bounds}
\end{table}

\begin{figure}[t]
\centering
\includegraphics[scale=0.51]{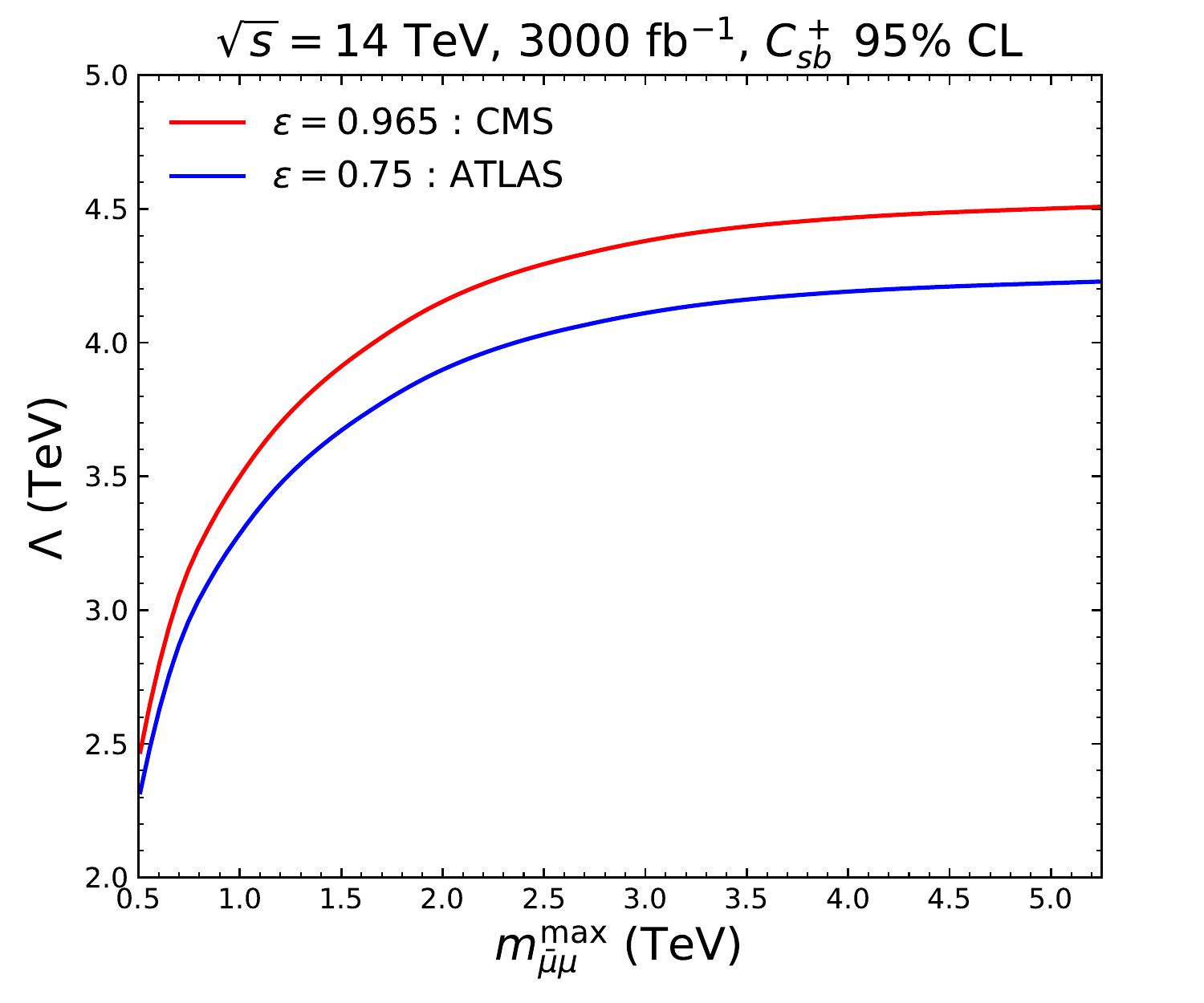}
\includegraphics[scale=0.51]{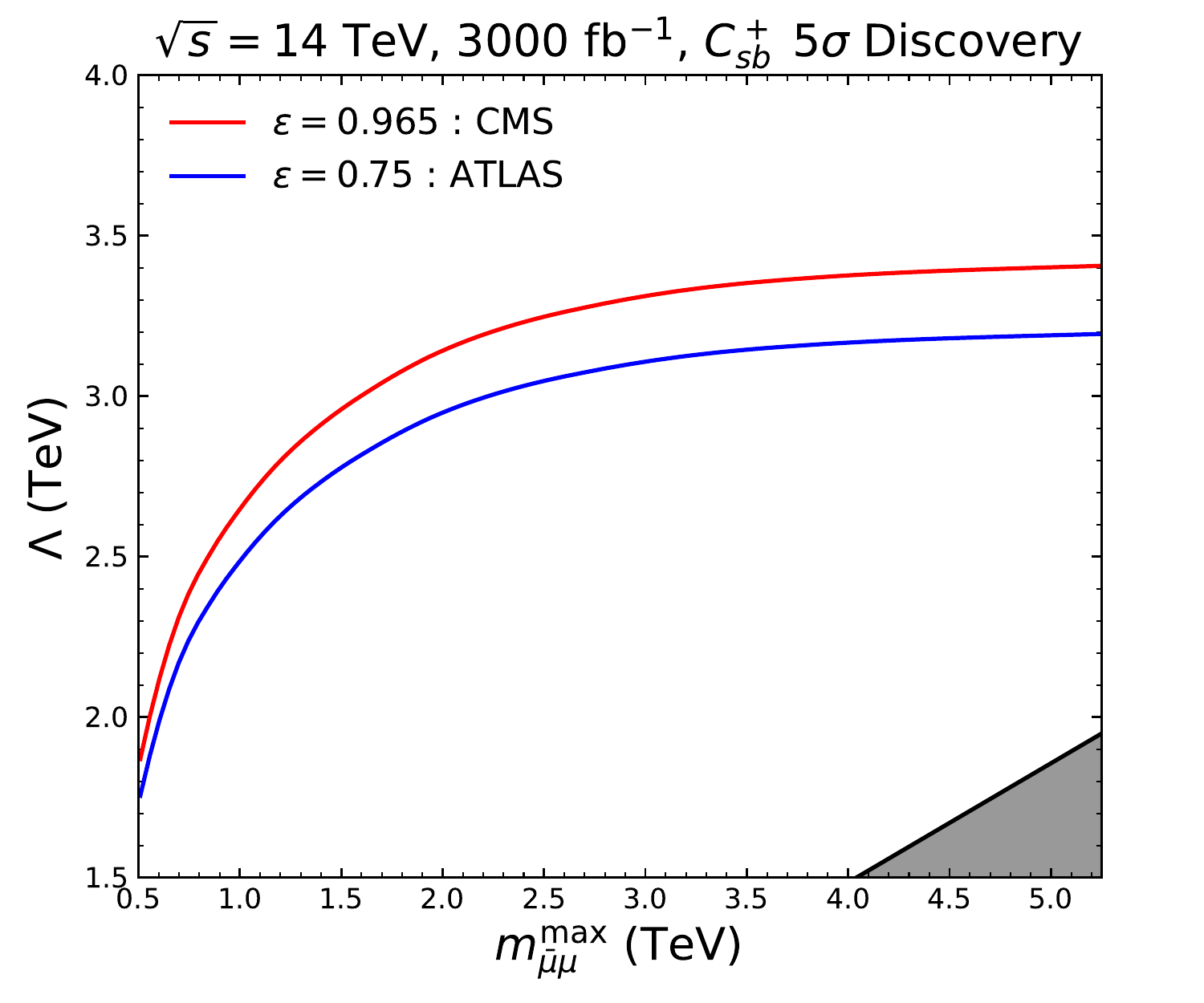}
\caption{$95\%$ exclusion limits and $5\sigma$ discovery sensitivities for $\Lambda$ as a function of $m^{\text{max}}_{\bar{\mu}\mu}$ at the $\sqrt{s}=14$~TeV HL-LHC using $L=3000$ fb$^{-1}$. The dark grey shaded region highlights the region in which tree-level unitary of the EFT is violated (see Eq.~\ref{UTB EFT}).} 
\label{Cbs 95CL and 5sig at 14TeV}
\end{figure}

In Table \ref{13 and 14 TeV bounds} it is seen that the $\sqrt{s}=14$ TeV HL-LHC can exclude $\Lambda=4.5$ TeV at the $95\%$~CL and can discover $\Lambda=3.4$ TeV at the $5\sigma$ level. A more detailed presentation of these results is shown in Fig. \ref{Cbs 95CL and 5sig at 14TeV} where the exclusion limits and discovery sensitivities for $\Lambda$ are given as a function of an upper cut of the dimuon invariant mass. We can see that the limits on $\Lambda$ lie well below the unitarity bound Eq.~\ref{UTB EFT} even when $m^{\text{max}}_{\bar{\mu}\mu}=5.25$ TeV.

In both the cases of $95\%$ exclusion and $5\sigma$ discovery, the sensitivity does not significantly change when dimuon events with $m_{\bar{\mu}\mu}\gtrapprox3$ TeV are discarded. Given this, if the underlying model of NP requires $m_{\bar{\mu}\mu}\gtrapprox3$~TeV in order for the EFT to remain valid, then valid limits can be obtained with minimal effect on the overall sensitivity. In a scenario such as this the EFT approach can both exclude and discover areas of parameter space for a given NP model without the need for a direct search. For example, $Z'$ bosons with $m_{Z'}>3$ TeV can be either excluded or discovered. Alternatively, if the underlining NP model requires a cut on the dimuon invariant mass stronger than $3$ TeV then the actual sensitivity of the $\sqrt{s}=14$ TeV HL-LHC can be significantly lower than that detailed in Table \ref{13 and 14 TeV bounds}. 

\subsection{Sensitivity at the FCC-hh}

Here we present $95\%$ exclusion limits and $5\sigma$ discovery sensitivities at a future $\sqrt{s}=100$ TeV proton-proton collider. The luminosity goal of each of the two experiments at the FCC-hh is intended to be $\sim1$ ab$^{-1}$ per year \cite{FCC:2018vvp}. This gives a total integrated luminosity of $\sim20$ ab$^{-1}$ over the lifetime of the collider for each experiment and $\sim40$ ab$^{-1}$ if both data sets are combined. Given this, we will use $L=\{1,20,40\}$ ab$^{-1}$ as benchmark luminosities to assess the sensitivity of a $\sqrt{s}=100$ TeV proton-proton collider to the $bs\mu\mu$ contact interaction. 

\begin{figure}[t]
\centering
\includegraphics[scale=0.51]{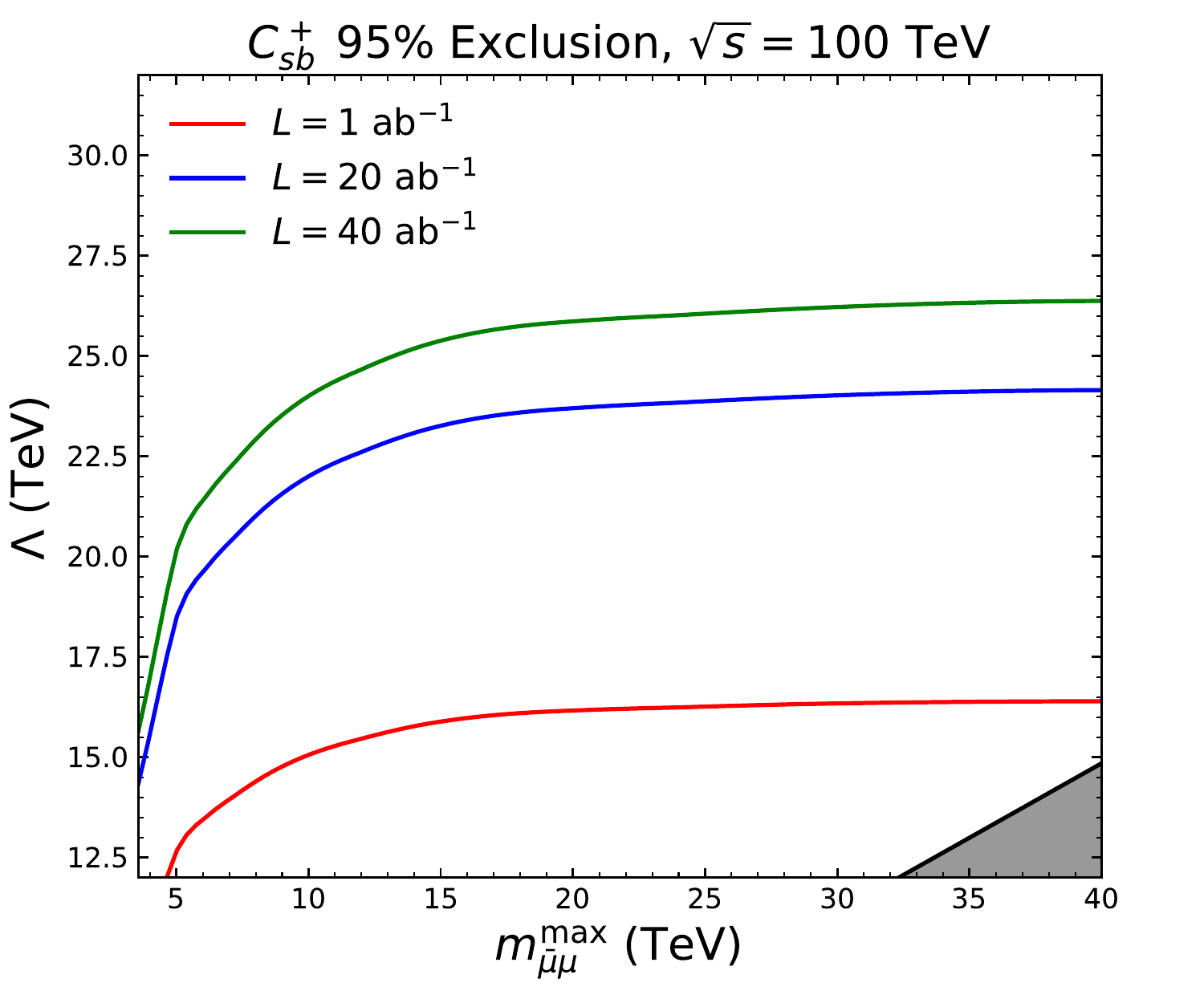}
\includegraphics[scale=0.51]{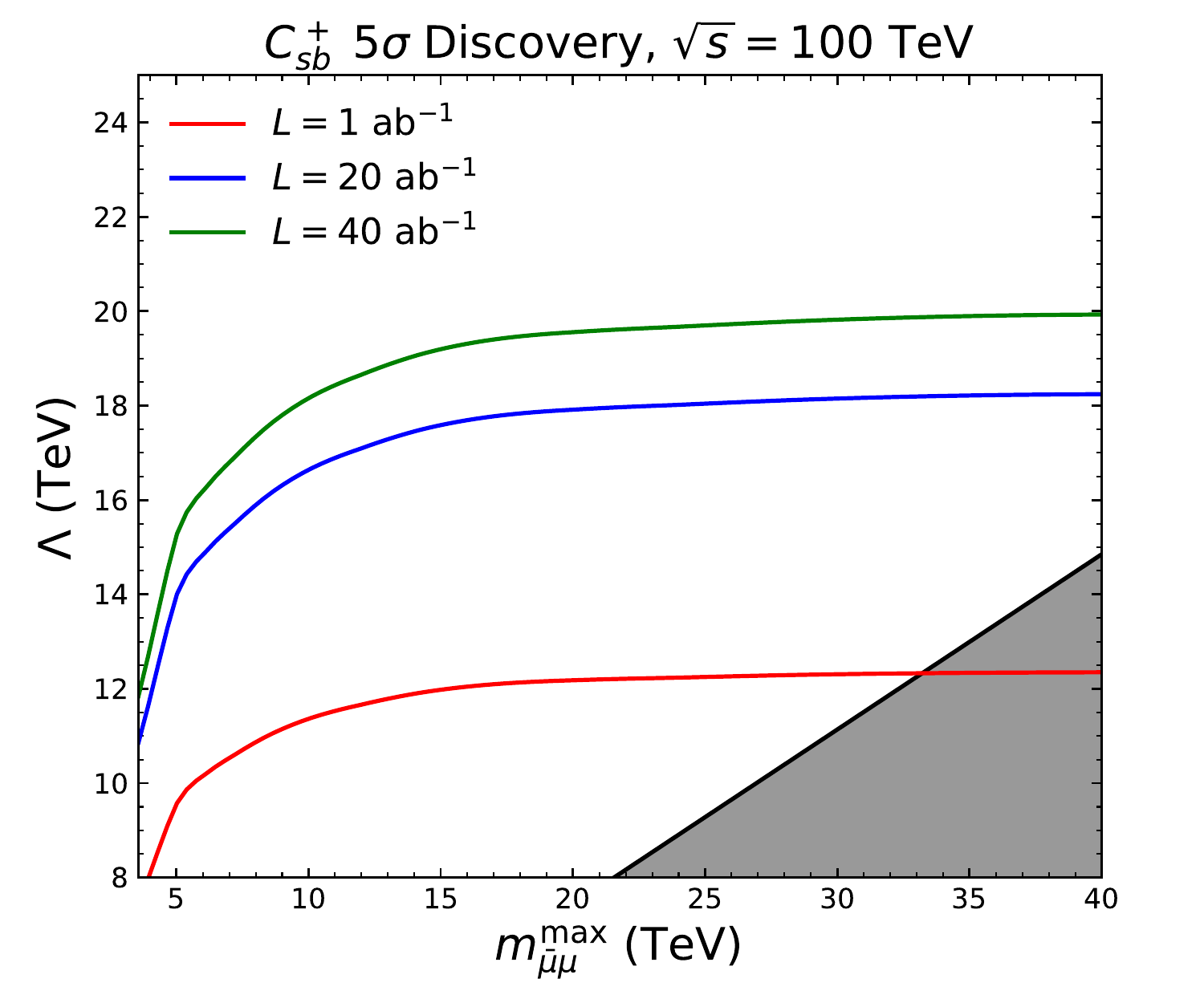}
\caption{$95\%$ exclusion limits and $5\sigma$ discovery sensitivities as a function of $m^{\text{max}}_{\bar{\mu}\mu}$ at $\sqrt{s}=100$ TeV. We plot these limits for three benchmark values of the luminosities, i.e., $L=\{1,20,40\}$~ab$^{-1}$. The dark grey shaded region highlights the region in which tree-level unitary of the EFT is violated (see Eq.~\ref{UTB EFT}).}
\label{Cbs 95CL and 5sig at 100TeV}
\end{figure}

In Table \ref{100 TeV bounds}, we give the $95\%$ exclusion limits on $\Lambda$ and values of $\Lambda$ needed for $5\sigma$ discovery at $\sqrt{s}=100$ TeV. It is seen that with a maximal luminosity of $L=40$ ab$^{-1}$ the FCC-hh can exclude $\Lambda\approx26$ TeV at the $95\%$ CL and discover signals of $\Lambda\approx20$ TeV at the $5\sigma$ level. A more detailed presentation of these limits is shown in Fig. \ref{Cbs 95CL and 5sig at 100TeV} where we again give the limits as a function of an upper cut on the dimuon invariant mass. It is noted that the tree-level unitary bound is more relevant at low luminosities.

\begin{table}[b]
	\centering
	{\tabulinesep=1.0mm
		\begin{tabu}{||c||c|c|c||c|c|c||}

			\hline
			~ & \multicolumn{3}{c||}{$95\%$ Exclusion} & \multicolumn{3}{c||}{$5\sigma$ Discovery} \\
			\hline
			$L$ (ab$^{-1}$) & $1$ & $20$ & $40$  & $1$ & $20$ & $40$ \\
			\hline
			$\Lambda$ (TeV) & $15.8$ & $24.1$ & $26.4$ & $12.0$ & $18.1$ & $19.8$ \\
			\hline
	\end{tabu}}
	\caption{$95\%$ exclusion limits and $5\sigma$ discovery sensitivities for $\Lambda$ at the $\sqrt{s}=100$ TeV FCC-hh. The event selection used to derive these bounds is detailed in Sec. \ref{Event Selection}. At a luminosity of $L=1$ ab$^{-1}$ $95\%$ exclusion limits and $5\sigma$ discovery sensitivities are obtained with $m^{\text{max}}_{\bar{\mu}\mu}=15$ TeV. For $L=20$ ab$^{-1}$ and $L=40$ ab$^{-1}$, we take  $m^{\text{max}}_{\bar{\mu}\mu}=40$ TeV for $95\%$ exclusion limits and $m^{\text{max}}_{\bar{\mu}\mu}=30$ TeV for $5\sigma$ discovery. The different values of $m^{\text{max}}_{\bar{\mu}\mu}$ have been chosen to maximise  sensitivity whilst still lying comfortably below the constraint imposed by tree-level unitarity of the EFT amplitude Eq.~\ref{UTB EFT} (as well as that of a simplified $Z'$ model Eq.~\ref{UTB Z'}).}
	\label{100 TeV bounds}
\end{table}

\begin{figure}[h!]
\centering
\includegraphics[scale=0.51]{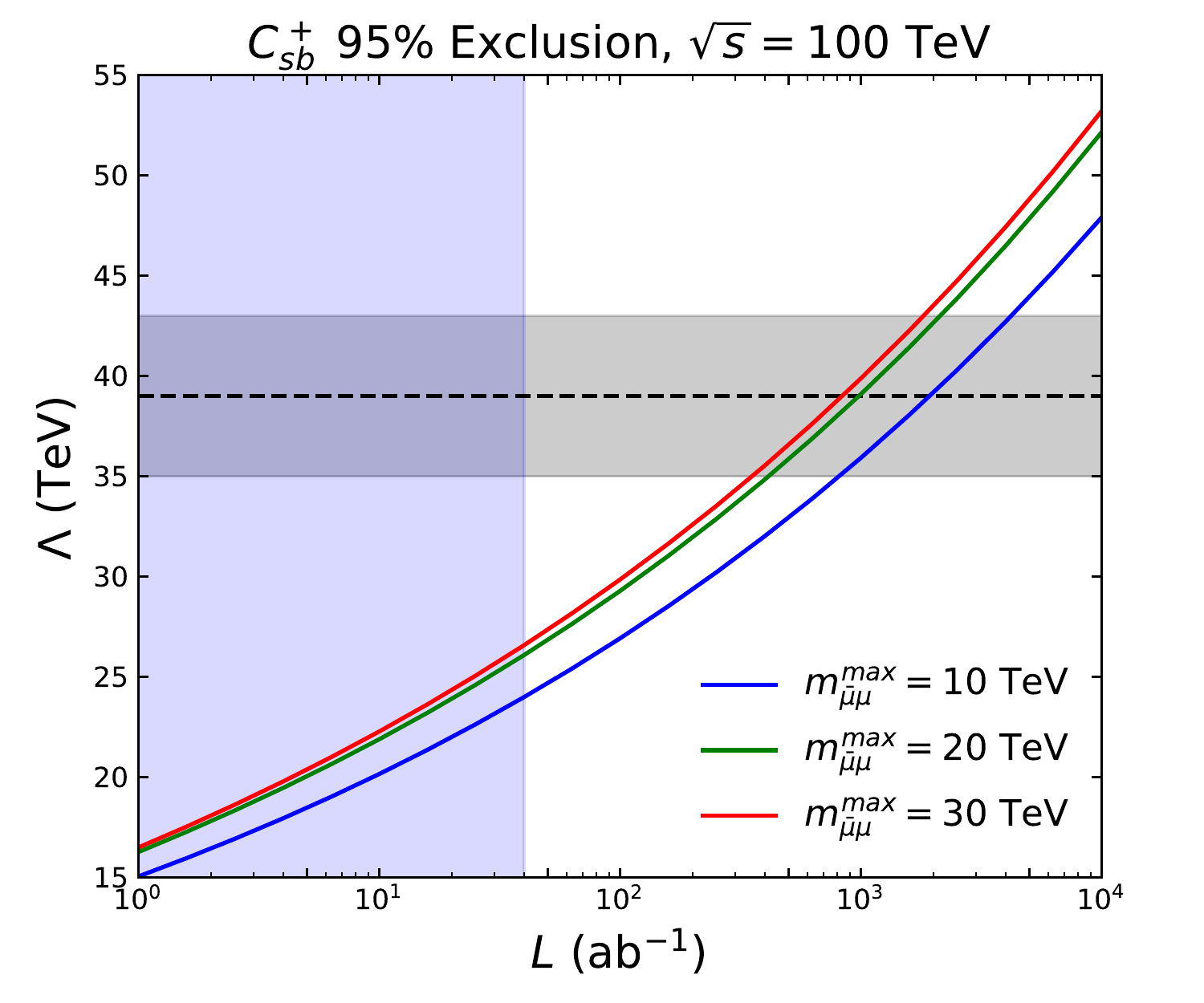}
\includegraphics[scale=0.51]{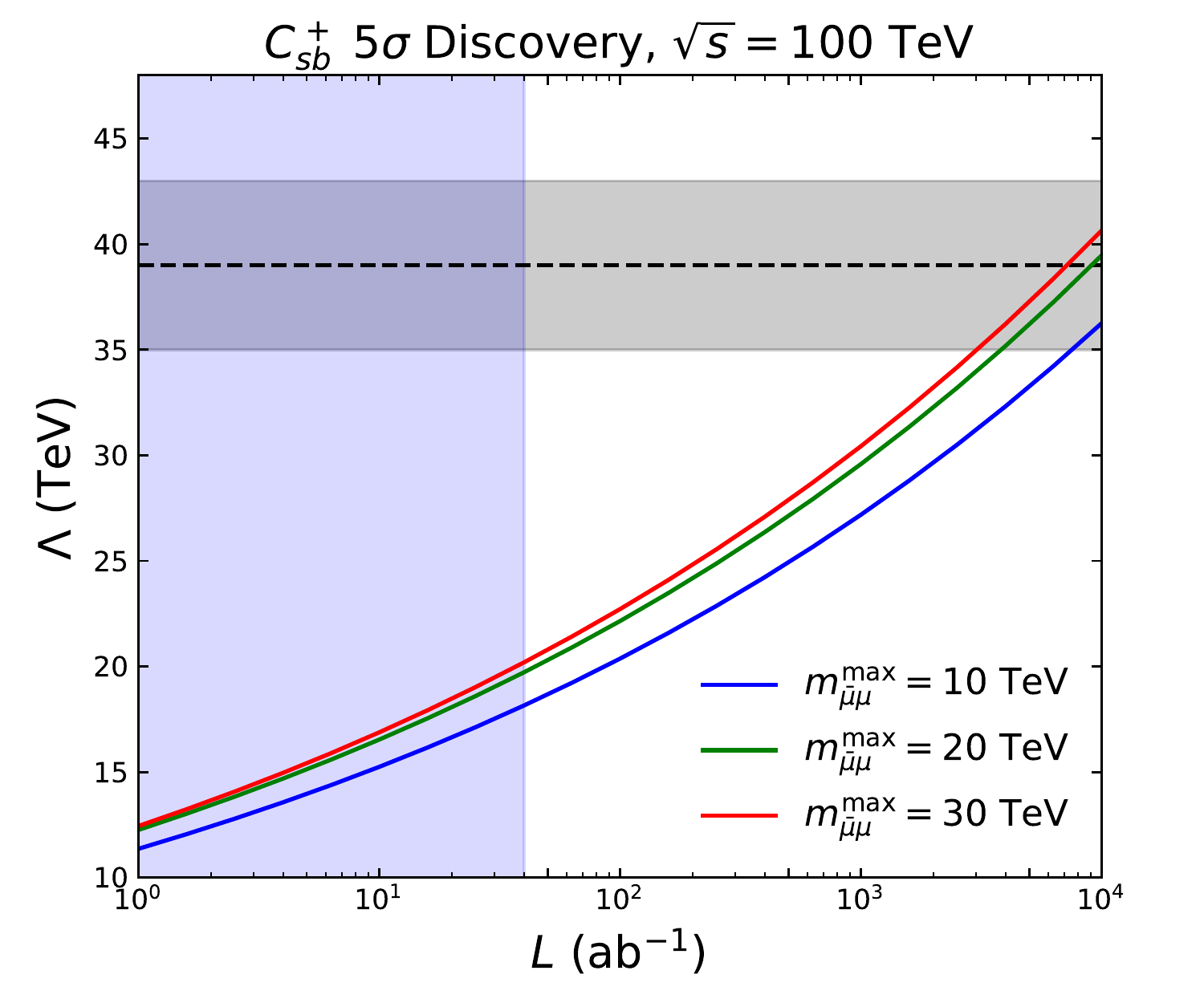}
\caption{$95\%$ exclusion limits and $5\sigma$ discovery sensitivities for $\Lambda$ as a function of total integrated luminosity $L$ at $\sqrt{s}=100$ TeV. We consider three values of $m^{\text{max}}_{\bar{\mu}\mu}$. The blue shaded region signifies $L\leq 40$ ab$^{-1}$, the design luminosity of the FCC-hh. The dashed black line corresponds to $\Lambda=39$ TeV with the grey shaded region corresponding to the uncertainty in Eq.~\ref{Lambda 39}.} 
\label{95CL and 5sig at 100TeV as function of lumi}
\end{figure}
Comparing Figs. \ref{Cbs 95CL and 5sig at 14TeV} and \ref{Cbs 95CL and 5sig at 100TeV} we see that the FCC-hh improves the sensitivity to the  $bs\mu\mu$ contact interaction significantly compared to that achievable at the HL-LHC, with sensitivity improving by $\approx500\%$. Having said this, the FCC-hh with its current design luminosity is not able to exclude or discover an EFT signal of strength of $\Lambda\approx40$ TeV as currently suggested by the B anomalies (see Eq.~\ref{Lambda 39}). Thus, in Fig. \ref{95CL and 5sig at 100TeV as function of lumi}, we extend beyond the design luminosity of the FCC-hh. We observe that roughly $20$ times more luminosity is needed for $95\%$ exclusion of a $\Lambda\approx40$ TeV signal and around $180$ times more for $5\sigma$ discovery. However, if new measurements involving the B-anomalies point towards a lower scale of new physics then the required luminosity decreases exponentially. Signal strengths up to $\Lambda\approx30$~TeV can be excluded or discovered with an order of magnitude less luminosity than is needed for $\Lambda\approx40$ TeV. 

\subsection{Beyond the FCC-hh}

\begin{figure}[t]
\centering
\includegraphics[scale=0.51]{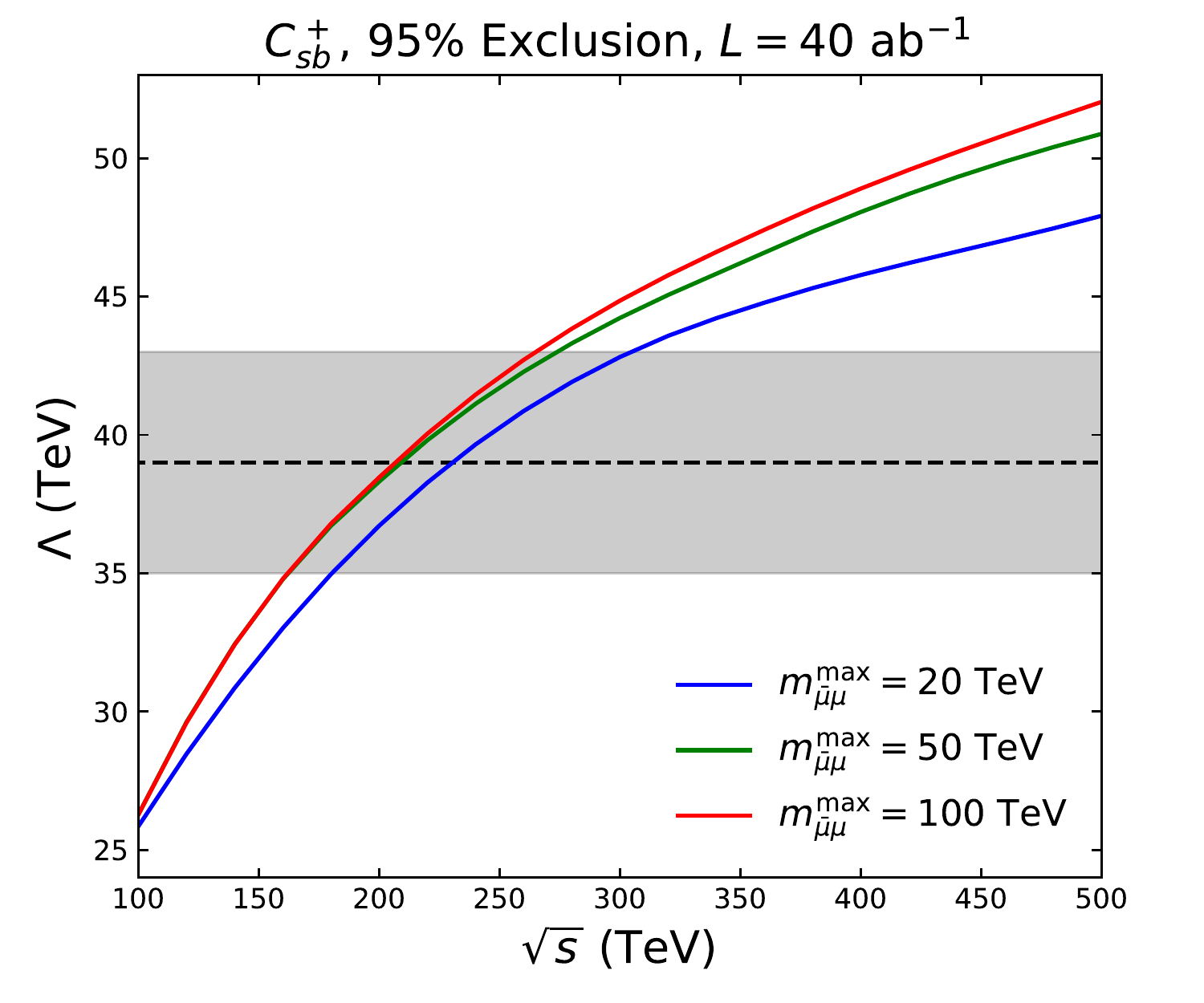}
\includegraphics[scale=0.51]{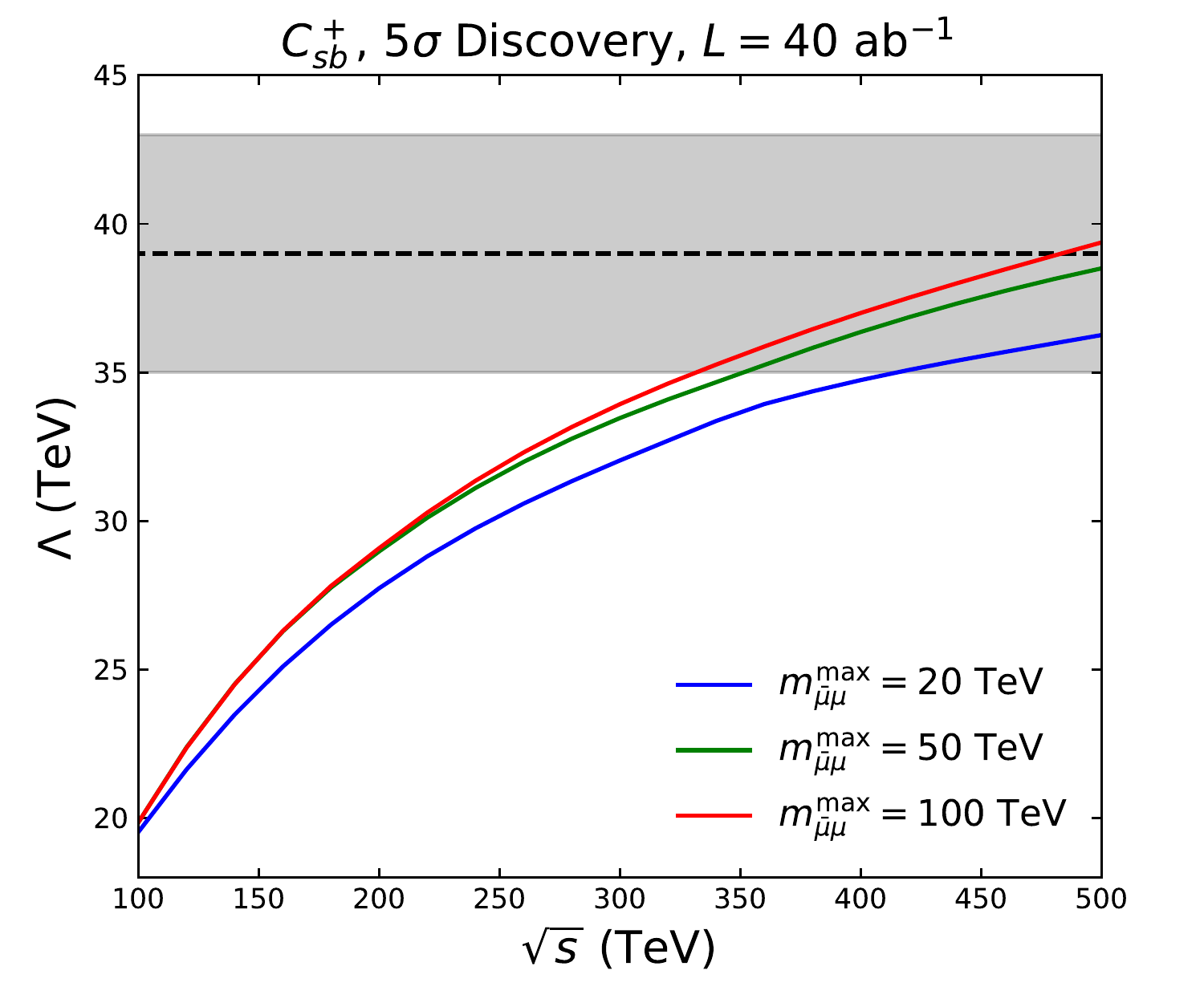}
\caption{$95\%$ exclusion limits and $5\sigma$ discovery sensitivities for $\Lambda$ as a function of collider c.o.m energy $\sqrt{s}$ with $L=40$ ab$^{-1}$. We consider three values of $m^{\text{max}}_{\bar{\mu}\mu}$. The dashed black line corresponds to $\Lambda=39$~TeV with the grey shaded region corresponding to the uncertainty in Eq.~\ref{Lambda 39}.}
\label{95CL and 5sig at 100TeV as function of energy}
\end{figure}

\begin{figure}[t]
\centering
\includegraphics[scale=0.51]{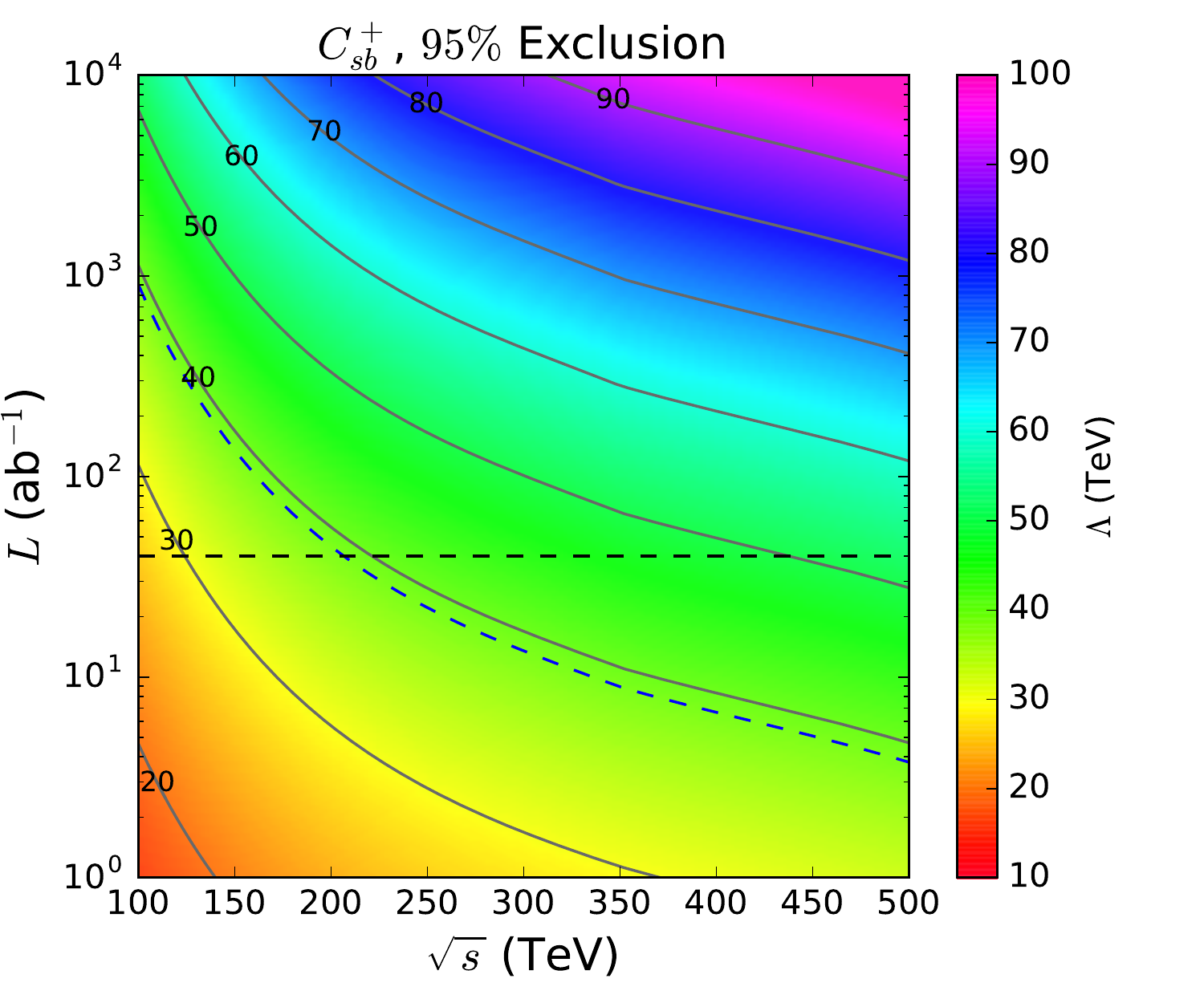}
\includegraphics[scale=0.51]{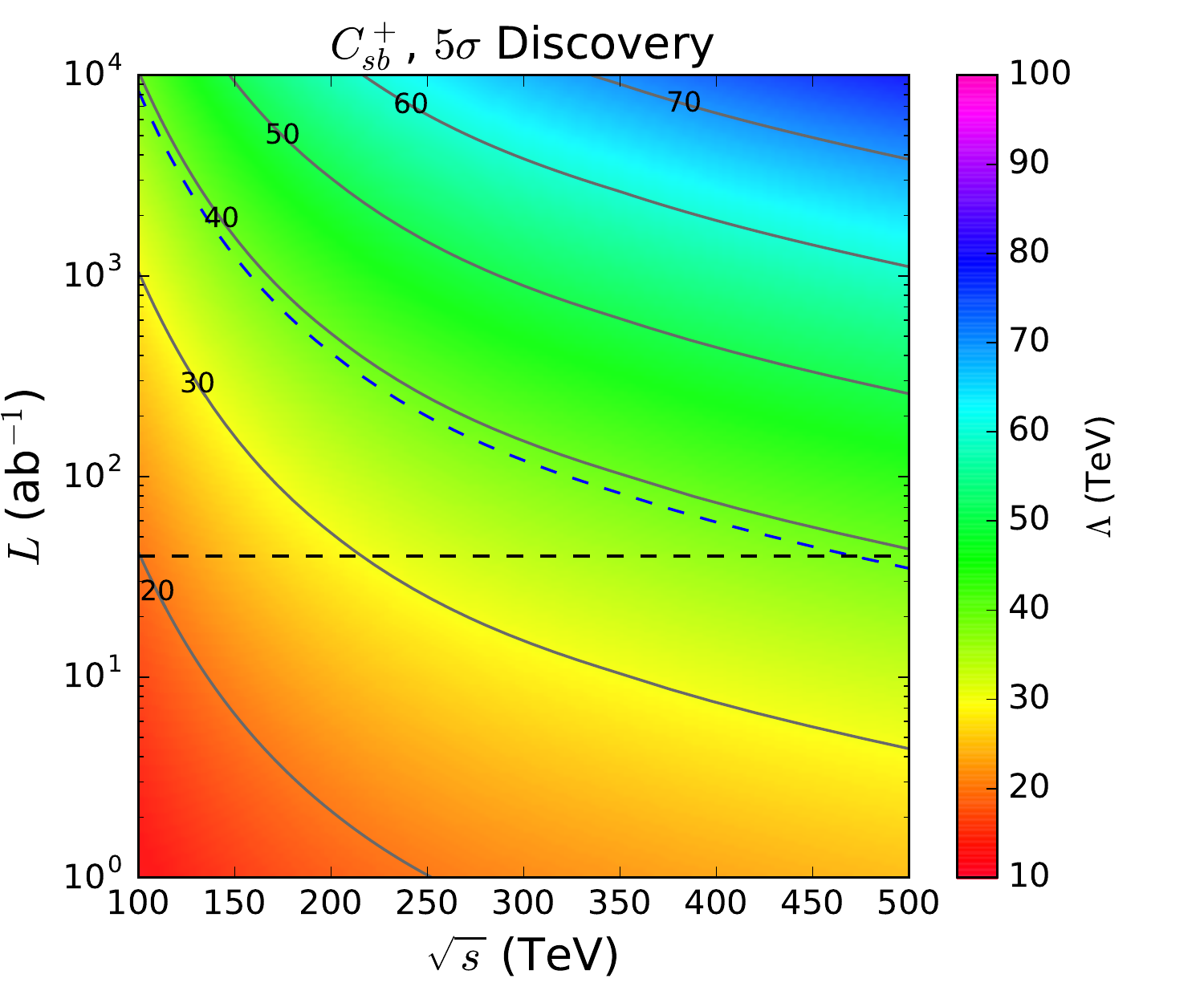}
\caption{$95\%$ exclusion limits and $5\sigma$ discovery sensitivities for $\Lambda$ as a function of both collider c.o.m energy $\sqrt{s}$ and luminosity $L=40$ ab$^{-1}$. Here $m^{\text{max}}_{\bar{\mu}\mu}$ is set to the tree level unitarity limit in Eq.~\ref{UTB EFT}.
The horizontal black dashed line gives  the maximal design luminosity at the FCC-hh. The blue dashed contour corresponds to $\Lambda=39$ TeV, the current value of $\Lambda$ suggested by the B-anomalies.}
\label{l vs e 95CL and 5sig}
\end{figure}

In the previous two subsections, it was seen that increasing the c.o.m energy of a collider from $\sqrt{s}=14$ TeV to $\sqrt{s}=100$ TeV dramatically increased sensitivity. Despite this, it was seen that the FCC-hh, with its design luminosity, can neither exclude nor discover signals of order $\Lambda\approx40$ TeV. Hence, in this section we explore the possibility of an FCC-hh with increased c.o.m energy. We first consider an FCC-hh with $\sqrt{s}>100$~TeV and $L=40$~ab$^{-1}$ before varying both the c.o.m energy and luminosity simultaneously. 

When deriving the $95\%$ exclusion limits on and $5\sigma$
discovery sensitivities for $\Lambda$ at a collider with $\sqrt{s}>100$ TeV we again use the event selection detailed in Sec. \ref{Event Selection}.
Given this, one must take care to choose a suitable value for $m^{\text{min}}_{\bar{\mu}\mu}$ at c.o.m energies higher than $100$~TeV. Linearly scaling the $m^{\text{min}}_{\bar{\mu}\mu}=0.4$ TeV cut used in ATLAS \cite{Aaboud:2017buh, Aad:2020otl} and CMS \cite{CMS:2021ctt} with $\sqrt{s}$, as was done at $\sqrt{s}=14$ and $100$ TeV, generally results in a cut that is too strong and one ends up throwing away events that have a meaningful impact on the sensitivity. The reason for this is that as c.o.m energy increases the DY process and, to lesser extent, the EFT signal is suppressed by increasingly larger negative Sudakov Double Logarithms. This effectively shifts the dimuon distribution to lower values of $m_{\bar{\mu}\mu}$. Hence, we find that a fixed minimum cut of $m^{\text{min}}_{\bar{\mu}\mu}=3$ TeV gives the optimised sensitivity in the region of $\sqrt{s}=[100,500]$ TeV.

Furthermore, it is seen that for a collider of $\sqrt{s}\geq100$~TeV the bins that give the biggest contribution to the overall sensitivity lie approximately in the interval $m_{\bar{\mu}\mu}=[7,30]$~TeV regardless of collider energy. Within this interval, around $\sqrt{s}=250$ TeV the top and diboson backgrounds become comparable in size to the DY backgrounds and start to dominate over the DY at c.o.m energies over $\sqrt{s}=300$ TeV. 

Fig. \ref{95CL and 5sig at 100TeV as function of energy} gives the $95\%$ exclusion limits and $5\sigma$ discovery sensitivities on $\Lambda$ as a function of collider c.o.m energy with $L=40$ ab$^{-1}$. It is seen that twice the c.o.m energy is needed to exclude signals of order $\Lambda\approx40$ TeV at the $95\%$ CL. Additionally, $5$ times the center of mass energy is needed for $5\sigma$ discovery. In light of this, we consider the exclusion and discovery potential of a future collider with variable luminosity and c.o.m energy. Our findings are presented in Fig.~\ref{l vs e 95CL and 5sig}.

\section{Conclusions\label{conclusions}}

In this work we investigated the sensitivity of a future proton-proton collider to new physics indicated by the long-standing rare $B$-decay anomalies. We kept our analysis model-independent and conservative by only considering the minimal effective contact interaction required by the data, $1/\Lambda^2 (\bar b_L \gamma_\mu s_L)(\bar \mu_L \gamma^\mu \mu_L)$. Current $B$-physics data indicates $\Lambda = 39 \pm 4$~TeV, which implies that the mass of the mediators could be above $100$ TeV, potentially putting them
out of reach for direct searches even at the FCC-hh. Given this, we have derived both exclusion limits and discovery sensitivities for the contact interaction itself at $pp$ colliders at various energies and luminosities, based on the tail of the inclusive dimuon invariant mass distribution. In deriving these results, we have included NLO QCD and EW corrections to the EFT signal, as well as to the dominant background component. We validated our modeling of the SM background against the current ATLAS and CMS searches and employed an optimized binning scheme. We take into account the requirements of unitarity bounds on our event selection.

We find that the $\sqrt{s}= 14$ TeV HL-LHC with ${\cal L} = 3000$ fb${}^{-1}$
can exclude EFT signals of strength $\Lambda=4.5$ TeV at $95\%$ CL. This compares
to a limit $\Lambda=3.9$ TeV obtained in \cite{Greljo:2017vvb} for ${\cal L} = 3000$ fb${}^{-1}$ and $\sqrt{s}=13$ TeV and
entails a cancellation between the NLO-QCD and EW corrections, as well as our inclusion of detector effects. The exclusion reach increases to $\Lambda=26$~TeV when considering the $\sqrt{s}=100$~TeV FCC-hh with its projected luminosity of 40~fb$^{-1}$. The discovery potential of the FCC-hh is also significantly stronger than that of the HL-LHC, $\Lambda=20$~TeV compared to $\Lambda=3.4$ TeV. All limits are subject to the validity of the EFT expansion, which depends on the UV physics underlying the contact interactions. For example, in a $Z'$ model the dimuon mass must simply be restricted to be sufficiently below the $Z'$ mass. We quantify the reduction of $\Lambda$ for various choices of upper limits on the dimuon mass. 

Probing a value of $\Lambda\approx39$ TeV as suggested by the B-anomalies would be possible with a machine with higher energy and/or
luminosity than the FCC-hh, as we have investigated for a wide range of energies and luminosities (See Fig. \ref{l vs e 95CL and 5sig}). For example, a $\sqrt{s}=100$~TeV collider would require a luminosity of  $L\gtrapprox 770$ ab$^{-1}$, for a $95\%$ exclusion of a $\Lambda=39$ TeV signal, whereas a $pp$ machine with $L = 40$~ab$^{-1}$ would require a c.o.m energy of $\sqrt{s}\gtrapprox200$~TeV. Discovery can be achieved
with $L\gtrapprox7100$~ab$^{-1}$ at $\sqrt{s}=100$ TeV or $\sqrt{s}\gtrapprox470$ TeV at a collider with $L = 40$~ab$^{-1}$.
While these results may appear disillusioning at first, we remind the reader that we have been maximally conservative in our
signal model: a concrete BSM model will invariably bring additional interactions, even if the mediators are too heavy for direct searches.
It may moreover be possible to improve the sensitivity by considering a more complex final state
with a reduced SM background. For example, for the (HL-)LHC, the authors of \cite{Afik:2018nlr} find that an exclusive $\mu^+\mu^-b$ search can improve sensitivity to a $b s\mu\mu$ contact interaction over that of an inclusive dimuon search. As such, we consider the question of whether
a ``no-lose'' case can be constructed for a 100 TeV collider based on the rare $B$-decay anomalies an open one. A comprehensive
case would combine contact interaction searches (for heavy mediators) with direct searches (for lighter mediators). We leave such studies
for future work.

\section*{Acknowledgements}
We thank Andrea Banfi and Jonas Lindert for many fruitful discussions throughout this project. CKK thanks Riccardo Torre for a helpful discussion regarding EFT validity. We thank Yoav Afik for clarifying several points regarding \cite{Afik:2018nlr, Afik:2019htr}.
SJ acknowledges support by the UK Science and Technology Facilities Council (STFC) under Consolidated Grants ST/P000819/1 and ST/T00102X/1.
BG acknowledges support by a PhD studentship from the UK STFC and the School of Mathematical and Physical Sciences at the University of Sussex. SK acknowledges support by DOE grant DE-SC0011784. CKK acknowledges support from the Royal Society and SERB (under the Newton International Fellowship programme Grant No. NF171488) during the part of this project. 
\begin{appendices}

\paragraph{Note Added:} After submission of this paper to the arXiv, A. Greljo informed us of an ongoing study of $b s\mu\mu$ contact interactions at future colliders. 

\section{\texorpdfstring{$R_2$}{R2} terms for the NLO EFT signal}\label{NLO-signal}
Over the last decade significant advancements have been made in the automated computation of one loop amplitudes. A key observation is the fact that any one loop amplitude $\mathcal{M}$ can be written as \cite{Ossola:2008xq}
\begin{equation}
    \mathcal{M}=\sum_id_i\text{Box}_i+\sum_ic_i\text{Triangle}_i+\sum_ib_i\text{Bubble}_i+\sum_ia_i\text{Tadpole}_i+R
    \label{one loop amp}
\end{equation}
where Box, Triangle, Bubble and Tadpole are known one-loop scalar integrals and $R$ is a rational term. The OPP method \cite{Ossola:2006us, Ossola:2008xq, Mastrolia:2008jb} offers an effective way to compute the coefficients $d_i,c_i,b_i,a_i$ and has been implemented in \textsc{MadLoop} \cite{Hirschi:2011pa} as a part of the \textsc{MadGraph} \cite{Alwall:2014hca, Frederix:2018nkq} framework. Despite this, the automated computation of one loop amplitudes still needs extra information that must be added to a given model by hand. The first of these missing pieces are the UV counter terms arising from the given renormalisation procedure. The second are the so called $R_2$ terms contained within the rational term $R$ in Eq. \ref{one loop amp}. Since the operators that generate the EFT signal are non-renormalisable, we are only concerned with the $R_2$ terms. 

To evaluate loop amplitudes we use dimensional regularisation and hence work in $d=4+\varepsilon$ space-time dimensions. Here, all quantities `in the loop' such as the loop momenta, the metric and Dirac matrices are $d$-dimensional whereas all external momentum vectors only have non-zero $4$-dimensional components. In $d$-dimensions any one loop amplitude can be expressed as a sum of amplitudes of the form \cite{Ossola:2008xq}
\begin{equation}
\mathcal{M}(q)=\int\frac{d^dq}{(2\pi)^d}\frac{N(q)}{D_0 D_1...D_n}
\label{single loop amp}
\end{equation}
where
\begin{equation}
    D_i=(q+\bar{p}_i)^2-m^2_i.
\end{equation}
Here, $q$ is the loop momenta, $\bar{p}_i$ is a sum of external momenta and $m_{i}$ are the masses of the particles in the loop. 

Any $d$-dimensional space-time vector $x$ can be split into a $4$-dimensional and $\varepsilon$-dimensional part such that $x=\bar{x}+\tilde{x}$. Here a bar indicates the $4$-dim part and a tilde the $\varepsilon$-dim. Given this, the numerator in Eq. \ref{single loop amp} can be split such that \cite{Ossola:2008xq}
\begin{equation}
    N(q)=\bar{N}(\bar{q})+\tilde{N}(\tilde{q}^2,\bar{q},\varepsilon).
\end{equation}
The $R_2$ terms are the finite parts of the loop amplitude stemming from the $\varepsilon$-dimensional part of the numerator and are defined for each loop integral of the form in Eq. \ref{single loop amp} such that \cite{Ossola:2008xq}
\begin{equation}
R_2=\lim\limits_{\varepsilon\to 0}\int\frac{d^dq}{(2\pi)^d}\frac{\tilde{N}(\tilde{q}^2,\bar{q},\varepsilon)}{D_0 D_1...D_n}.
\end{equation}

When working within dimensional regularisation, a scheme for dealing with the $d$-dimensional Dirac matrices must be chosen. Given this, we work in the naive dimensional regularisation (NDR) scheme. This is done to be consistent with the default SM UFO model used by MadGraph to which we add the relevant EFT operators. In the NDR scheme, $\gamma^5$ is defined to anticommute with all of the Dirac matrices such that
\begin{equation}
    \left\{\bar{\gamma}_\mu,\gamma^5\right\}=0\quad\quad\quad\left\{\tilde{\gamma}_\mu,\gamma^5\right\}=0.
\end{equation}
Whilst computations are generally simpler in the NDR scheme compared to others, this scheme does contain algebraic inconsistencies which, at first glance, seem to be troublesome when considering the EFT signal. In particular, one of the main shortcomings of the NDR scheme is the fact that the expression $\text{tr}(\gamma_\mu\gamma_\nu\gamma_\rho\gamma_\sigma\gamma^5)$ is ambiguous. Traces of this form occur whenever a Feynman diagram contains a closed fermion loop. Whilst it is the case that the NLO EW corrections to the EFT signal do indeed involve closed fermion loops, see Tab. \ref{R2 ddmumu} and \ref{R2 ppmumu}, it is also the case that two of the space-time indices on the Dirac matrices within the trace are always contracted during the calculation. This contraction of indices leads to unambiguous results in all instances and allows consistent use the NDR scheme. 

To fix our conventions and notations we give the Feynman rules used in our NLO calculations:
\begin{equation*}
\centering
\adjincludegraphics[valign=c, scale=0.28]{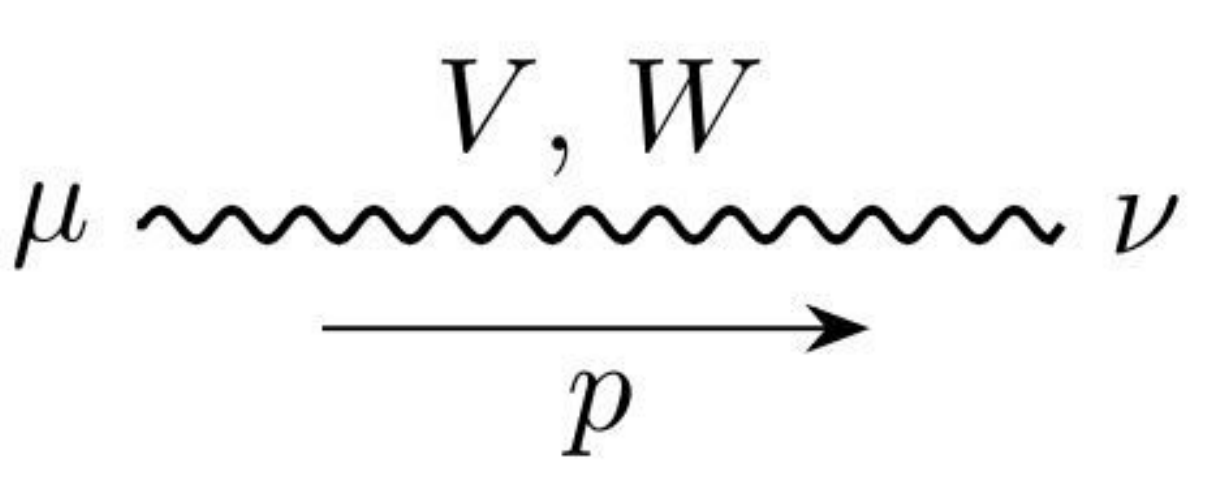}
\;=\;\dfrac{-i\eta_{\mu\nu}}{p^2-m^2_{V,W}},
\quad\quad\quad\quad
\adjincludegraphics[valign=c, scale=0.28]{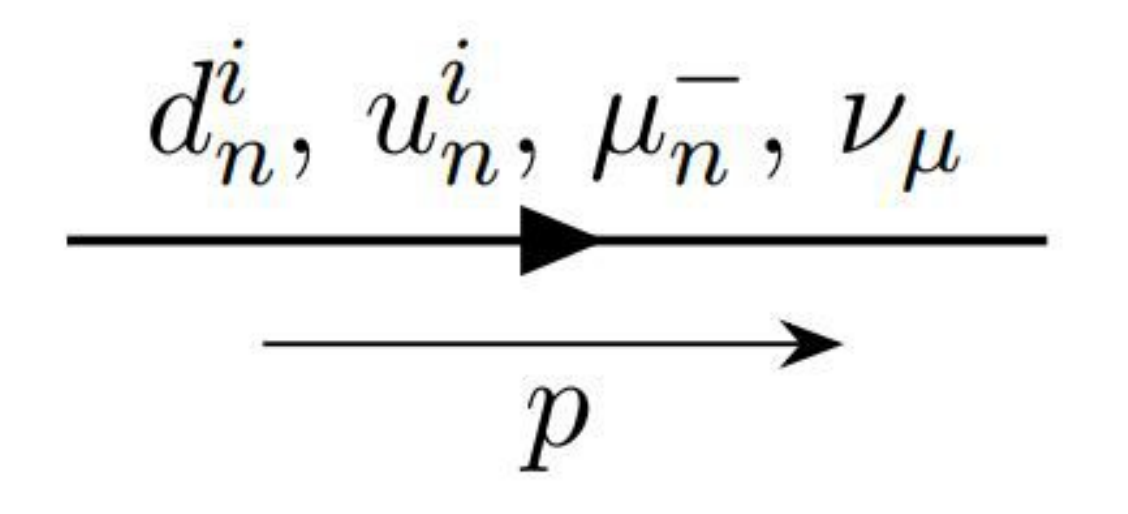}
\;=\;\frac{i\slashed{p}^2}{p^2},
\end{equation*}
\begin{equation*}
\centering
\adjincludegraphics[valign=c, scale=0.28]{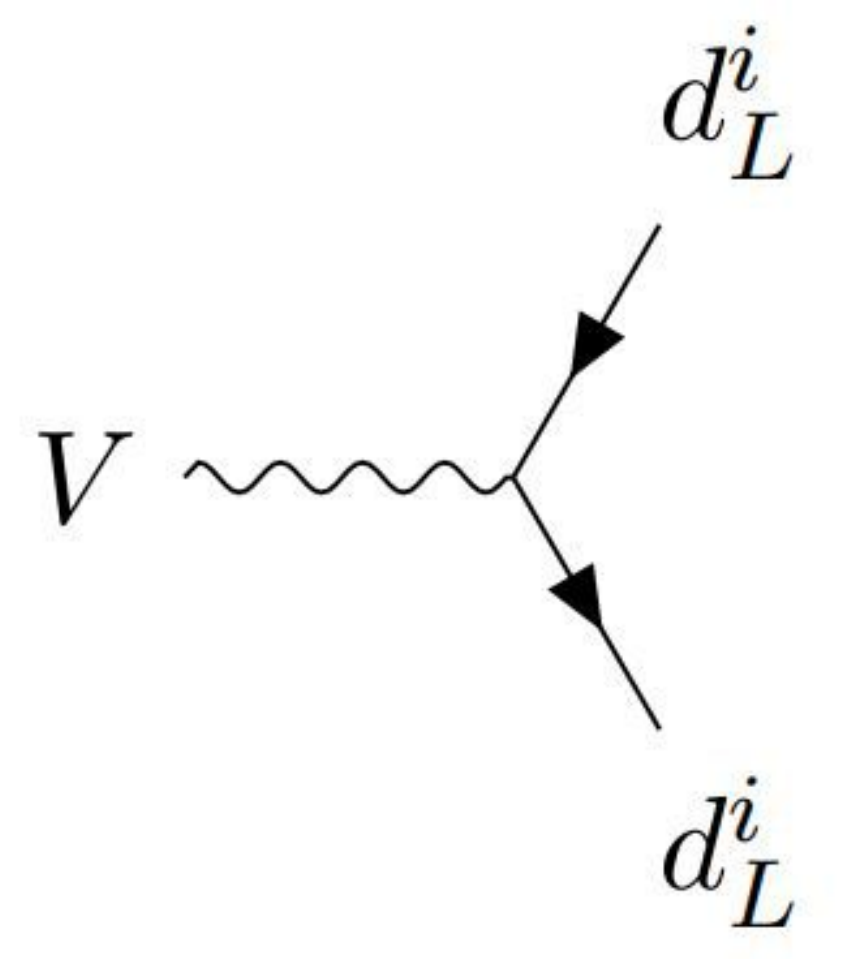}
\;=\;i\gamma_\mu(v_{d}+a_{d}\gamma^5),
\quad\quad\quad\quad
\adjincludegraphics[valign=c, scale=0.28]{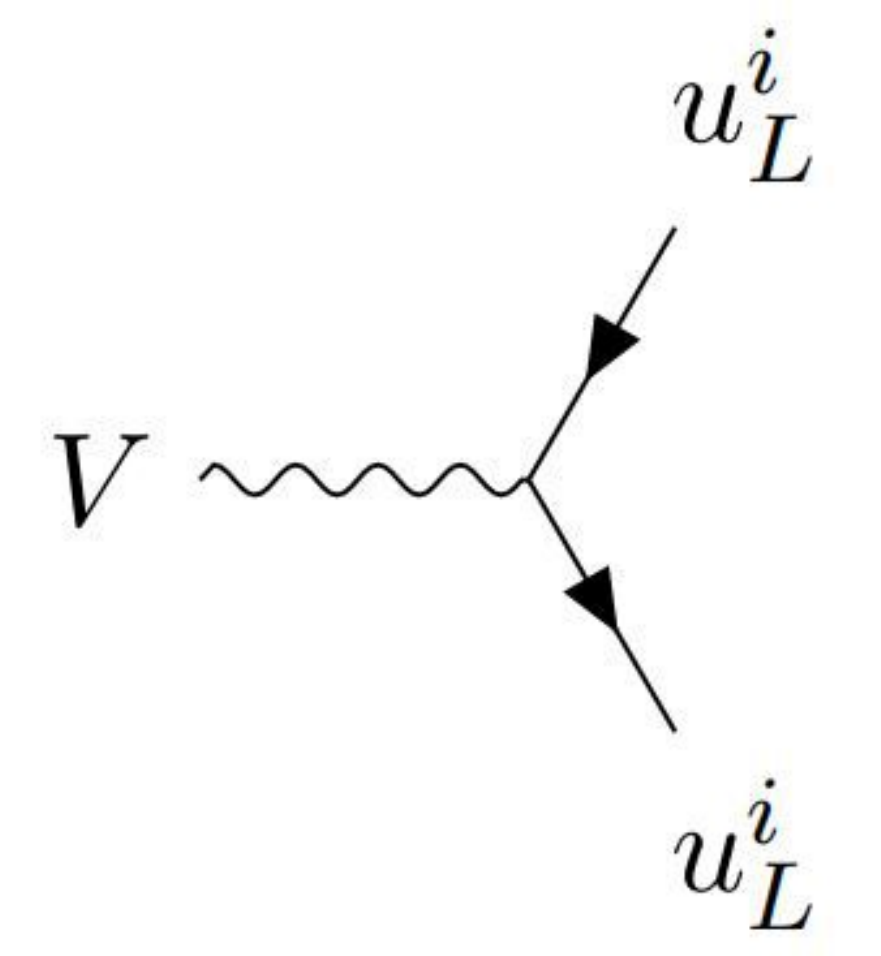}
\;=\;i\gamma_\mu(v_{u}+a_{u}\gamma^5),
\end{equation*}
\begin{equation}
\centering
\adjincludegraphics[valign=c, scale=0.28]{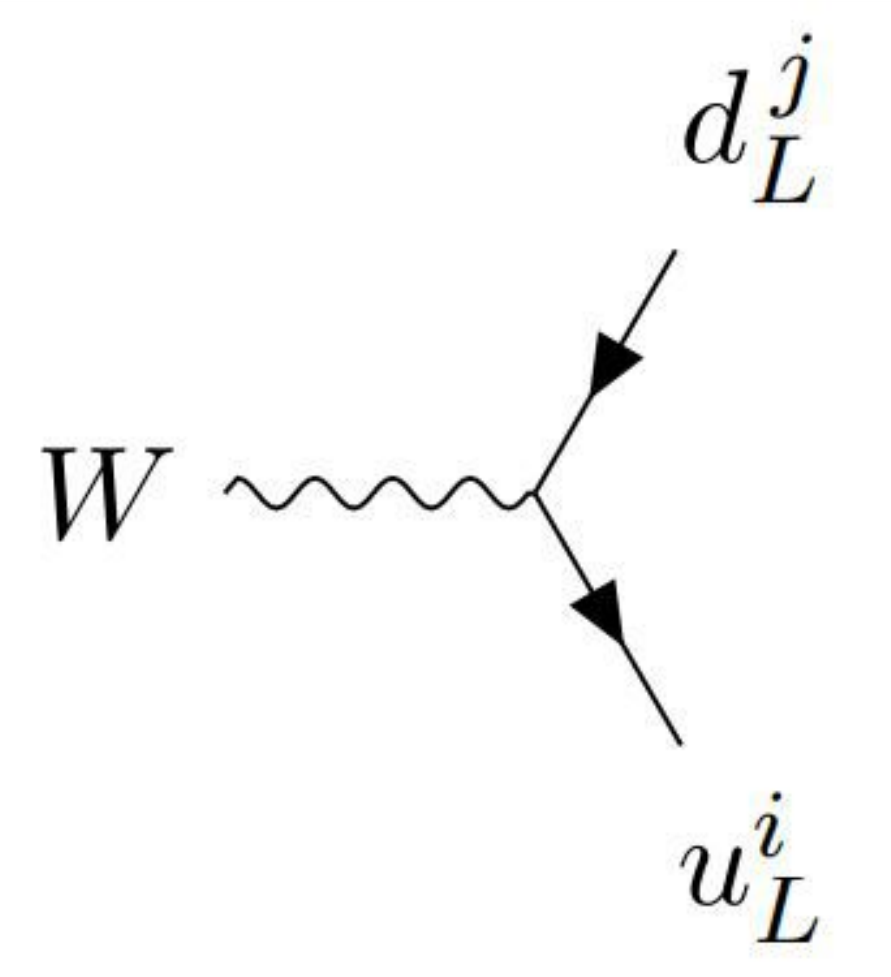}
\;=\;\frac{ie}{2\sqrt{2}\sin{\theta_W}}\gamma_\mu(1-\gamma^5)V_{ij},
\end{equation}
\begin{equation*}
\centering
\adjincludegraphics[valign=c, scale=0.28]{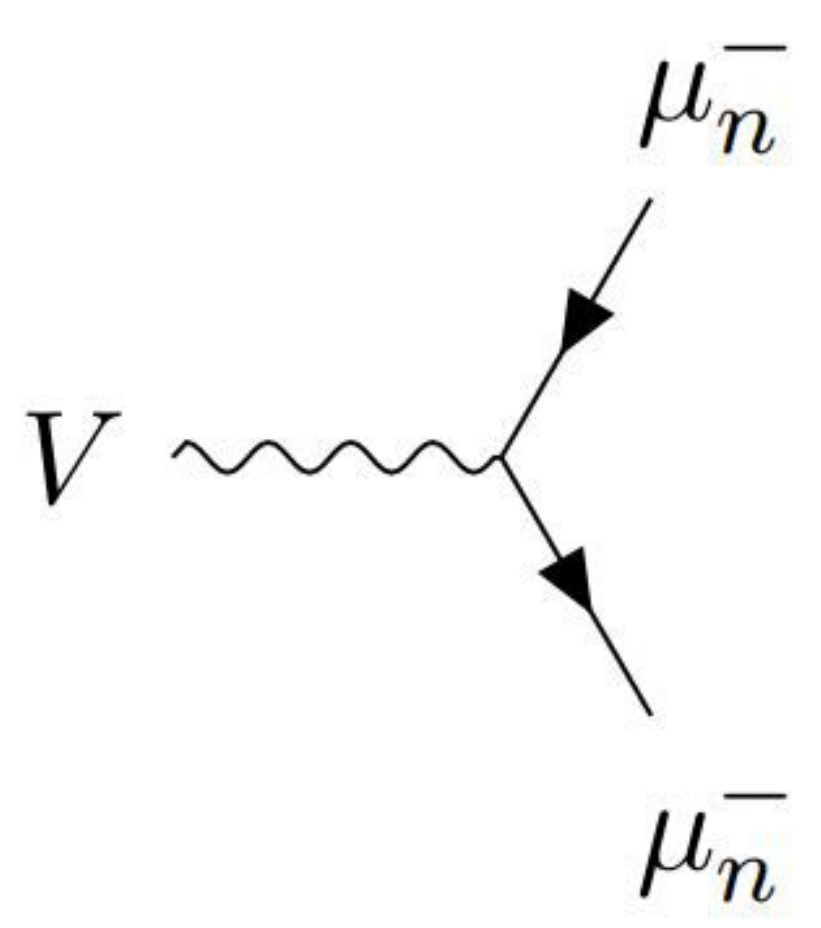}
\;=\;i\gamma_\mu(v_{\mu,n}+a_{\mu,n}\gamma^5),
\quad\quad\quad\quad
\adjincludegraphics[valign=c, scale=0.28]{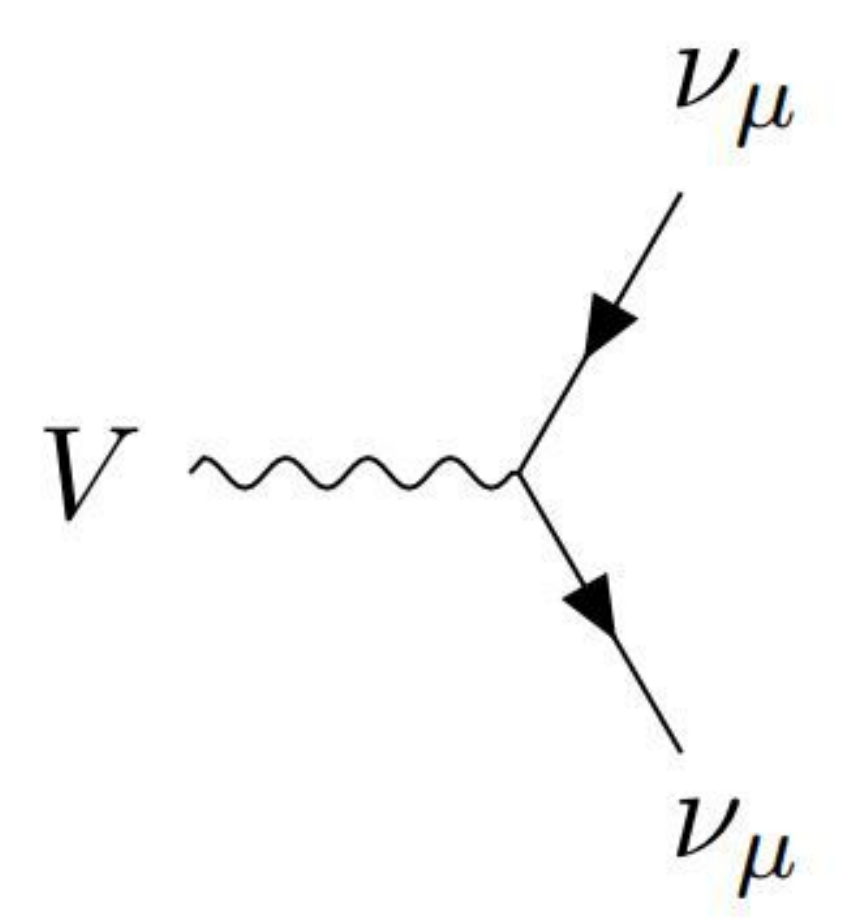}
\;=\;i\gamma_\mu(v_{\nu}+a_{\nu}\gamma^5),
\end{equation*}
\begin{equation*}
\centering
\adjincludegraphics[valign=c, scale=0.28]{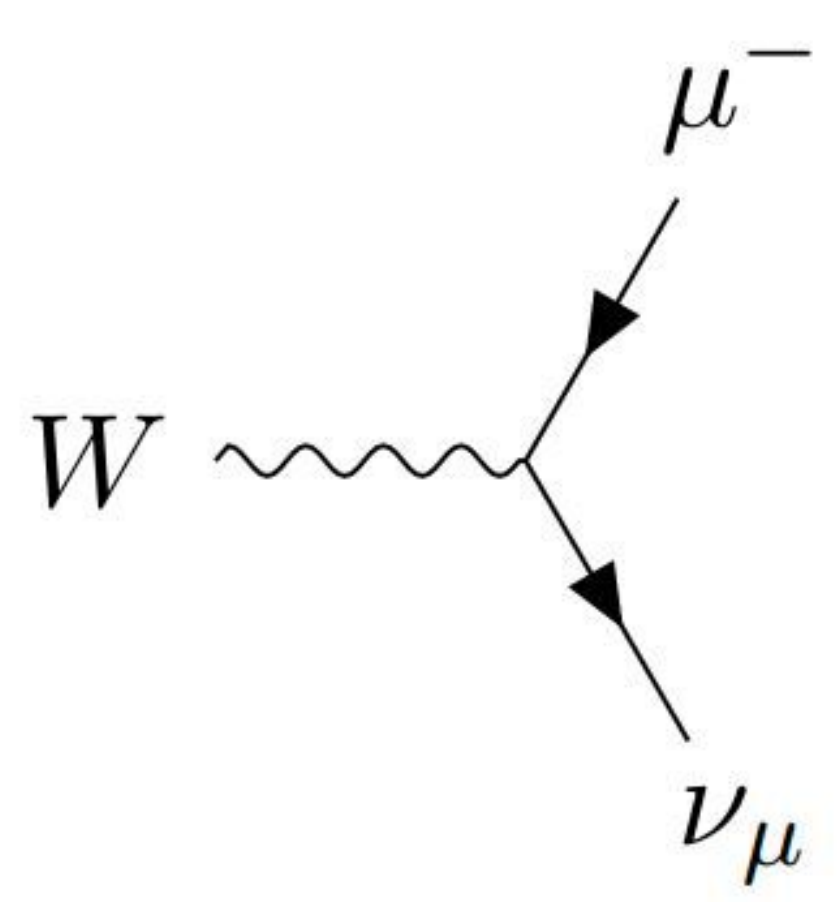}
\;=\;\frac{ie}{2\sqrt{2}\sin{\theta_W}}\gamma_\mu(1-\gamma^5).
\end{equation*}
Above, $V$ is a neutral SM gauge boson, i.e., $V=\{\gamma,g,Z\}$, $W$ is one of the charged SM $W^{\pm}$ bosons and $m_{V,W}$ is the mass of the $V$ or $W$. In regards to indices, $n=\{L,R\}$, indicating the chirality of a given fermion, and $i$ and $j$ run over the generations of quarks. It is noted that all quarks are taken to be massless except for the top quark which we do not consider here.

The LO Feynman diagram for the $\bar{d}^{\:j}_Ld^{\:i}_L\to\mu^+_L\mu^-_L$ transition is given by
\begin{equation}
\centering
\adjincludegraphics[valign=c, scale=0.28]{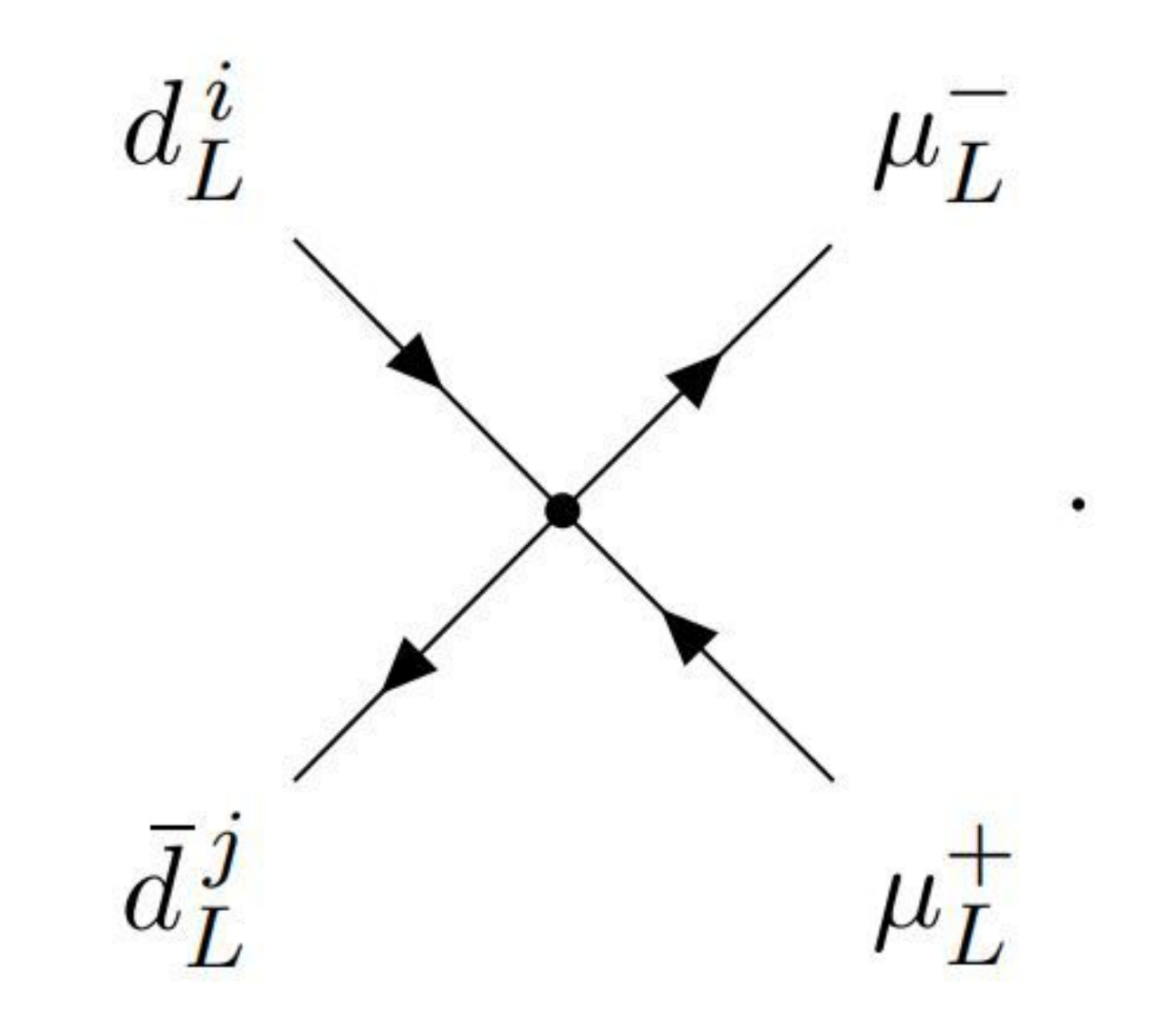}
\end{equation}  
The corresponding LO amplitude $\mathcal{M}_{\text{LO}}$ can be explicitly written as
\begin{equation}
    i\mathcal{M}_{\text{LO}}=iC^+_{ij}\bar{d}^{\:j}_L(p_2)\gamma^\nu d^{\:i}_L(p_1)\bar{\mu}_L(p_4)
    \gamma_\nu\mu_L(p_3).
\end{equation}
The one loop corrections to the EFT signal can be categorized into four types \footnote{It is noted that if $i=j$ then there are additional diagrams that are not presented here.}. The four types of diagrams and the resulting $R_2$ terms are detailed in Tab. \ref{R2 ddmumu}. It is noted however, that there are actually four equivalent diagrams of type C for both $V=\gamma$ and $V=Z$ for a given choice of $i$ and $j$. These four diagrams correspond to the four possible configurations in which $V$ exchange can take place between one of the initial state quarks and one of the final state muons. Even though only one of the four diagrams is shown here, the $R_2$ terms arising from each of these fours diagrams are equivalent.

Finally, it is noted that there are one loop EW diagrams proportional to $C^+_{sb}$ that do not contribute to $\bar{d}^{\:j}_Ld^{\:i}_L\to\mu^+_L\mu^-_L$ but still contribute to $pp\to\mu^+\mu^-$. For completeness, we give the $R_2$ terms arising from these diagrams. The first of these diagrams arises from diagrams of type~D in Tab. \ref{R2 ddmumu} in which the final state muons are right-handed. This diagram is proportional to 
\begin{equation}
    i\mathcal{M}_{\text{LR}}=iC^+_{ij}\bar{d}^{\:j}_L(p_2)\gamma^\nu d^{\:i}_L(p_1)\bar{\mu}_R(p_4)
    \gamma_\nu\mu_R(p_3).
\end{equation}
We label this a diagram of type E and its $R_2$ term is given in Tab. \ref{R2 ppmumu}.

The remaining diagrams involve initial state up-type quarks and are proportional to  either
\begin{equation}
    i\mathcal{M}^{\text{up}}=iC^+_{ij}\bar{u}^{l}_L(p_2)\gamma^\nu u^{k}_L(p_1)\bar{\mu}_L(p_4)
    \gamma_\nu\mu_L(p_3)
\end{equation}
or
\begin{equation}
    i\mathcal{M}^{\text{up}}_{\text{LR}}=iC^+_{ij}\bar{u}^{l}_L(p_2)\gamma^\nu u^{k}_L(p_1)\bar{\mu}_R(p_4)
    \gamma_\nu\mu_R(p_3).
\end{equation}
The first of these diagrams, type F, arise from the flavour changing interaction of the $W$. The remaining diagrams involve the presence of neutrinos in loops and originate from the operator structure seen in Eq. \ref{L smeft pm basis}. Again, the $R_2$ terms for all diagrams are given in Tab. \ref{R2 ppmumu}.

\begin{table}[H]
	\centering
	{\tabulinesep=1.0mm
		\begin{tabu}{|c|c|c|}
		\hline
        Type & Diagram & $R_2$ \\
        \hline 
        A &      
        \adjincludegraphics[valign=c, scale=0.28]{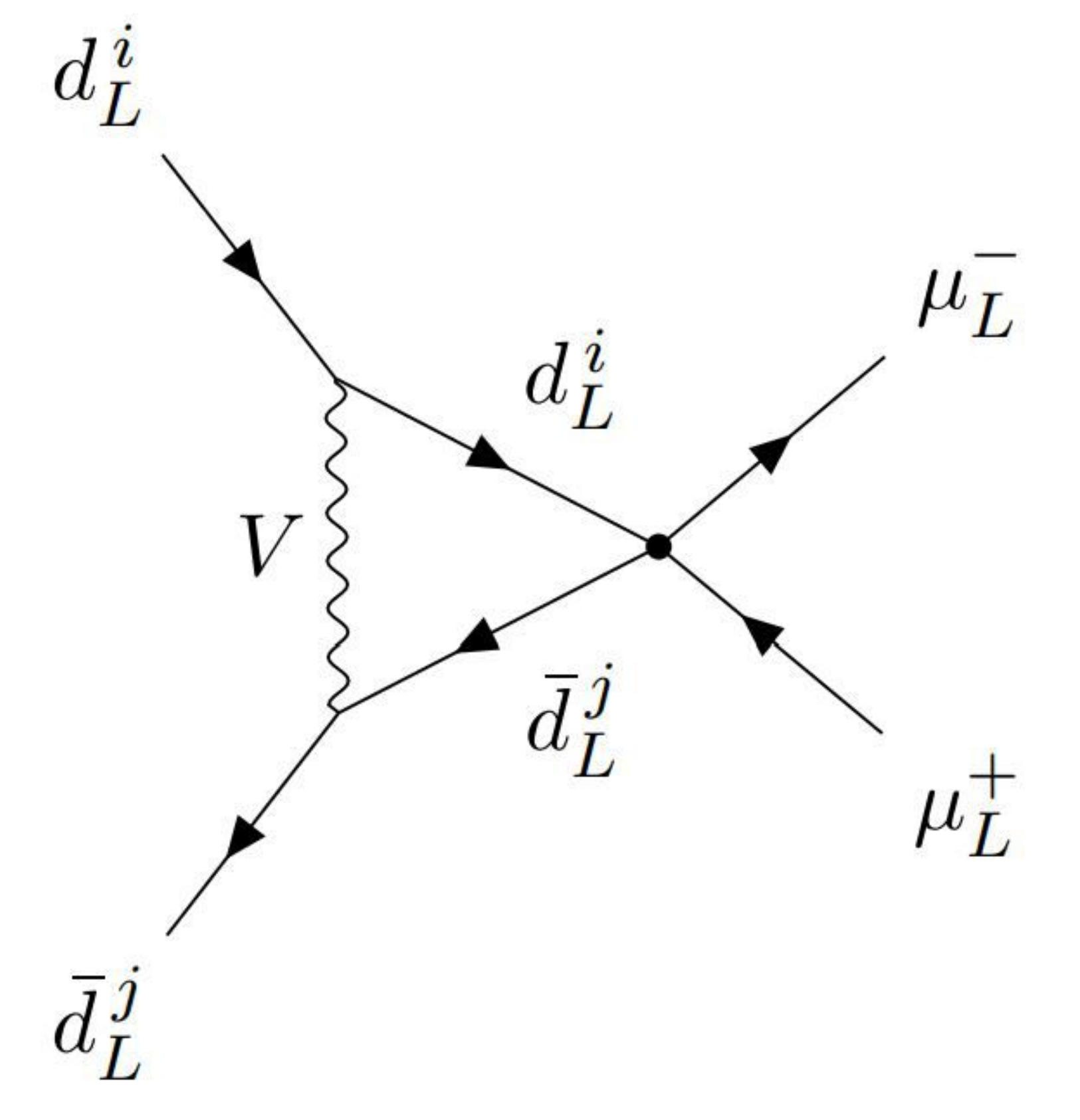}
        & 
        $\dfrac{i}{8\pi^2}(v_d-a_d)^2\mathcal{M}_{\text{LO}}$\\
        \hline
        B & 
        \adjincludegraphics[valign=c, scale=0.28]{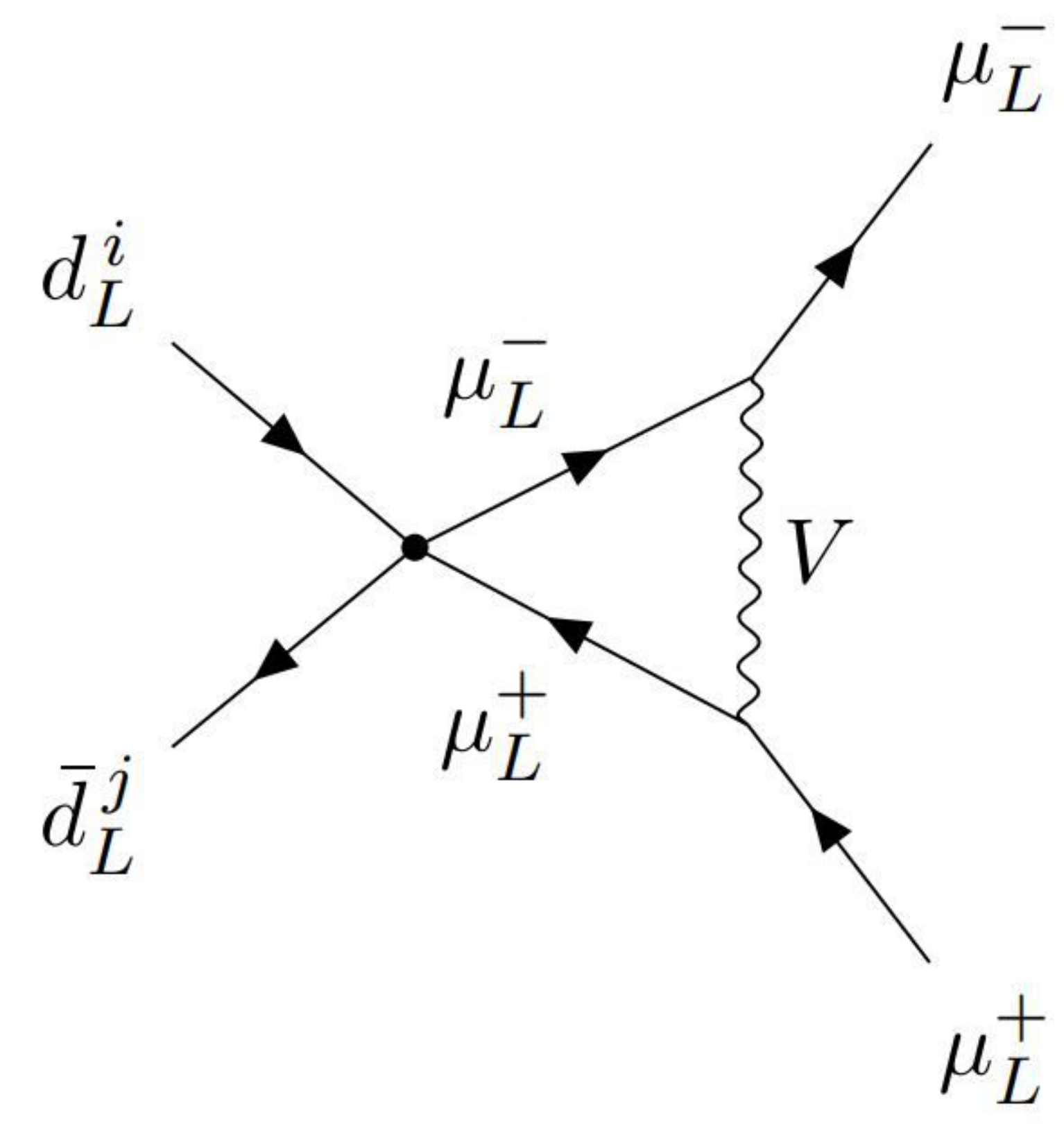}
        & 
        $\dfrac{i}{8\pi^2}(v_{\mu,L}-a_{\mu,L})^2\mathcal{M}_{\text{LO}}$\\
        \hline
        C & 
        \adjincludegraphics[valign=c, scale=0.28]{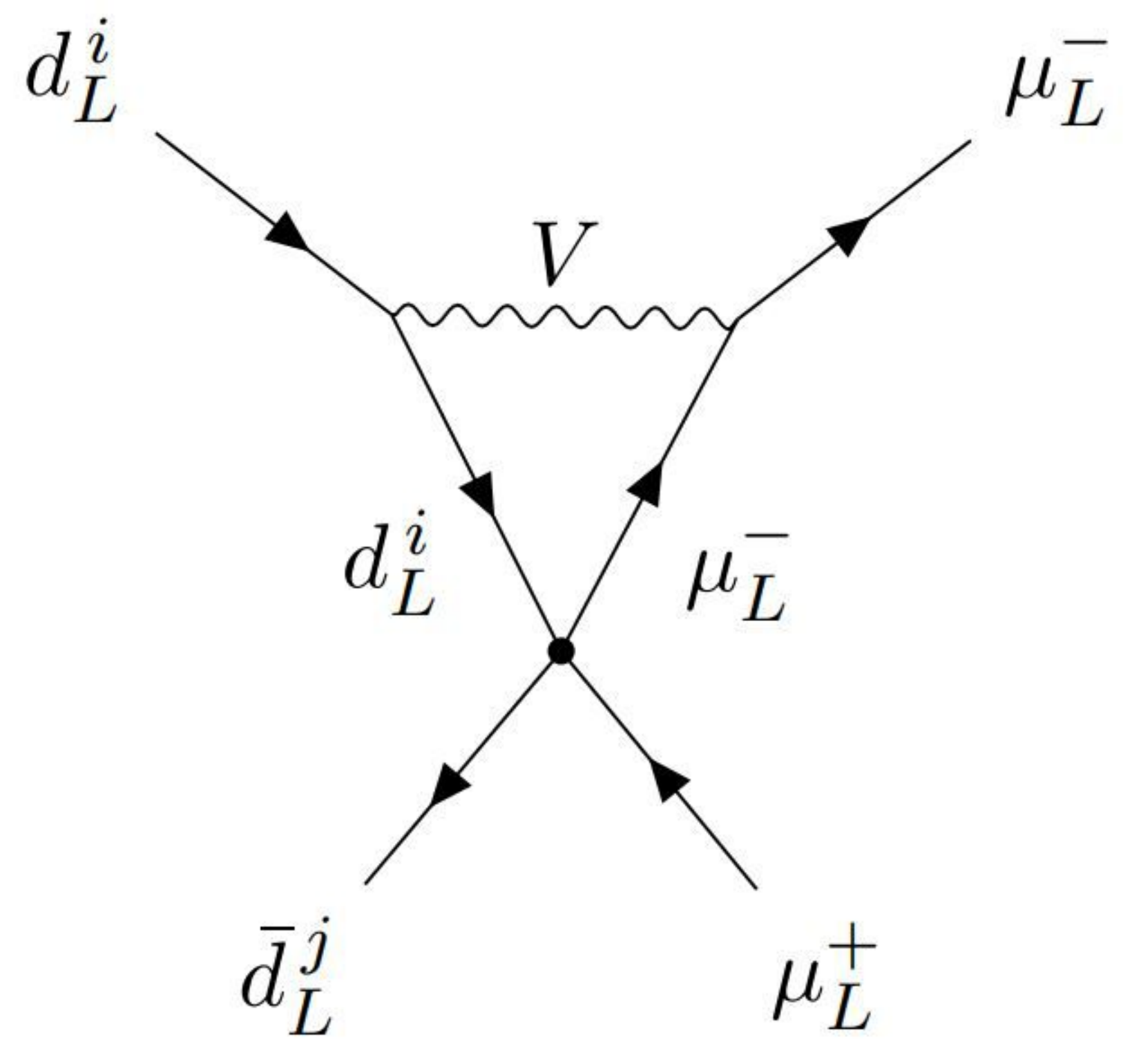}
        & 
        $\dfrac{i}{8\pi^2}(v_{\mu,L}-a_{\mu,L})(v_d-a_d)\mathcal{M}_{\text{LO}}$\\
        \hline
        D & 
        \adjincludegraphics[valign=c, scale=0.28]{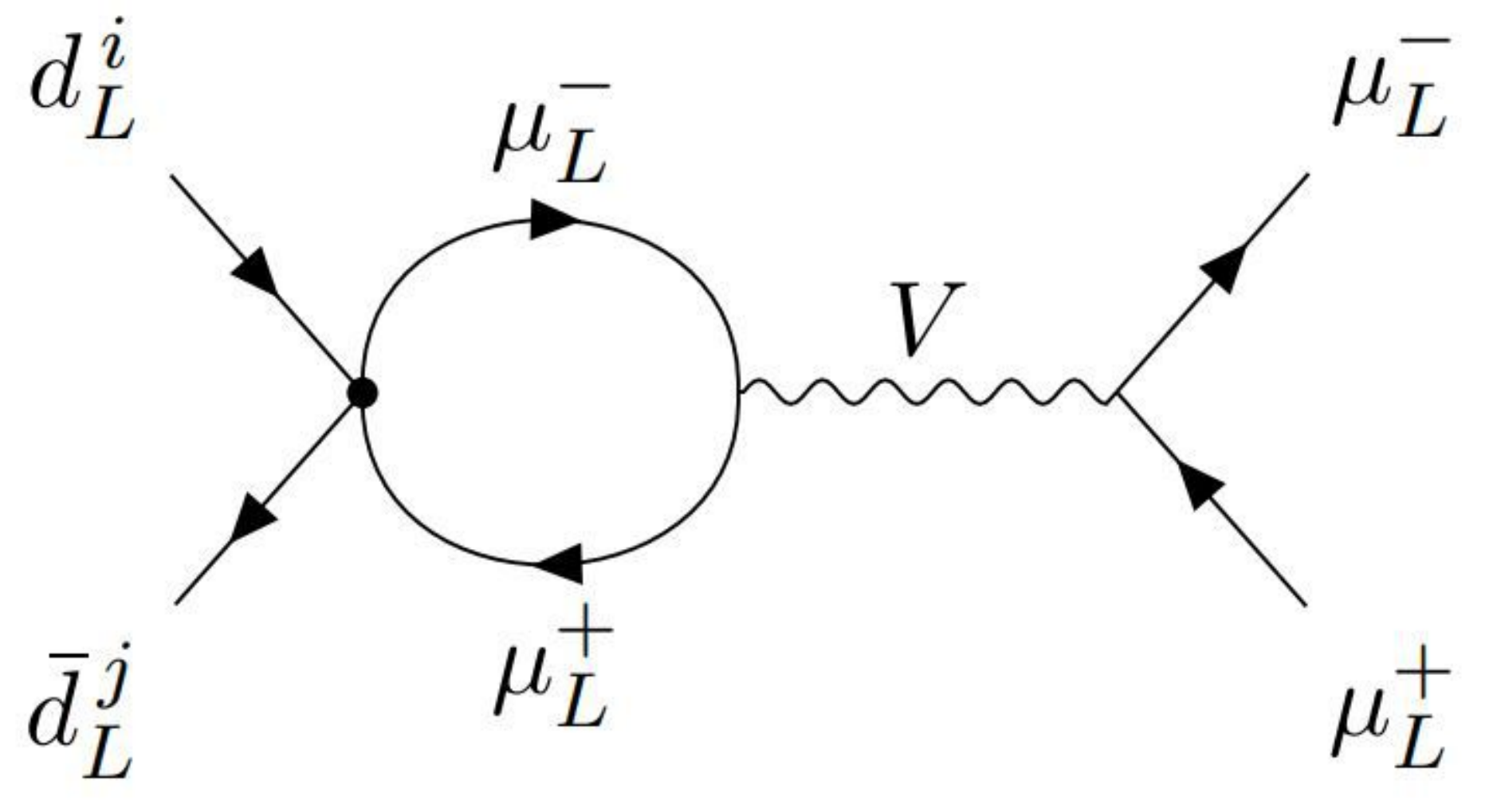}
        & 
        $\dfrac{-i}{48\pi^2}\dfrac{m_{\bar{\mu}\mu}^2}{(m_{\bar{\mu}\mu}^2-m^2_V)}(v_{\mu,L}-a_{\mu,L})^2\mathcal{M}_{\text{LO}}$\\
		\hline
	\end{tabu}}
	\caption{The one loop corrections to the EFT signal that give a contribution to the $\bar{d}^{\:j}_Ld^{\:i}_L\to\mu^+_L\mu^-_L$ scattering amplitude. The $R_2$ terms corresponding to these diagrams are given in the final column. There are four equivalent diagrams of type C but only one is shown here.}
	\label{R2 ddmumu}
\end{table}

\begin{table}[H]
	\centering
	{\tabulinesep=1.0mm
		\begin{tabu}{|c|c|c|}
		\hline
        Type & Diagram & $R_2$ \\
        \hline 
        E &      
        \adjincludegraphics[valign=c, scale=0.28]{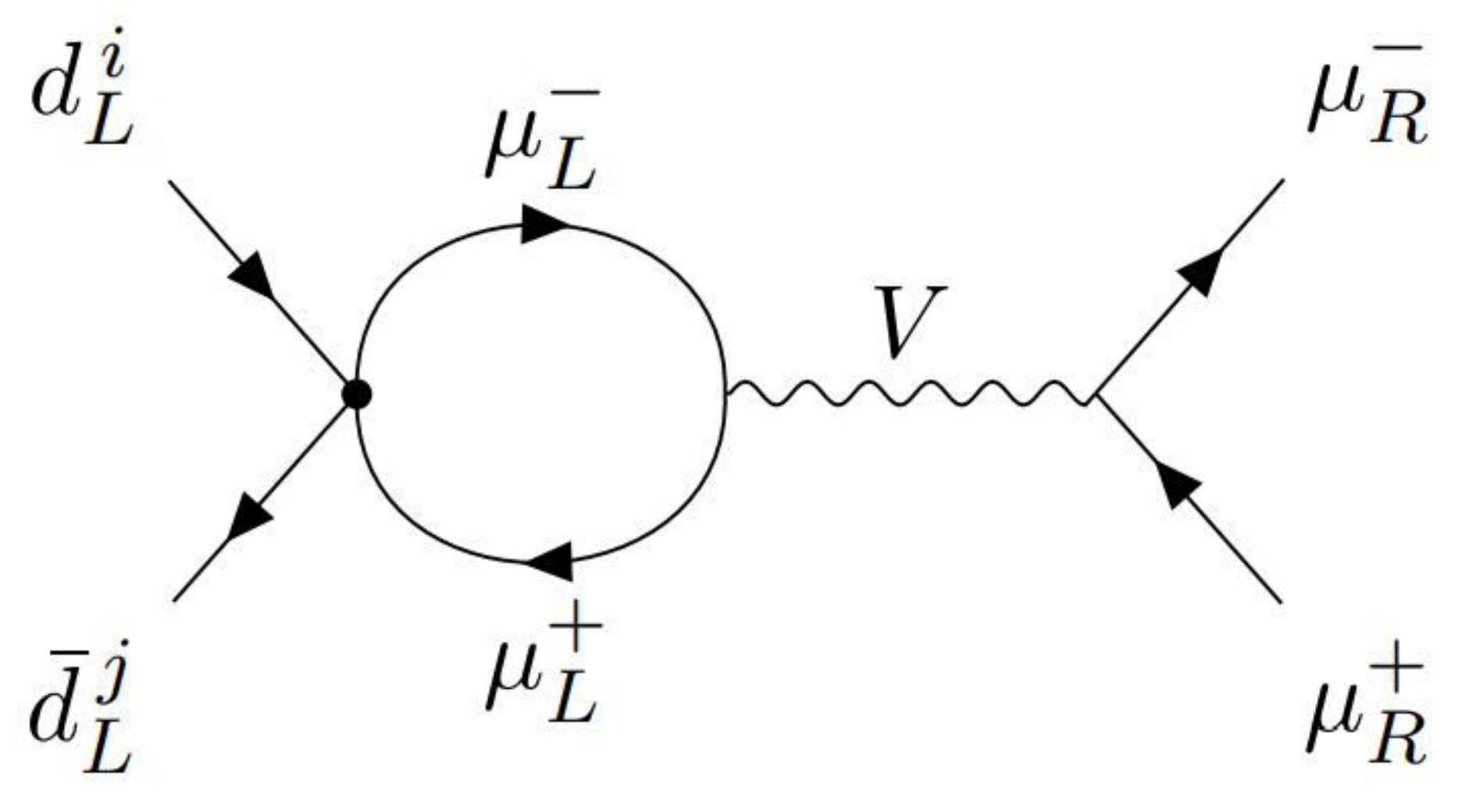}
        & 
        $\dfrac{-i}{48\pi^2}\dfrac{m_{\bar{\mu}\mu}^2}{(m_{\bar{\mu}\mu}^2-m^2_V)}v_{\mu,R}(v_{\mu,L}-a_{\mu,L})\mathcal{M}_{\text{LR}}$\\
        \hline
        F & 
        \adjincludegraphics[valign=c, scale=0.28]{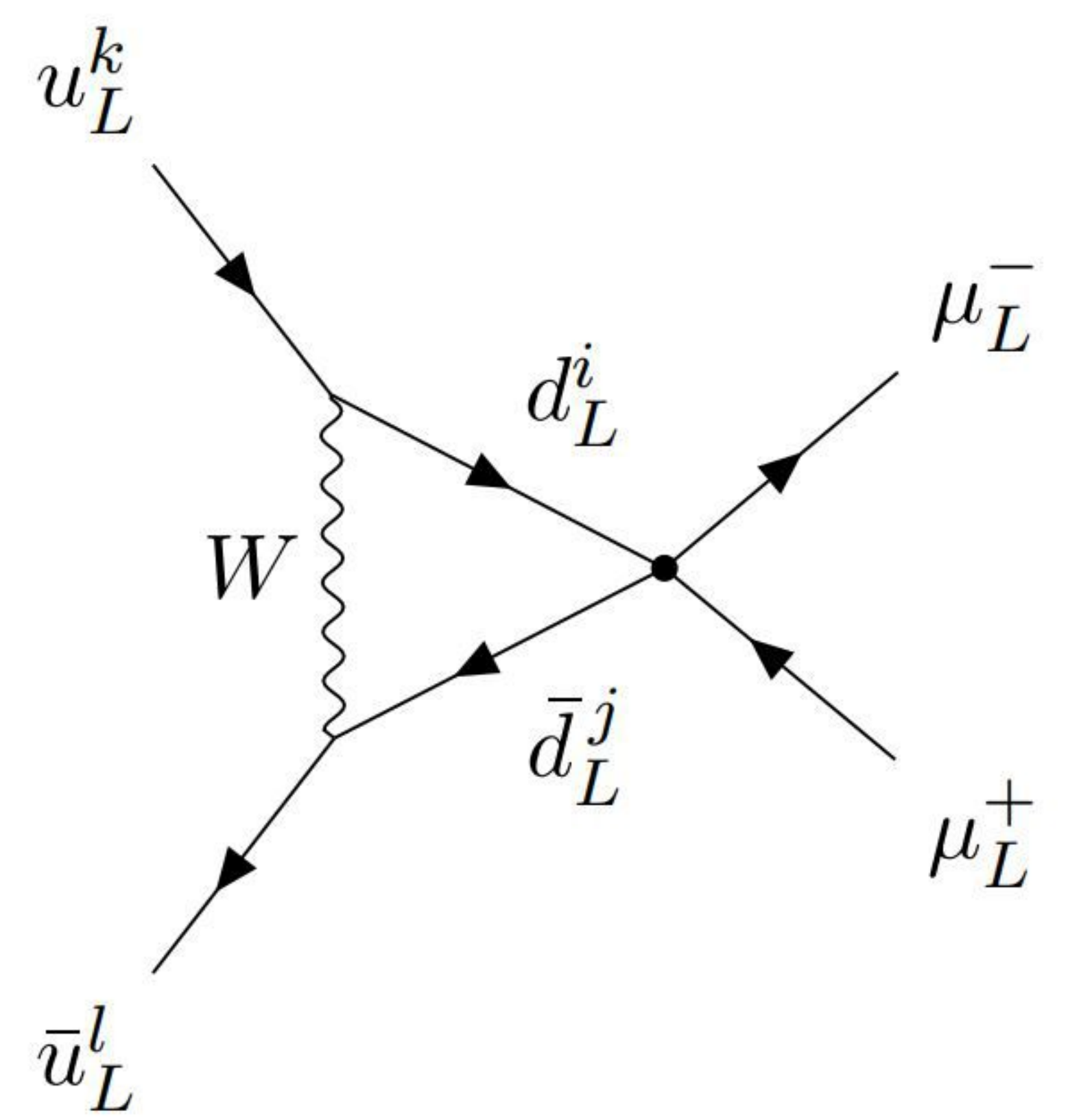}
        & 
        $\dfrac{-ie^2}{8\pi^2\sin^2\theta_W}V^*_{ki}V_{lj}\mathcal{M}^{\text{up}}$\\
        \hline
        G & 
        \adjincludegraphics[valign=c, scale=0.28]{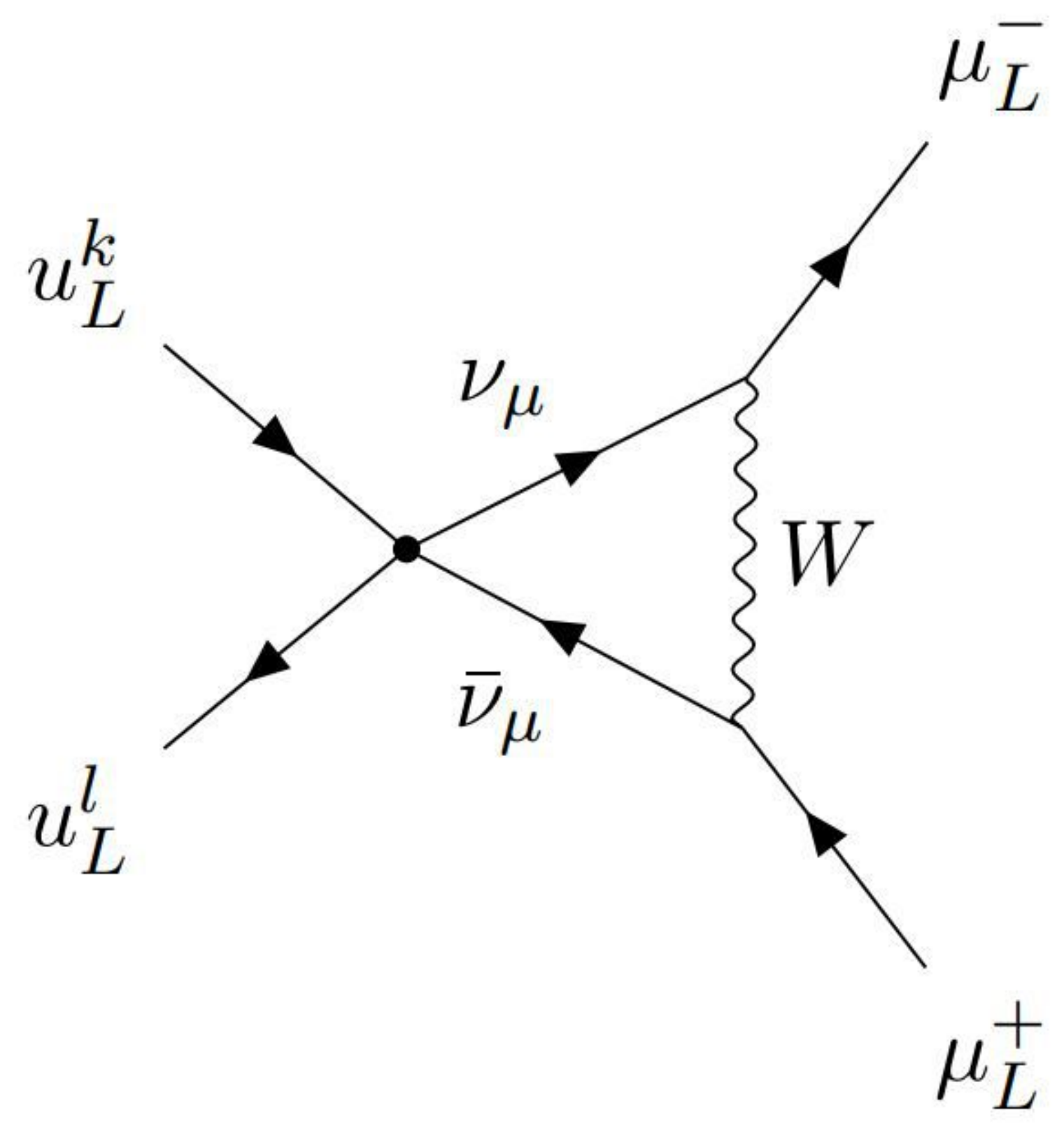}
        & 
        $\dfrac{-ie^2}{8\pi^2\sin^2\theta_W}V^*_{ki}V_{lj}\mathcal{M}^{\text{up}}$\\
        \hline
        H & 
        \adjincludegraphics[valign=c, scale=0.28]{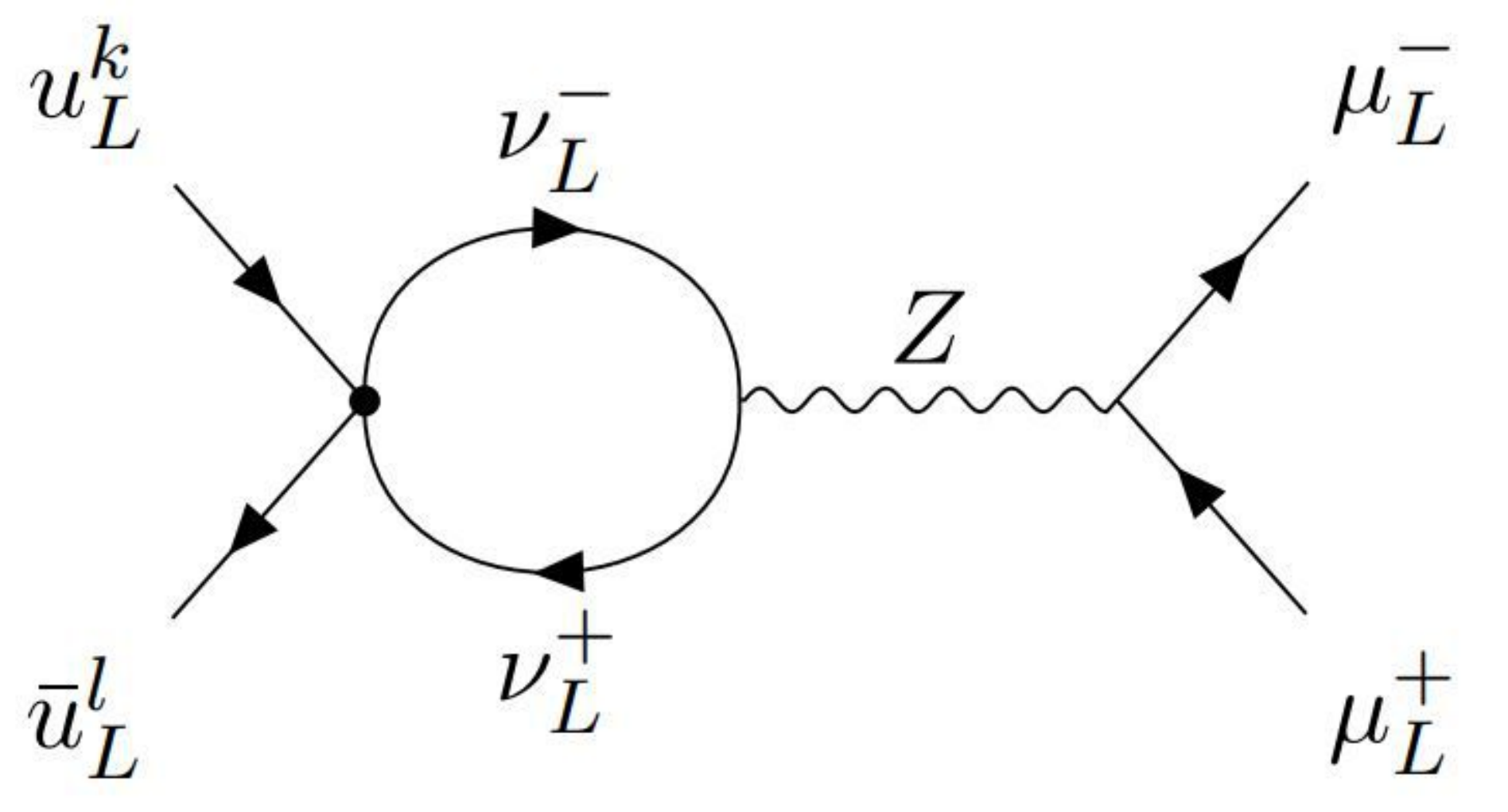}
        & 
        $\dfrac{-i}{48\pi^2}\dfrac{m_{\bar{\mu}\mu}^2}{(m_{\bar{\mu}\mu}^2-m^2_Z)}(v_{\nu}-a_{\nu})(v_{\mu,L}-a_{\mu,L})V^*_{ki}V_{lj}\mathcal{M}^{\text{up}}$\\
		\hline
		I & 
        \adjincludegraphics[valign=c, scale=0.28]{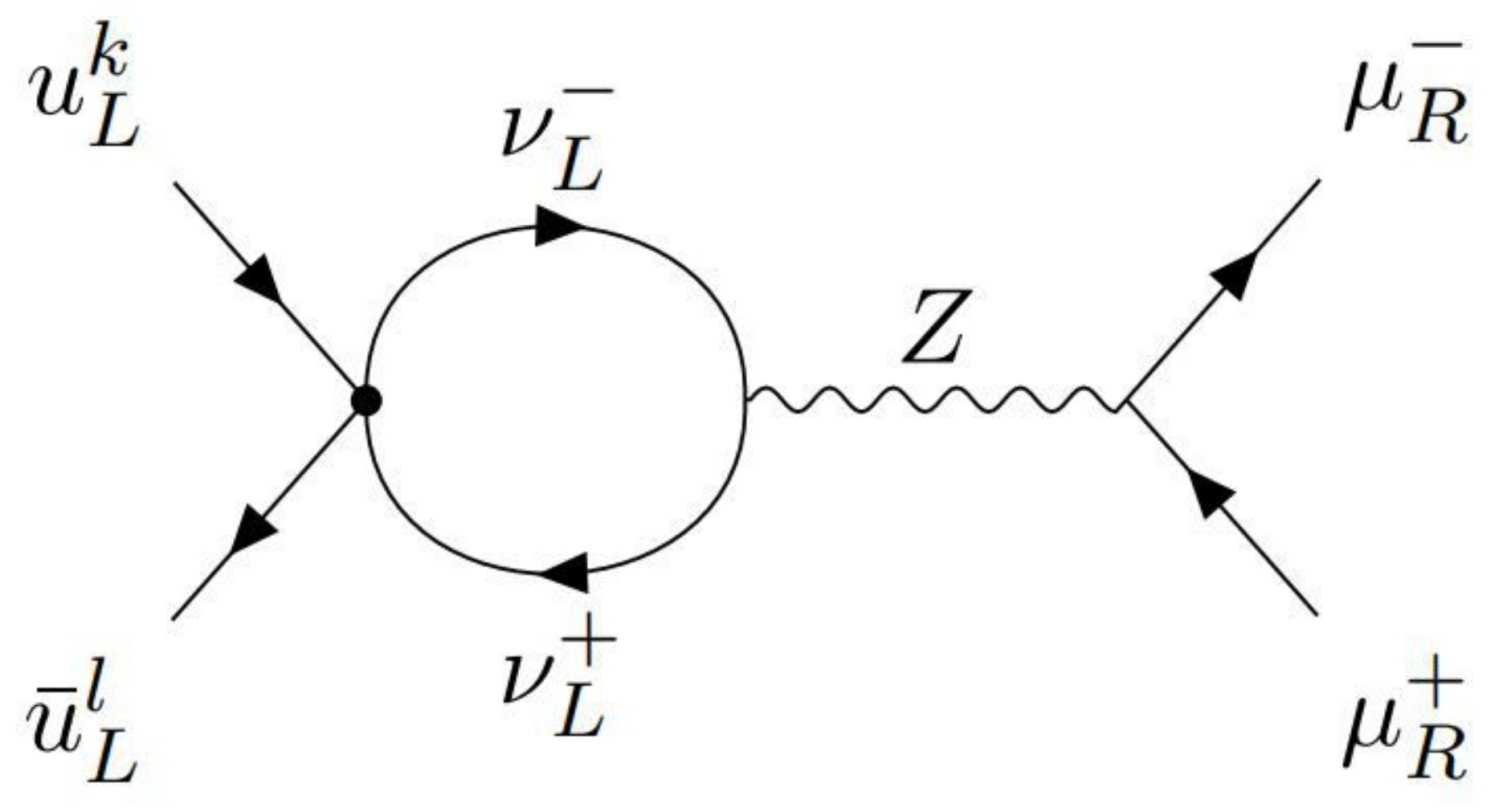}
        & 
        $\dfrac{-i}{48\pi^2}\dfrac{m_{\bar{\mu}\mu}^2}{(m_{\bar{\mu}\mu}^2-m^2_Z)}v_{\mu,R}(v_{\nu}-a_{\nu})V^*_{ki}V_{lj}\mathcal{M}^{\text{up}}_{\text{LR}}$\\
		\hline
	\end{tabu}}
	\caption{The one loop corrections to the EFT signal that give a contribution to the $pp\to\mu^+\mu$ scattering amplitude but are not proportional to $\mathcal{M}_{\text{LO}}$. The $R_2$ terms corresponding to these diagrams are given in the final column.}
	\label{R2 ppmumu}
\end{table}

\section{EFT Validity: Simplified \texorpdfstring{$Z'$}{Z'} Model} \label{appendix b}

In this section we work out the EFT expansion and partial-wave unitarity bound for a $Z'$ mediator (cf Sec. \ref{EFT Validity}).
The minimum terms needed for a $Z'$ to mediate an interaction between left-handed $b$ and $s$ quarks and muons are\begin{equation}
\mathcal{L}_{\text{int}}=g_{\mu\mu}Z'_\nu\bar{\mu}_L\gamma^\nu\mu_L-g_{sb}Z'_\nu\bar{b}_L\gamma^\nu s_L-g^*_{sb}Z'_\nu\bar{s}_L\gamma^\nu b_L.
\label{Z' lag}
\end{equation}
For simplicity, we take the couplings $g_{\mu\mu}$ and $g_{sb}$ to be real such that $g_{\mu\mu}>0$ and $g_{sb}<0$. Given Eq. \ref{Z' lag}, the LO amplitude for the process $b_Ls_L\mu^+_L\mu^-_L$ is given by
\begin{equation}
\adjincludegraphics[valign=c, scale=0.28]{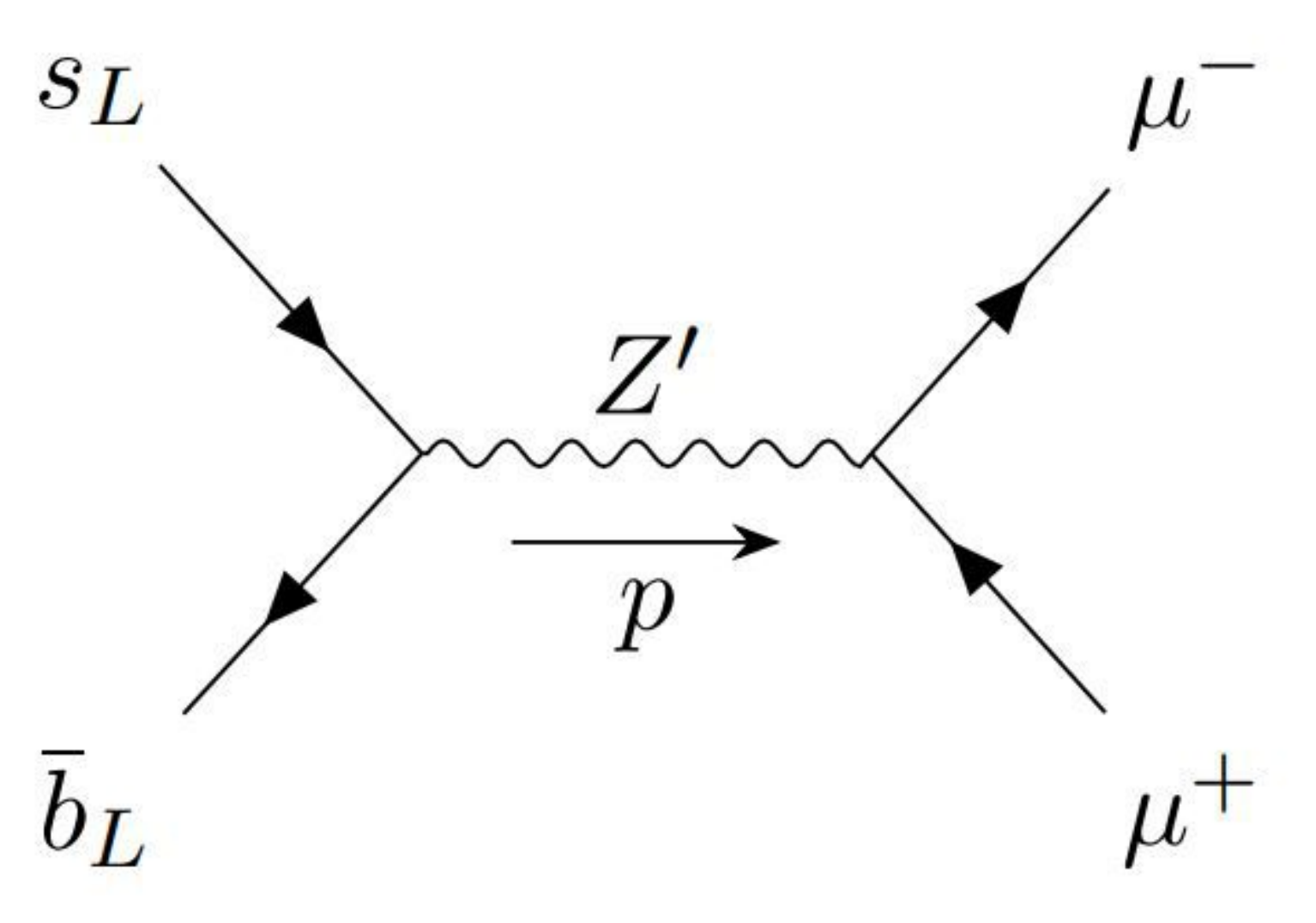}
\;=\;\frac{-ig}{p^2-m_{Z'}^2}\bar{b}_L(p_1)\gamma^\nu s_L(p_2)\bar{\mu}_L(p_3)\gamma_\nu\mu_L(p_4)
\label{Z' amp}
\end{equation}
where $g=g_{sb}g_{\mu\mu}$.  

Expanding the factor multiplying the spinor structure in Eq. \ref{Z' amp} in powers of $p^2$, we have
\begin{equation}
\frac{-ig}{p^2-m_{Z'}^2}=\frac{ig}{m_{Z'}^2}+\frac{igp^2}{m_{Z'}^4}+\frac{igp^4}{m_{Z'}^6}+\frac{igp^6}{m_{Z'}^8}+\cdots.
\label{Expansion}
\end{equation}
We see that the higher-order terms can be neglected provided $p^2 \ll m_{Z'}^2$, which considering $p^2 = m_{\bar{\mu}\mu}^2$
is equivalent to
$m_{\bar{\mu}\mu} \ll m_{Z'}$. Hence, in the region where $m_{Z'}\gg m_{\bar{\mu}\mu}$, the signal amplitude
is accurately described by the leading term, which is reproduced by the dim-$6$ effective operator in Eq. \ref{L smeft Csb}, with 
\begin{equation}
\frac{g}{m_{Z'}^2}=C_{sb}^+ = \frac{1}{\Lambda^2}.
\label{z' to eft dim6}
\end{equation}
Expressed in terms of $\Lambda$, the EFT is applicable provided $m_{\bar{\mu}\mu} \ll \sqrt{g}\Lambda$. Together with the
partial wave unitarity bound Eq. \ref{UTB Z'}, we see that $g$ cannot be greater than $2\pi/\sqrt{3}$. 

We can reproduce the higher-order terms in the expansion (Eq. \ref{Expansion}) in the EFT by matching them
to effective operators of dimension greater than $6$.
Using notation analogous to that in Sec. \ref{SMEFT}, the relevant operators at dimension-8 are  \cite{Alioli:2020kez,Murphy:2020rsh,Li:2020gnx}
\begin{equation}
\begin{split}
\mathcal{L}^\text{SMEFT}\supset&\frac{c^{(3,1)}_{Q_{ij}L_{kl}}}{\hat{\Lambda}^4}\left(\partial_\nu(\bar{Q}_i\gamma_\mu\sigma^aQ_j)\right)\left(\partial^\nu(\bar{L}_k\gamma^\mu\sigma_aL_l)\right)+\frac{c^{(1,1)}_{Q_{ij}L_{kl}}}{\hat{\Lambda}^4}\left(\partial_\nu(\bar{Q}_i\gamma_\mu Q_j)\right)\left(\partial^\nu(\bar{L}_k\gamma^\mu L_l)\right) \\
&
+\frac{c^{(3,2)}_{Q_{ij}L_{kl}}}{\hat{\Lambda}^4}(\bar{Q}_i\gamma_\mu\overleftrightarrow{\partial}^{a}_\nu Q_j)(\bar{L}_k\gamma^\mu\overleftrightarrow{\partial}^{a,\nu} L_l)+\frac{c^{(1,2)}_{Q_{ij}L_{kl}}}{\hat{\Lambda}^4}(\bar{Q}_i\gamma_\mu \overleftrightarrow{\partial}_\nu Q_j)(\bar{L}_k\gamma^\mu\overleftrightarrow{\partial}^{\nu} L_l) ,
\end{split}
\label{l smeft dim8}
\end{equation} 
and we can change to an operator basis analogous to the $(C^+_{ij},C^-_{ij})$ basis (cf Sec. \ref{SMEFT}),
\begin{equation}
\begin{split}
\mathcal{L}^{\text{SMEFT}}\supset&\;\frac{f^{1}_{ij}}{\hat{\Lambda}^4}\left[\left(\bar{d}^i_L(x_1)\gamma_\mu \left(\partial_\nu d^j_L(x_2)\right)+\left(\partial_\nu\bar{d}^i_L(x_1)\right)\gamma_\mu d^j_L(x_2)\right)+ \right.\\
& \left. \left(\bar{\mu}_L(x_4)\gamma^\mu\left(\partial_\nu\mu_L(x_3)\right)+\left(\partial_\nu\bar{\mu}_L(x_4)\right)\gamma^\mu\mu_L(x_3) \right) \right] \\
&\;+\frac{f^{2}_{ij}}{\hat{\Lambda}^4}\left[\left(\bar{d}^i_L(x_1)\gamma_\mu \left(\partial_\nu d^j_L(x_2)\right)-\left(\partial_\nu\bar{d}^i_L(x_1)\right)\gamma_\mu d^j_L(x_2)\right)+ \right.\\
& \left. \left(\bar{\mu}_L(x_4)\gamma^\mu\left(\partial_\nu\mu_L(x_3)\right)-\left(\partial_\nu\bar{\mu}_L(x_4)\right)\gamma^\mu\mu_L(x_3) \right) \right] ,
\end{split}
\label{dim 8 nb}
\end{equation}
where 
\begin{equation}
f^{k}_{ij}=c^{(1,k)}_{Q_{ij}L_{\mu\mu}}+c^{(3,k)}_{Q_{ij}L_{\mu\mu}}.
\end{equation}
The $2\to2$ scattering amplitude $\mathcal{M}$ calculated in the EFT through dimension-8 becomes
\begin{equation}
i\mathcal{M}=iA_{ij}\bar{d}^{\;i}_L(p_1)\gamma_\nu d^{\;j}_L(p_2)\bar{\mu}_L(p_3)\gamma^\nu \mu_L(p_4) ,
\end{equation}
where 
\begin{equation}
A_{ij}=\frac{g^+_{ij}}{\hat{\Lambda}^2}+\frac{f^1_{ij} s}{\hat{\Lambda}^4}+\frac{f^2_{ij}(-s-2t)}{\hat{\Lambda}^4}.
\label{A_ij}
\end{equation}
Setting $\hat{\Lambda} = \Lambda$ and comparing to Eqs.~\ref{Z' amp} and \ref{Expansion}, we obtain
$g^+_{sb}=f^1_{sb}=g$ and $f^2_{sb}=0$, for which
\begin{equation}
i\mathcal{M}=\left(\frac{ig}{m_{Z'}^2}+\frac{ig m^2_{\bar{\mu}\mu}}{m^4_{Z'}}\right)\bar{b}_L(p_1)\gamma_\nu s_L(p_2)\bar{\mu}_L(p_3)\gamma^\nu \mu_L(p_4)
\end{equation}
reproduces Eq.~\ref{Z' amp} through the second term in the expansion Eq.~\ref{Expansion}. To conclude this section, let us note that
Eq.~\ref{A_ij} can capture the case of a leptoquark mediator, too, in which case $f^1_{sb} = f^2_{sb}$, such that the dimension-8 term is
proportional to $t$, reflecting the fact that the mediator is in the $t$-channel. (In this case, the EFT is valid provided
$|t| < m_{\rm LQ}^2$, which is not a simple cut on the dimuon mass.) This gives a concrete demonstration of
the universal applicability of the EFT, which further extends to loop-level mediators.

\end{appendices}

\bibliographystyle{unsrturl}
\bibliography{bphysrefs}

\end{document}